\documentclass[]{aa}

\usepackage[utf8]{inputenc}
\usepackage{graphicx}
\usepackage{hyperref}
\hypersetup{colorlinks=false, urlcolor=blue, linkcolor=red}
\usepackage{xcolor}
\usepackage{longtable}
\usepackage[para]{threeparttable}
\usepackage{booktabs}        
\usepackage{comment}
\usepackage{chngcntr} 
\usepackage{pdflscape}

\newcommand{\mum}{$\mu$m}
\newcommand{\eg}{e.g.,~}
\newcommand{\ie} {i.e.,~}
\newcommand{\kms}{~km~s$^{-1}$}

\newcommand{\jybeam}{Jy\,beam$^{-1}$}
\newcommand{\mjybeam}{mJy\,beam$^{-1}$}

\newcommand{\msun}{M$_{\odot}$~}

\newcommand{\aka}{a.k.a.~}
\newcommand{\sigb}{$\Sigma_{B}$}

\title{Magnetic threads and gravity: ALMA Observations of IRDC G14.225-0.506}
\titlerunning{B-fields and Gravity in the IRDC G14.2}

\author{
Nacho A\~nez-L\'opez \inst{1,2,3} 
\and Gemma Busquet \inst{5,6,4} 
\and Josep Miquel Girart \inst{3,4} 
\and Junhao Liu \inst{7}  
\and Qizhou Zhang\inst{8} 
\and Patrick M. Koch \inst{9}
\and Anaëlle Maury \inst{3}
\and Hauyu Baobab Liu \inst{10} 
\and Zhi-Yun Li  \inst{13} 
\and Keping Qiu \inst{11,12} 
\and Shanghuo Li \inst{11,12} 
\and Huei-Ru Vivien Chen \inst{14}
\and Ya-Wen Tang \inst{9} 
\and Shih-Ping Lai \inst{14}
\and Ramprasad Rao \inst{8} 
\and Paul Ho \inst{9}
}

\institute{
Institut de radioastronomie millimétrique (IRAM), 300 rue de la piscine, 38406 Saint Martin d’Hères, France
\and Universit\'e Paris-Saclay, Universit\'e Paris Cit\'e, CEA, CNRS, AIM, 91191, Gif-sur-Yvette, France 
\and Institut de Ci\`encies de l'Espai (ICE, CSIC), Can Magrans s/n, E-08193 Cerdanyola del Vall\`es, Catalonia 
\and Institut d'Estudis Espacials de Catalunya (IEEC), Campus del Baix Llobregat—UPC, Esteve Terradas 1, E-08860 Castelldefels, Catalonia, Spain 
\and Departament de Física Quàntica i Astrofísica (FQA), Universitat de Barcelona (UB), Martí i Franquès 1, E-08028, Barcelona, Catalonia, Spain
\and Institut de Ciències del Cosmos (ICCUB), Universitat de Barcelona, Martí i Franquès 1, E-08028 Barcelona, Catalonia, Spain
\and National Astronomical Observatory of Japan, 2-21-1 Osawa, Mitaka, Tokyo 181-8588, Japan
\and Center for Astrophysics $\vert$ Harvard \& Smithsonian, 60 Garden Street, Cambridge, MA, 02138, USA
\and Academia Sinica, Institute of Astronomy and Astrophysics, Taipei 10617, Taiwan
\and Department of Physics, National Sun Yat-sen University, No. 70, Lien-Hai Road, Kaohsiung City 80424, Taiwan, R.O.C.
\and School of Astronomy and Space Science, Nanjing University, Nanjing 210023, P. R. China
\and Key Laboratory of Modern Astronomy and Astrophysics (Nanjing University), Ministry of Education, Nanjing 210023, P. R. China
\and Astronomy Department, University of Virginia, Charlottesville, VA 22904, USA 
\and Institute of Astronomy \& Department of Physics
Institute of Astronomy and Department of Physics, National Tsing Hua
University, Hsinchu 300044, Taiwan
}

\begin{document}

\abstract
{In the star formation process, the interplay between gravity, turbulence, and magnetic fields is significant, with magnetic fields apparently serving a regulatory function by opposing gravitational collapse. Nonetheless, the extent to which magnetic fields are decisive relative to turbulence and gravity, in what environments, and under which conditions, remains uncertain.}{This study aims to ascertain the role of magnetic fields in the fragmentation of molecular clouds into clumps, down to core scales.}{We examine the magnetic field as observed with ALMA at core scales (approximately 10000~AU/0.05~pc) towards the infrared dark cloud G14.225-0.506, focusing on three regions with shared physical conditions, and juxtapose it with prior observations at the Hub-filament system scale (approximately 0.1~pc).}{Our findings indicate a similar magnetic field strength and fragmentation level between the two hubs. However, distinct magnetic field morphologies are identified across the three regions where polarized emission is detected. In the region N (northern Hub), the large-scale magnetic field, which is perpendicular to the filamentary structure, persists at smaller scales in the southern half but becomes distorted near the more massive condensations in the northern half. Notably, these condensations exhibit signs of impending collapse, as evidenced by supercritical mass-to-flux values. In the region S (southern Hub), the magnetic field is considerably inhomogeneous among the detected condensations, and we do not observe a direct correlation between the field morphology and the condensation density. Lastly, in an isolated dust clump located within a southern filament of the northern hub, the magnetic field aligns parallel to the elongated emission, suggesting a transition in the field geometry.}
{The magnetic field shows a clear evolution with spatial scales. We propose that the most massive condensations detected in the northern Hub are undergoing gravitational collapse, as revealed by the relative significance of the magnetic field and gravitational potential (\sigb~) and mass-to-flux ratio. The distortion of the magnetic field could be a response to the flow of material due to the collapse.}

\maketitle
\nolinenumbers

\section{Introduction}

The evolution of filamentary molecular clouds, the site of formation of massive stars, is a complex process involving gravitational force, turbulence, and magnetic fields. These processes operate across a vast range of spatial and density scales. Among the key physical agents involved, the magnetic field plays a significant role in the formation and fragmentation of filaments, as well as in the collapse of dense cores, and hence it is an essential ingredient to develop a comprehensive model of star formation \citep{Maury2022,Pattle2023}.  

Several models have been proposed to explain the collapse and fragmentation process of molecular clouds in the frame of high-mass star formation. 
For example, massive stars form in massive and turbulent cores according to \cite{McKee&Tan2003}; also, turbulence is what triggers fragmentation in the frame of the inertial inflow model \citep{Padoan2020}. 
Star mergers have also been proposed as a possible explanation \citep{Bonnel1998}.
Another possible scenario is Global Hierarchical Collapse (GHC) \citep{Vazquez-Semadeni2019}, where the cloud would collapse into filaments that would channel the material toward hubs, sites where the filaments converge. 
In addition, \cite{Gomez2018} suggests an MHD model and explores the morphology of the magnetic field. The authors predict a filament evolution very similar to that of the GHC model, where the magnetic field presents a U-shaped due to the accretion towards and along the filament.

In order to measure the magnetic field in the interstellar medium, we must resort to the interaction of the magnetic field with its environment \citep{Hildebrand2000}.
It is thought that interstellar grains are partially aligned with the magnetic field \citep{Davis1951}, specifically with their major axis perpendicular to it, producing linearly polarized thermal emission perpendicular to the magnetic field direction. 
In addition, several techniques have been developed to estimate the magnetic field strength from polarized dust thermal emission, such as the Davis-Chandrasekhar-Fermi (DCF) method \cite{Davis1951, ChandrasekharFermi1953} and the Intensity Gradient (IG) techniques \cite{Koch2012}.

Over the past decade, numerous observational campaigns have been conducted, targeting both low- and high-mass sources \citep[\eg][]{Zhang2014, Palau2021}. In parallel, studies have focused on filamentary regions across a range of spatial scales—from several parsecs down to 0.01~pc—such as those toward NGC\,6334 \citep{Li2015, Cortes2021, Arzoumanian2021, Wu2024A}, G34.43$+$0.24 \citep{Tang2019}, and SDC18.624-0.070 \citep{Lee2025}.

Large surveys have also explored the role of magnetic fields in the fragmentation process, revealing a tentative correlation between fragmentation and the mass-to-flux ratio \citep{Palau2021, Huang2025}. This result further supports the idea that magnetic fields play a significant role in the fragmentation process.

Understanding the evolution of magnetic fields across spatial scales—from molecular clouds to dense cores—is essential for building a comprehensive picture of star formation. Various studies have aimed to trace this evolution \citep[\eg][]{Wu2024A}. For example, \cite{Pillai2020} observed a change in magnetic field orientation: from perpendicular to the filament in low-density regions to parallel in high-density regions, a scheme similar to the one proposed by \cite{Gomez2018}. A similar transition was reported by \cite{Kwon2022} in the Serpens Main molecular cloud. At even smaller scales, \cite{Koch2022ApJ} resolved a network of dust lanes and streamers connecting dense cores, finding that the magnetic field is generally aligned with these structures.

In this context, we focus on the IRDC G14.225$-$0.506 (hereafter IRDC G14.2), located southwest of the \ion{H}{II} region M17. 
While its distance was initially estimated at 1.98 kpc \citep{Xu2011}, more recent Gaia DR2 data suggest a range of 1.49–1.57~kpc \citep{Zucker2020}. 
For the present work, we adopted a distance of 1.6~kpc as in \cite{DiazMarquez2024}.
The IRDC G14.2 exhibits a prominent network of filamentary structures with two main hubs, named Hub-N and Hub-S.
\citep{Busquet2013, Chen2019}, both associated with a rich population of protostars and young stellar objects (YSOs) \citep{PovichWhitney2010, Povich2016, Ohashi2016, Busquet2016,
DiazMarquez2024, Zhao2025arXiv250216413Z}.
In addition, \cite{Chen2019} identified a third hub candidate, referred to as Hub-C, located at the intersection of two filamentary structures.
These hubs are warmer, denser, and more massive than the surrounding filaments, and they exhibit larger velocity dispersions \citep{Busquet2013}.

Previous polarimetric observations towards the region include the work of \cite{Santos2016}. They carry out optical and near-infrared polarimetric observations to map the magnetic field surrounding the filamentary structures. 
They find the magnetic field morphology in the POS to be perpendicular to both the large-scale molecular cloud and the filaments, deriving the magnetic field strength, Alfvén Mach numbers, and mass-to-flux ratio that point out toward a sub-Alfvénic and supercritical regime, suggesting that magnetic fields cannot prevent gravitational collapse. 

\cite{Busquet2016} presents 1.3~mm continuum emission observation toward G14.2 
carried out with the Submillimeter Array with angular resolution in the range (1''—3'') as well as 870~\mum~(APEX) 
at 18''.6 and 350~\mum~ (Caltech Submillimeter Observatory, CSO) ones at 9''.6 resolution. 
They find that while Hub-N, being understood as regions of greater density where the filaments intersect, presents 4 fragments, Hub-S is more fragmented, showing up to 13 fragments. Interestingly, authors find that all derived physical properties, such as the density and temperature profiles, the level of turbulence, the magnetic field around the hubs, and the rotational-to-gravitational energy, are remarkably similar in both hubs. 

The magnetic field strength was estimated from CSO polarimetric observation (350~\mum) at hub and filament scales ($\sim$ 0.1~pc, $\sim$10'' angular resolution). Using three different methods, the authors obtained a magnetic field strength of~0.6-0.8~mG in the Hub north and 0.1–0.2~mG in the Hub south, which is consistent with the lower fragmentation level observed in the N-hub by \cite{Busquet2016}. 
In addition, the B-field-to-gravity force ratio was estimated toward the Hub north, finding a gravitational collapse domain \citep{AnezLopez2020}. 

In this paper, we present new ALMA polarimetric observations at 1.39~mm with $\sim$2'' resolution ($\sim\,3200$~AU) toward the two hubs and three additional fields along the filament structure of G14.225-0.506, to investigate the magnetic field down to core scales. 
This cloud is part of  17 massive protostellar cluster forming clumps observed in full polarization mode with the ALMA main array under project 2017.1.00793.S (P.I.
Qizhou Zhang). The overview of the survey can be found in \citep{Zhang2025}, and detailed analyses of the infrared dark cloud G28.34+0.06  and NGC 6334; can be found in \citep{Junhao2020, Junhao2023, Junhao2023b, Junhao2024}, respectively. 
The paper is structured as follows. Section~\ref{sec:observation} describes the observation set up, and we present the results of the continuum emission in Section~\ref{sec:continuum}. 
In Section~\ref{sec:bfield} we present the magnetic field morphology and strength, and its analysis through the magnetic field significance (\sigb~maps) is presented in Section \ref{sec:sigmab}. Finally, Section~\ref{sec:discussion} provides a discussion and Section \ref{sec:conclusion} the conclusions of this work.  

\section{Observations}\label{sec:observation}
 
\begin{figure}
    \centering
    \includegraphics[trim = 0cm 0cm 0cm 0cm, clip,width=0.4\textwidth]{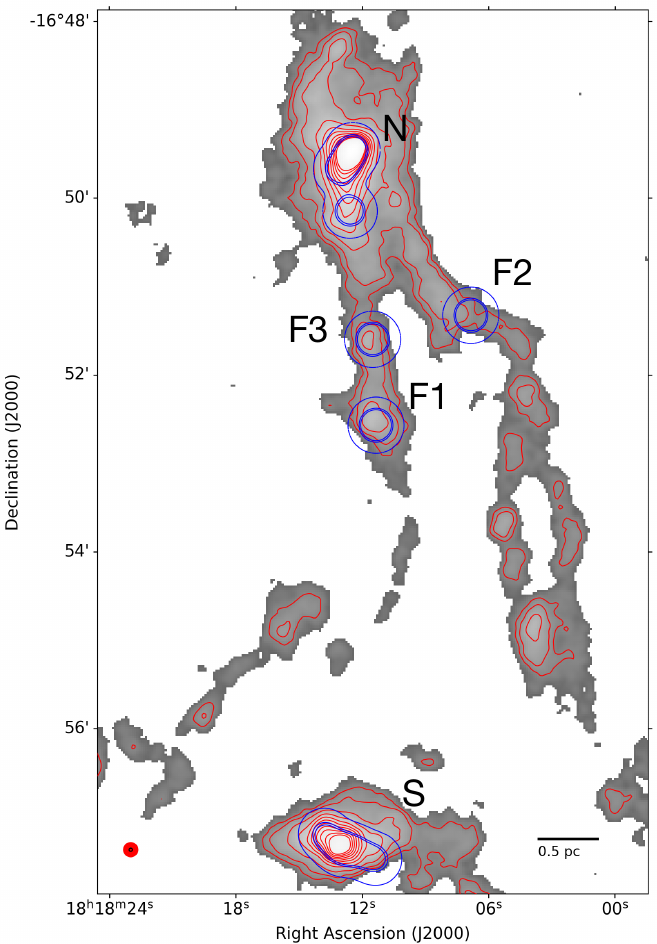}
    \caption{CSO continuum emission at 350~\mum~(gray map) and red contours which depict [3, 6, 9, 15, 25, 35, 45, 55, 65, 75, 85, 95] times rms noise (80 \mjybeam).
    The red and black circles in the lower left corner show the CSO and ALMA beam size, respectively. 
   Blue contours represent the 80\%, 40\%, and 33\% levels of the ALMA primary beam for each target region, labeled N, F1, F2, F3, and S.}
    \label{fig:AlmaRegObserved}
\end{figure}

The data presented here toward G14.2 is part of the ALMA project 2017.1.00793.S. 
The observations were done in full polarization mode at 1.39~mm (band 6) with the C43-4 and C43-1 configurations. 
The observations were performed between March and September 2018 for the C43-4 configuration and in July 2018 for the C43-1 configuration. 
We observed three and four contiguous single fields in the North and South Hubs, named N and S, respectively. 
We also observed three pointings in the filaments associated with the North Hub, named F1, F2, and F3. Notably, the F1 field coincides with the Hub-C candidate identified by \cite{Chen2019}.
The observed primary beams are shown in Figure~\ref{fig:AlmaRegObserved}. The correlator was configured to observe three wideband spectral windows of 1875~MHz each, centered at 216.4, 218.5 and 233.5~GHz. 
In addition, four narrowband spectral windows of 58.6~MHz and a spectral resolution of 0.32~\kms\ were used to observe the CO\,(3-2), N$_2$D$^+$\,(3-2), $^{13}$CS\,(5-4), and OCS\,(19-18) spectral lines. 
The Stokes~I continuum emission along with line data from C43-1 are presented in \cite{Zhao2025arXiv250216413Z}. 
In this paper, we present the linear polarization of the continuum emission.

The imaging process was made using the CASA software, specifically with the 6.5.5 version. 
The maps presented here were done using Natural weighting with a Gaussian taper to the visibilities with a width of 50k$\lambda$. 
In addition, we only used visibilities corresponding to baselines shorter than 500~k$\lambda$. This maximizes the polarization signal at scales of a few thousand AU (2--10$''$). 
The resulting synthesized beam  is $2\farcs37\times2\farcs10$ with a position angle of $-78.6^{\circ}$.
The rms noise of the maps ($\sigma_{qu}$) was 0.048 \mjybeam\ for the Stokes $Q$ and $U$. For Stokes $I$ ($\sigma_I$) was higher, 0.12~\mjybeam, probably due to the limited dynamic range. The signal-to-noise ratio achieved is around 1800.
The final images were corrected for the primary beam response using the CASA task impbcor to the 20$\%$ sensitivity level of the Stokes~I primary beam model.
To calculate the polarization intensity, we have applied Ricean debiasing, and we have also applied a threshold of 10 times the rms in Stokes~I.
The polarization images were computed using a cutoff of 3 times the $Q$ and $U$ rms. 
Position angles were computed as $\theta = 1/2 \arctan2(U/Q)$. 
Only magnetic field segments within 40$\%$ of the primary beam were considered toward mosaics, which represents a relaxation of the suggested limit for a single point by ALMA (33.33$\%$), which we have applied in the individual point, that is justified by the overlap of pointing that occurs in the mosaics \citep{Hull2020}. 
We obtained magnetic field morphology in the POS, assuming radiative alignment torques (RAT), and it was traced by linearly polarized emission and rotating polarization segments by 90$^{\circ}$ \citep[see review by][]{Lazarian2015}. 

\section{Dust continuum emission} \label{sec:continuum}

\begin{figure*}[!ht]
    \centering
    \includegraphics[trim= 0cm 0cm 0cm 0cm, clip, width=0.85\textwidth]{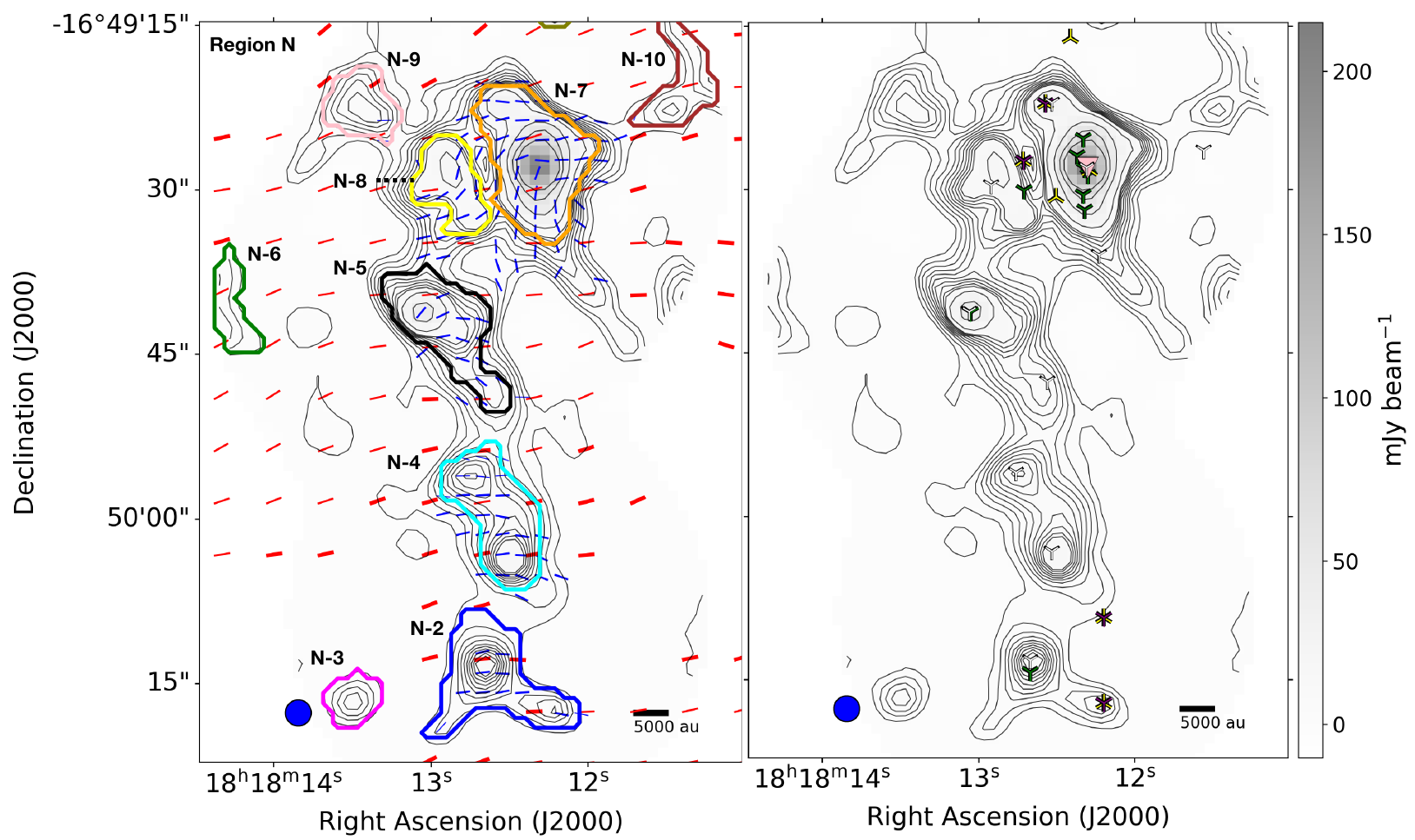}
    \includegraphics[trim= 0cm 0cm 0cm 0cm ,clip,width=0.99\textwidth]{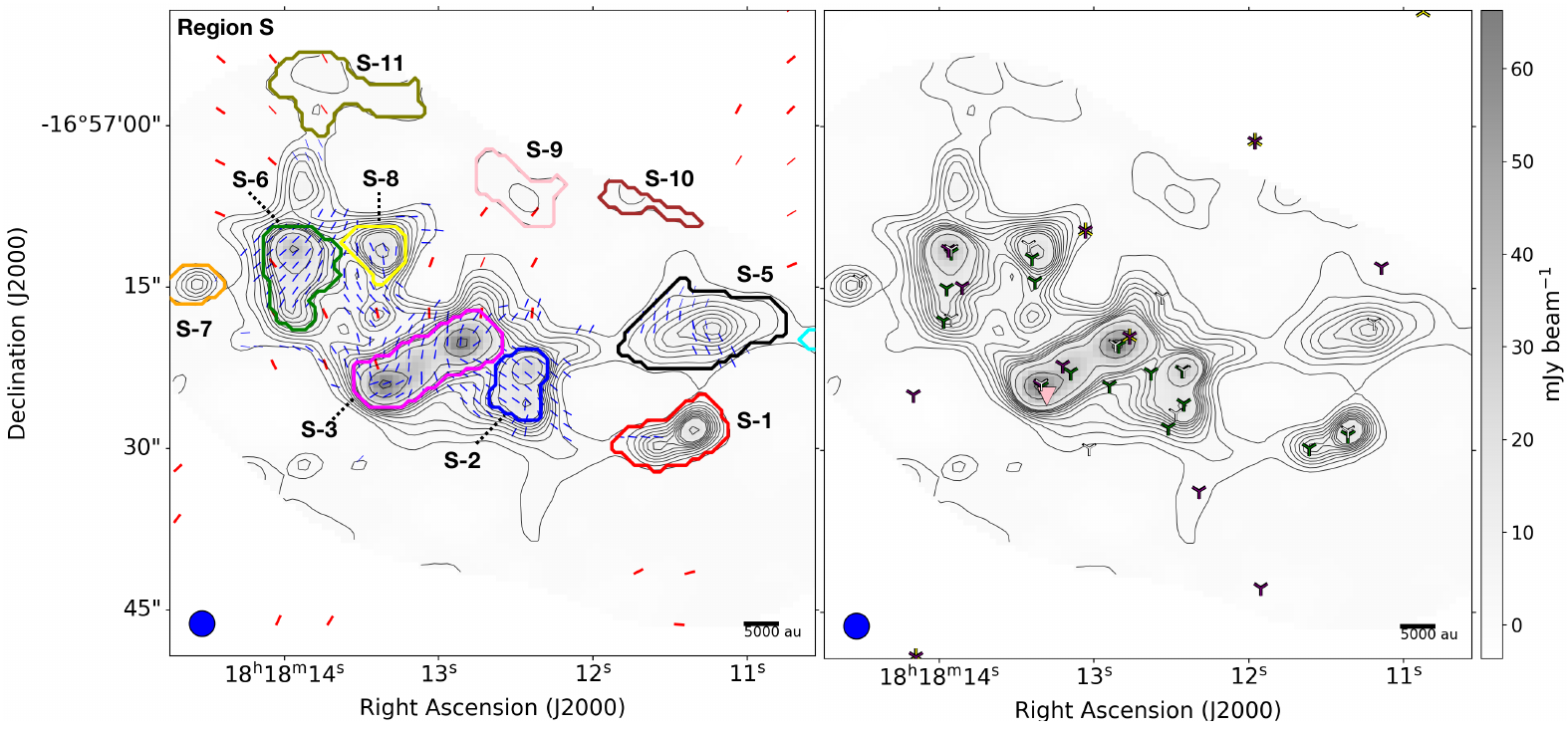}
    \caption{Continuum emission at 1.39~mm  toward region N (upper panel) and region S (bottom panel). Contours depict [10, 20, 30, 40, 50, 60, 70, 80, 90, 100, 150, 300, 500, 700] times rms noise (0.12 \mjybeam), overlapped with magnetic field segments in blue, where thin and thick segments depict segments within the inner 50$\%$ and 40$\%$ of the primary beam model respectively. 
    Red segments show the CSO magnetic field at 350~\mum\, where thin and thick segments indicate polarization intensity between 2 and 3 sigmas, and larger than 3 times sigma respectively \citep[CSO sigma = 0.02~\jybeam][]{AnezLopez2020}. Color contours show leaves from dendrogram analysis, and the black labels present the corresponding ID label. The blue solid circle in the bottom left corner of each panel depict the ALMA beam size. The Pink triangle shows the H$_2$O maser detected in \cite{Wang2006}. Green tripods show cores detected with the SMA telescope at 1.3~mm \citep{Busquet2016}. Yellow and purple tripods depict radio sources detected at 6~cm (C-band) and at 3.6~cm (X-band) by \cite{DiazMarquez2024} respectively. White tripods show NH$_3$ cores detected by \cite{Ohashi2016}.}
    \label{fig:leavesN}\label{fig:leavesSouth}\label{fig:bfieldCSOALMA_S}
\end{figure*}

\begin{figure}
    \centering
    \includegraphics[trim= 0cm 0cm 0cm 0cm, clip, width=0.5\textwidth]{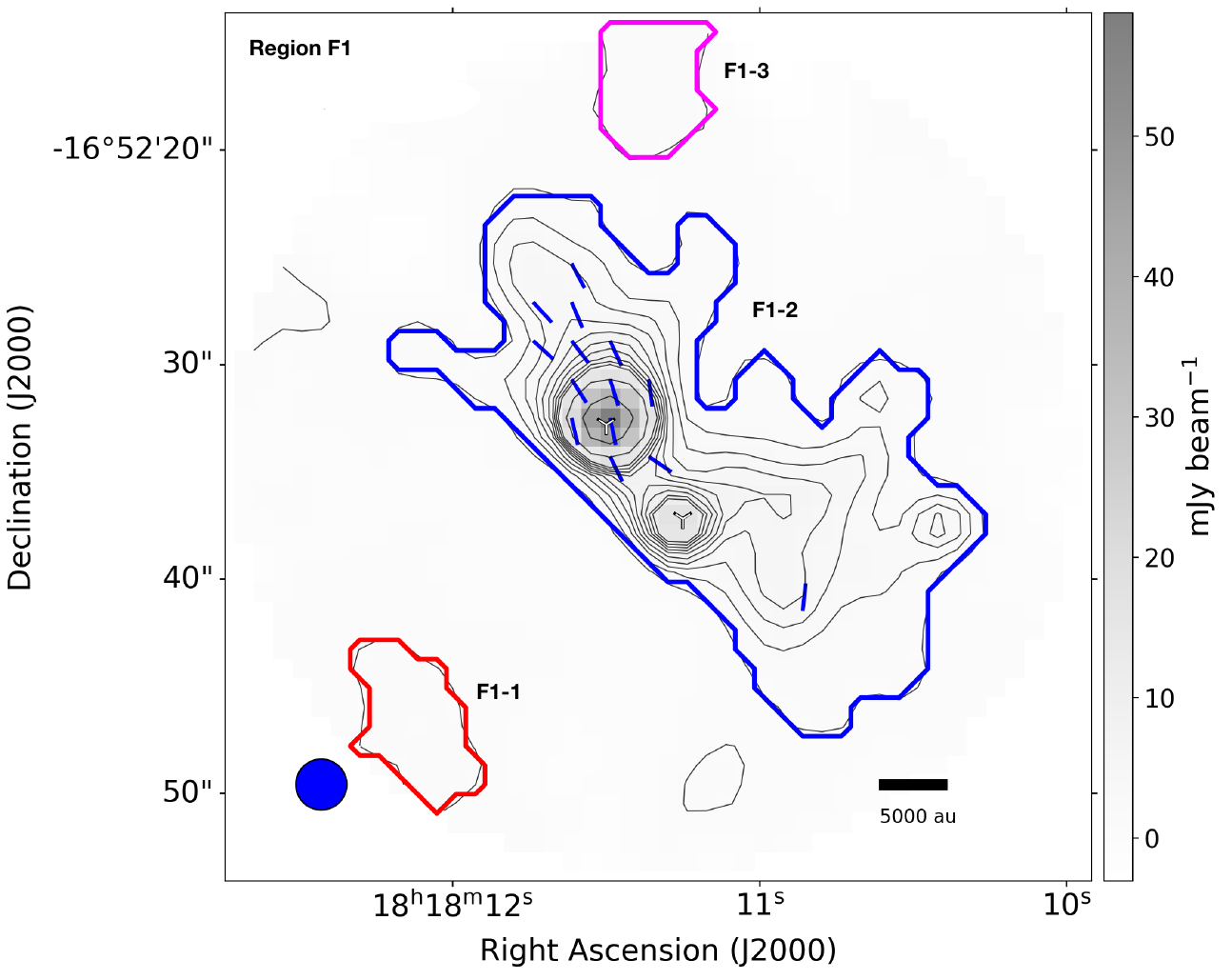}
    \caption{similar to Fig. \ref{fig:leavesN}-left, but for region F1.}
    \label{fig:leavesF1}
\end{figure}

Figures~\ref{fig:leavesN} and \ref{fig:leavesF1} show the 1.39 mm dust continuum emission toward the three observed regions where polarized emission is detected (N, S, and F1). The continuum emission, which an enormous signal-to-noise ratio (1800) allows tracing both the denser cores and the weaker filamentary emission connecting them, is consistent with previous observations at similar scales \citep{Busquet2016, Zhao2025arXiv250216413Z}. 

The continuum emission toward region N shows a main elongated distribution with north-south orientation, along the dense gas filament seen in NH$_3$ and N$_2$H$^+$ \citep{Busquet2013, Chen2019}, and 8 emission peaks along it (see Fig.~\ref{fig:leavesN}-upper panels).
Toward region S, the observations resolve multiple emission enhancements distributed along a generally elongated and curved structure extending from North to South-East (see Fig.~\ref{fig:leavesSouth}-bottom).
The location and previous detection within each region are detailed in Table~\ref{tab:loc-cores} in the appendix.

The continuum emission toward region F1 also exhibits an elongated morphology oriented from northeast to southwest (see Fig.~\ref{fig:leavesF1}), consistent with the main axis of the filament traced by CSO observations (see Fig.~\ref{fig:AlmaRegObserved}). This region is considered a candidate hub and displays both warm and compact ammonia emission \citep{Busquet2013, Chen2019}.
The continuum emission toward region F2 reveals a rotated L-shaped structure, consisting of two elongated features that converge at an angle of approximately 90$^{\circ}$ (see Fig. \ref{fig:leavesF2}). Three intensity enhancements are observed: one on each arm of the structure and a third at their intersection.
The continuum emission in region F3 also presents a filamentary morphology, oriented North to South, and contains a single prominent emission peak.

To avoid biases as much as possible in the selection of the areas, we have performed the analysis by dendrogram \citep{Rosolowsky2008} to extract the hierarchical structure in the present ALMA observation after re-binning the maps by a factor of 3 in order to achieve a pixel / beam = 1 / 3 where beam = $2\farcs37\times2\farcs10$ and pix = $0\farcs8\times0\farcs7$. Thus, we can study the B field in selected regions with enough independent polarization measurements. 
We adopted a minimum significance threshold of 1 $\sigma_I$ ($\sigma_I$ = 0.12 \mjybeam) for each leaf, along with a requirement that each leaf span an area equivalent to at least three beams to be considered independent. These criteria represent a relaxation of those used in the dendrogram analysis by \cite{Zhao2025arXiv250216413Z}, in order to better align with the objectives outlined above.
We consider, for the purpose of analysis, structures that have no substructure, \aka leaves in the dendrogram vocabulary, within each field (see Fig.~\ref{fig:leavesN}-left and \ref{fig:leavesF1} for the regions N and S, and F1, respectively). From now on, we will refer to these leaves as condensations.

\section{Magnetic field}\label{sec:bfield}

In this section we present the magnetic field morphology toward each region where polarized emission is detected at core scales ($<$10000~AU/0.05~pc), i.e., N, S, and F1 regions.

Figures~\ref{fig:leavesN}-upper-left, \ref{fig:leavesN}-bottom-left and Fig.~\ref{fig:leavesF1} show the morphology of the plane-of-sky (POS) magnetic field in region N, S and F1 respectively,  where only segments with a signal greater than 3$\sigma_{qu}$ are displayed. Figures also include magnetic field observations at 350~\mum~from the CSO at scales of $\sim$0.1~pc, previously presented in \cite{AnezLopez2020}. 

\begin{figure}
    \centering
    \includegraphics[trim= 0cm 0cm 0cm 0cm, clip, width=0.6\linewidth]{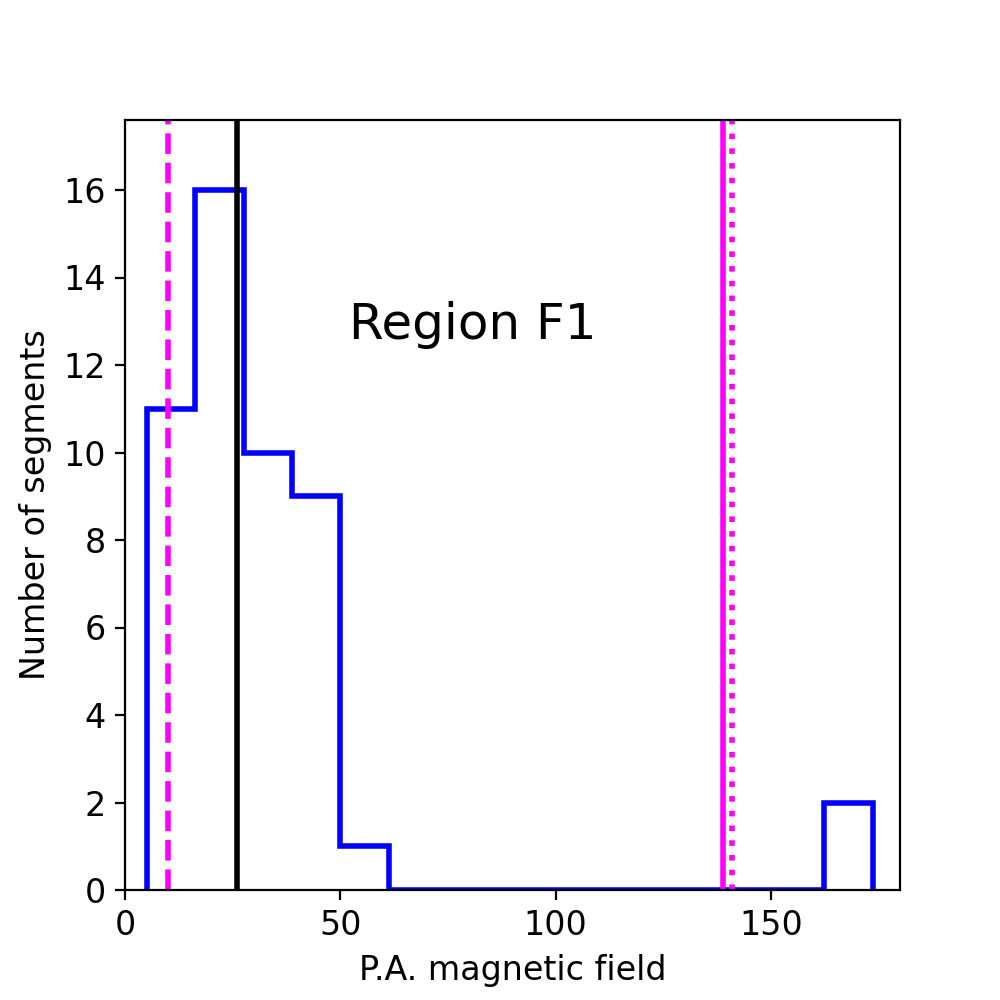}
    \caption{Magnetic field position angle histogram, toward region F1. 
    Black solid line depict the median value over the distribution ($\sim$26º).
    Magenta dashed lined shows the orientation of main filament orientation ($\sim$10º) \cite{Busquet2013}.
    Magenta solid and dotted lines indicate the overall orientation of the polarization in H-band and R-band, respectively \citep{Santos2016}.}
    \label{fig:paHistoALMA1}
\end{figure}

Figures~\ref{fig:paHistoALMA}-upper, \ref{fig:paHistoALMA}-bottom, and \ref{fig:paHistoALMA1} show the histogram of the position angle distribution for the regions N, S, and F1, respectively, where the median value is indicated by a black line. We also show the polarization overall orientation detected in H-band and R-band by \cite{Santos2016}.

\subsection{Hub-N}

\begin{figure}
    \centering
    \includegraphics[trim= 0cm 0cm 0cm 0cm, clip, width=0.25\textwidth]{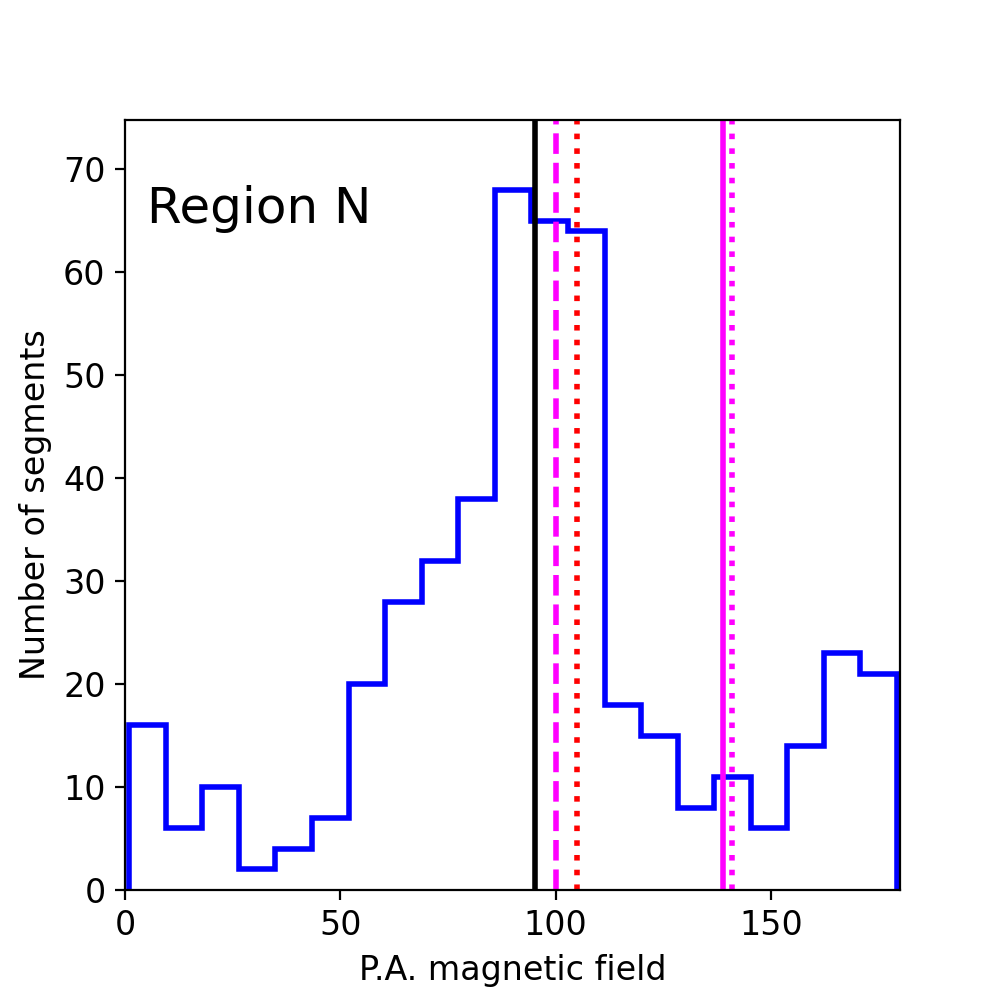}
    \includegraphics[trim = 0cm 0cm 0cm 0cm, clip, width=0.25\textwidth]{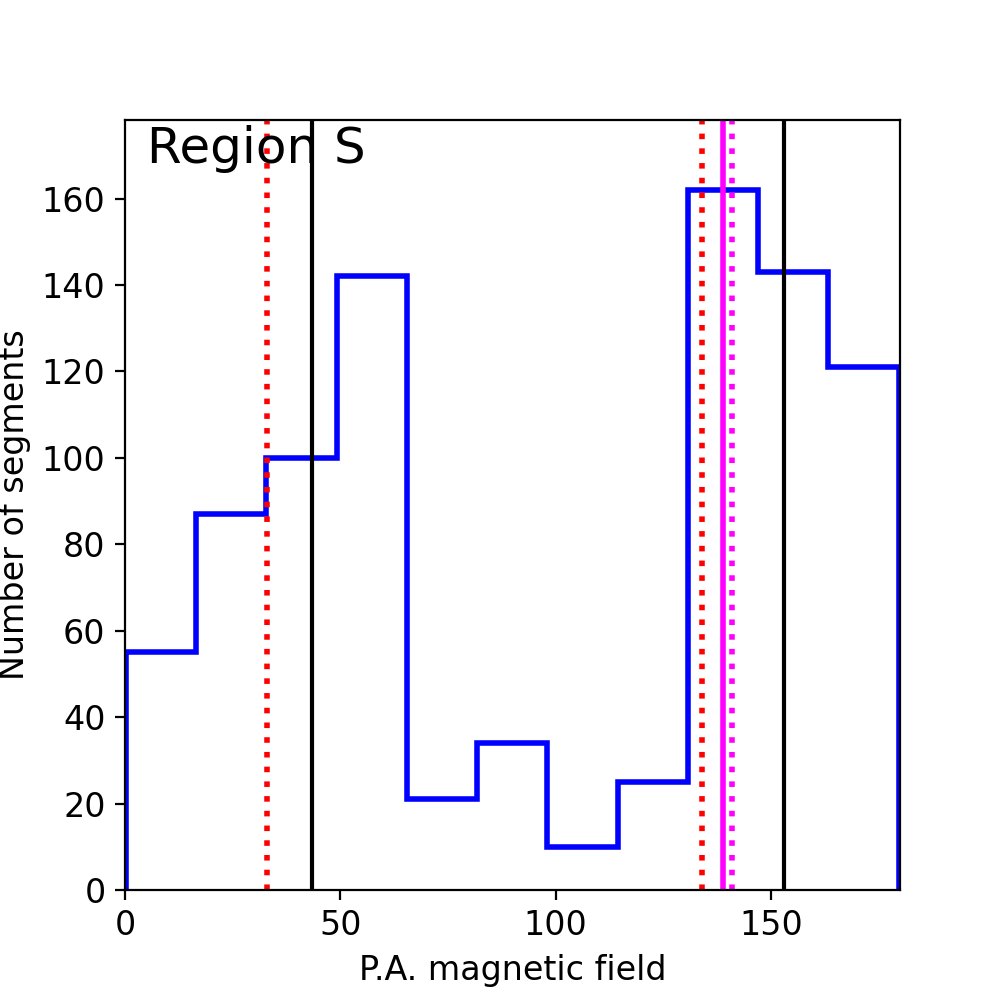}
    \caption{Magnetic field position angle histogram toward region N (top) and S (bottom).  
    Black solid line depict the angle median value \ie 95$^{\circ}$ for region N and (44$^{\circ}$, 153$^{\circ}$) for region S. 
    Red dotted line shows the main direction of the CSO magnetic field orientation, 105$^{\circ}$ and (33º, 134º) for region N and S respectively, as showed in \cite{AnezLopez2020}. 
    Magenta dashed line shows perpendicular to main filament orientation (\ie F10-E, 100$^{\circ}$,  \cite{Busquet2013}.
    Magenta solid and dotted lines indicate the overall orientation of the polarization in H-band and R-band, respectively \citep{Santos2016}.}
    \label{fig:paHistoALMA}\label{fig:paHistoALMA_S}
\end{figure}

The magnetic field segments in region N are predominantly oriented perpendicular to the elongated continuum emission (Fig.~\ref{fig:leavesN}-upper-left), where the median value is 95$^{\circ}$ (Fig. \ref{fig:paHistoALMA}-upper). 
The median position angle closely matches the value reported by the CSO (105$^{\circ}$) and is perpendicular to the main filament orientation (10$^{\circ}$) reported by \cite{Busquet2013}. 

However, the main magnetic field orientation is altered in the vicinity of the main peak emission, coinciding with the locations of the condensations N-5, N-7, and also N-8.
This behavior is reflected in the bimodal distribution shown in the histogram (Fig.~\ref{fig:paHistoALMA}-upper), with a secondary peak around 0$^{\circ}$/180$^{\circ}$.
Therefore, while at larger scales (CSO, angular resolution $\sim10''$) the magnetic field appears to be homogeneous and perpendicular to the filament orientation, at higher angular resolution ($\sim2''$), the ALMA observations show distortions in the environments of the most massive condensations (see Table \ref{tab:dendrogramN}). 

\begin{figure*}
    \centering
    \includegraphics[width=0.7\textwidth]{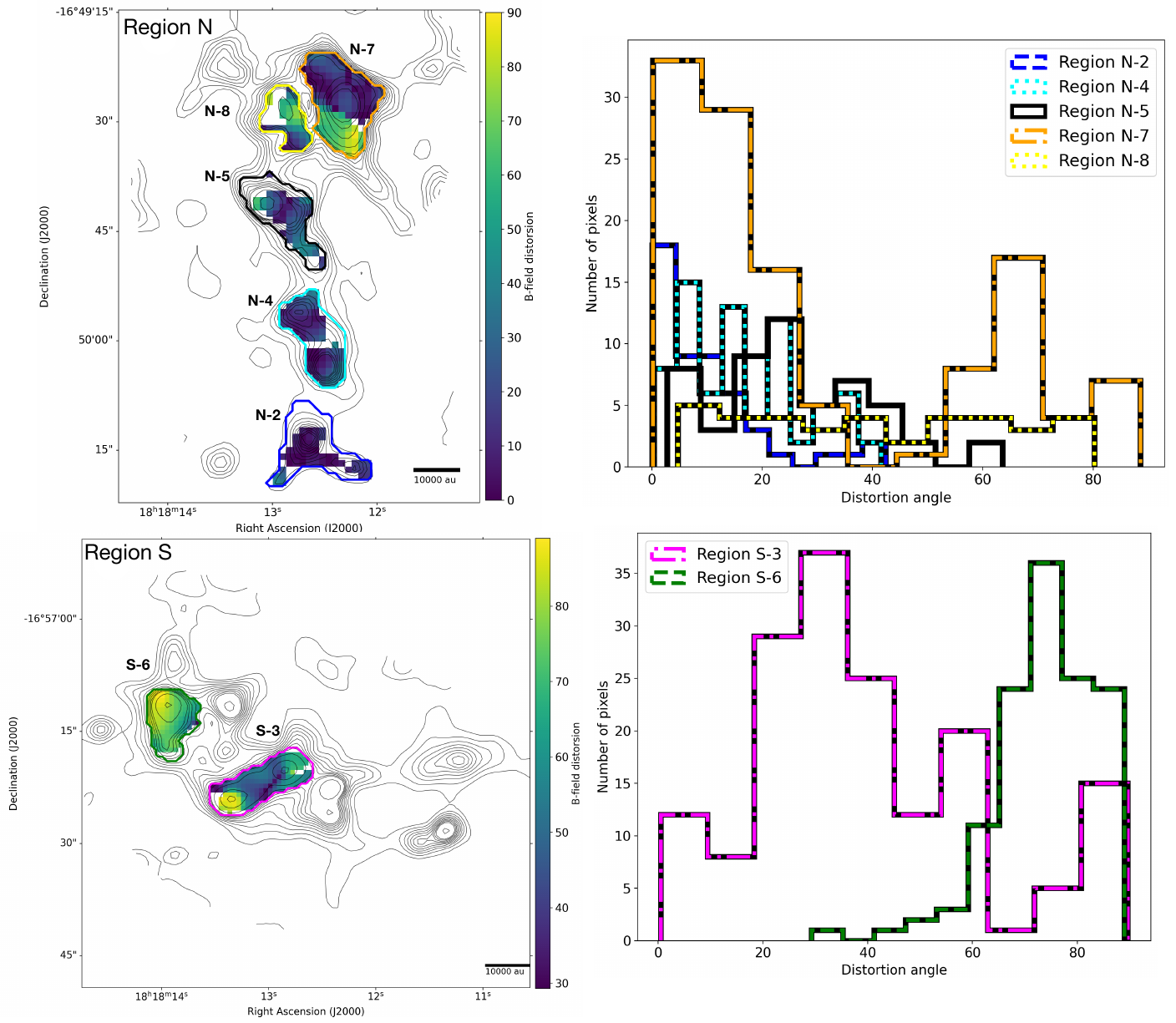}
    \caption{The relative position angle map and histogram (left and right respectively) of the magnetic field observed by current ALMA $\sim$2'' ar and by former CSO~$\sim$10'' ar toward region N (upper) and S (bottom).}
    \label{fig:distorsionN}\label{fig:distorsionS} \label{fig:distorsionHistoN}\label{fig:distorsionHistoS}
\end{figure*}

To illustrate the distortion of the magnetic field, we present in Fig.~\ref{fig:distorsionN}-upper-left the relative position angle with respect to the larger scale, computed within every condensation in the regions.
The angular resolution inequality prevents us from comparing pixel by pixel. We compute the difference between the ALMA position angle for each pixel and the corresponding averaged CSO value for each condensation. 
Fig.~\ref{fig:distorsionN}-upper-left, and also Fig.~\ref{fig:distorsionHistoN}-upper-right that present the relative angle histogram, show that the CSO magnetic field prevails at ALMA scales throughout the elongated continuous emission, basically in N-2 and N-4 condensations, but it is distorted when we approach N-8 and especially in the N-7 environment. 

Figure \ref{fig:polpercent}-left in the appendix presents the polarization fraction map toward region N, whose averaged value is 3.68\%. We found the lowest polarization values ($<$2\%) towards the main emission peaks, N-7 (orange) and N-8 (yellow). 
These results are summarized in Table \ref{tab:dendrogramN}.

\subsection{Hub-S}

Figure~\ref{fig:leavesSouth}-bottom-left presents the POS magnetic field morphology toward the region S. Region S displays a less regular angle distribution than region N (see Fig.~\ref{fig:paHistoALMA_S}-bottom). 
It also shows a bimodal distribution, with two main peaks at 44$^{\circ}$ and 153$^{\circ}$ that agree with the distribution previously detected at filament scales with the CSO telescope \citep[33$^{\circ}$ and 134$^{\circ}$][]{AnezLopez2020}. Table~\ref{tab:dendrogramS} summarizes average angles and standard deviation for each condensation.

We also present the relative angle map of the magnetic field relative to the field previously detected by the CSO in Fig.~\ref{fig:distorsionS}-bottom-left. 
Unfortunately, previous CSO detection towards this region is very scarce, so it is only possible to calculate the relative angle in two condensations, S-3 (magenta) and S-6 (green).
High values, around $90^{\circ}$, are present at the peak emission in both S-3 and S-6 condensations as shown in Fig.~\ref{fig:distorsionHistoS}-bottom-right. 
Toward S-6, the vast majority of the vectors are distorted by more than $60^{\circ}$, whereas S-3 exhibits a more spread distribution, with a peak around $30^{\circ}$. 

Figure~\ref{fig:polpercent}-right in the appendix shows the polarization percent map toward the region S, with an average value of $\sim$5$\%$. Table~\ref{tab:dendrogramS} summarizes the results of the mean values of the polarization percentage on each condensation, along with their dispersion. 
The lower values are found in S-3 and S-2 (1.29 and 1.41 $\%$, respectively), while the larger values (5.39$\%$ and 5.33$\%$) are found toward S-1 and S-5. 

\subsection{Region F1}

Figure~\ref{fig:leavesF1} presents the POS magnetic field morphology toward region F1, which has been identified as a Hub candidate \citep{Chen2019}. The magnetic field exhibits a predominant northeast to southwest orientation, aligned with the continuum emission.  
However, we note that the magnetic field in the vicinity of the peak emission ($\sim$ 18h18m11.30,  -16º52'33'') deviates to a North-South-like orientation. 
Fig.~\ref{fig:paHistoALMA1} presents the histogram of the position angles toward region F1, where the predominant orientation is about 26$^{\circ}$, which is in good agreement with the main orientation of the filament \citep[10$^{\circ}$][]{Busquet2013}. 
Unfortunately, the CSO polarization observations used for comparison in previous sections do not cover this specific field of view, preventing a direct comparison.
Table~\ref{tab:dendrogramN1} summarizes the main results of the magnetic field position angle analysis within the condensation where the magnetic field is detected.

In Figure~\ref{fig:polpercent}-bottom in the appendix, we show the polarization percent map toward the region F1. Values range between 0.6 and 4.9\%, with an averaged value of 2.1\%. 
We found the lowest polarization ($<2$\%) percent towards the peak emission. Table~\ref{tab:dendrogramN1} presents the averaged polarization percent, 2.87$\%$, and its standard deviation, 1.50$\%$. 

\begin{table*}
\centering
\tiny

\caption{Results of the condensation analysis.}
\begin{threeparttable}
\begin{tabular}{c|c|c|c|c|c|c|c|c|c|c|c|c|c}
\toprule
ID & Flux\tnote{a}  & T\tnote{b} & {\scriptsize Mass} & n(H$_2$)  &$\Delta$v\tnote{c} & 
$<$PA$>$\tnote{d} & std\tnote{e} &{\scriptsize $<$pol$>$\tnote{f} } & {\scriptsize std\tnote{g}} &{\scriptsize $<$\sigb$>$\tnote{h}} & { \scriptsize std\tnote{i} } & {\scriptsize \sigb$<$ 1 } & {\scriptsize b \tnote{j}} \\ 

& {\scriptsize (mJy) } & \scriptsize{(K)} & (\msun) & {\scriptsize (10$^6$ cm$^{-3}$)}  &{\scriptsize (km s$^{-1}$) } & ($^{\circ}$) & ($^{\circ}$) & ($\%$) & ($\%$) & & & ($\%$) & ($^{\circ}$) \\ 
\hline
\hline

N-2 &  60.0 & 15.0 &  7 & 2.34 & 0.49 &  94 &  8 & 5 & 2  & 0.5 & 0.5 & 72.7 & 6.2\\
N-4 & 100.0 & 16.0 & 10 & 4.13 & 0.77 &  84 & 11 & 4 & 2  & 0.4 & 0.4 & 88.6 & 7.0\\
N-5 & 160.0 & 19.5 & 13 & 5.17 & 1.29 &  85 & 27 & 2 & 1  & 0.7 & 0.4 & 75.5 & 5.7\\
N-7 & 760.0 & 22.5 & 50 & 12.4 & 1.56 & 106 & 39 & 3 & 2  & 0.3 & 0.4 & 89.6 & 33.3\\
N-8 & 100.0 & 19.0 &  8 & 7.57 & 2.00 & 134 & 35 & 2 & 1  & 0.6 & 0.4 & 75.8 & 37.3\\

\hline
\hline
F1-2 & 260.0 & 26.5 & 14 & 0.64 & 1.47 & 28 & 25 & 3 & 2 & 0.3 & 0.5 & 74.5 & 6.5\\
\hline
\hline
S-2 & 80.0  & 20.5 &  6  & 10.2 & 1.06 &  53  & 25 & 3  & 1 & 1.0  & 0.4  & 67.2 & 26.9 \\
S-3 & 310.0 & 21.5 & 22  & 9.79 & 1.86 & 132  & 51 & 2  & 1 & 0.4  & 0.4 & 89.1 & 50.8\\
S-6 & 130.0 & 16.8 & 13  & 8.72 & 1.48 & 143  & 13 & 4  & 1 & 0.3  & 0.4  & 90.8 & 12.2\\
S-8 & 50.0  & 15.0 &  6  & 13.7 & 2.00 &  22  & 48 & 3  & 1 & 0.7  & 0.3  & 82.9 & 57.7\\
\bottomrule
\end{tabular}
\begin{tablenotes}\footnotesize
\item[a] Integrated flux within the leaf, 
\item[b] Kinetic temperature (averaged) \cite{Ohashi2016, Busquet2013}
\item[c] the linewidth are measured from  N$_2$H$^+$ (1-0) \cite{Ohashi2016,Chen2019},
\item[d] B-field position angle circular mean,
\item[e] B-field position angle circular standard deviation,
\item[f] median polarization percent, 
\item[g] polarization percent standard deviation,
\item[h] mean \sigb,
\item[i] \sigb~standard deviation.
\item[j] the turbulent dispersion about the large-scale field \cite{Hildebrand2009}
\end{tablenotes}
\end{threeparttable}
\label{tab:dendrogramS}\label{tab:dendrogramN1}\label{tab:dendrogramN}
\end{table*}

\begin{table*}
\centering
\tiny

\caption{Results of the condensation analysis.}
\begin{threeparttable}
\begin{tabular}{c|c|c|c|c|c|c|c|c|c|c}
\toprule
ID & {\scriptsize B}$_{pos}^{IG}$ \tnote{a}  & $\lambda^{IG}$& {\scriptsize B}$_{pos}^{C}$  \tnote{b}& $\lambda^{C}$ &  {\scriptsize B}$_{pos}^{H}$ \tnote{b}& $\lambda^{H}$ & {\scriptsize B}$_{pos}^{FG}$ \tnote{b} & $\lambda^{FG}$ & {\scriptsize B}$_{pos}^{SF}$& $\lambda^{SF}$\\

& {\scriptsize (mG)} & & {\scriptsize (mG) } & & {\scriptsize (mG)}& &{\scriptsize (mG)} & &{\scriptsize (mG)}&\\
\hline
\hline
N-2 & 3.2 &  1.1 &   0.5 &  0.3 &   0.5 & 0.3 &   0.4 &   0.3 &   0.5 & 0.3 \\
N-4 & 3.8 &  1.0 &   0.7 &  0.4 &   0.6 & 0.4 &   0.6 &   0.4 &   1.0 & 0.3 \\
N-5 & 6.0 &  0.9 &   - &  - &   0.5 & 0.6 &   0.4 &   0.7 &  2.3  & 0.2  \\
N-7 & 9.9 &  1.4 &   - &  - &   0.9 & 1.0 &   0.5 &   1.7 &  0.7 &  1.2 \\
N-8 & 5.9 &  0.8 &   - & - &  1.0 & 0.3 &   0.6 & 0.6 &  
0.7 & 0.5\\
\hline
\hline
F1-2 & 0.8 & 1.3 &   0.2 &  0.3 &   0.4 & 0.2 &   0.2 & 0.4 &   0.8 & 0.1\\
\hline
\hline
S-2 & 15.9 & 0.8 & 0.7 & 0.5 &   0.8 & 0.5 &   0.6 & 0.7 & 
0.6 & 0.7 \\
S-3 & 9.0 & 1.0 & - & - & 0.9 & 0.6 & 0.4 & 1.5 & 
0.5 & 1.1 \\
S-6 & 10.2 & 1.3 & 1.8 & 0.2 & 2.0 & 0.2 &  1.5 & 0.3 &  
1.6 & 0.3\\
S-8 & 11.9 & 0.7 & - & - & 2.5 & 0.2 & 0.5 & 0.9 & 
0.6 & 0.8 \\ 
\bottomrule
\end{tabular}
\begin{tablenotes}\footnotesize
\item[a] Uncertainty is inherited from the calculation of the average value of \sigb~and is therefore related to its standard deviation. See column h in Table \ref{tab:dendrogramN}. 
\item[b] The ratio error/value for the magnetic field strength is a factor $<$ 3. 
\end{tablenotes}
\end{threeparttable}
\label{tab:Bpos}
\end{table*}

\section{Magnetic field versus gravitational potential: the IG-method} \label{sec:sigmab}

\begin{figure*}
    \centering 
    \includegraphics[trim = 0cm 0cm 0cm 0cm,clip,width=0.4\textwidth]{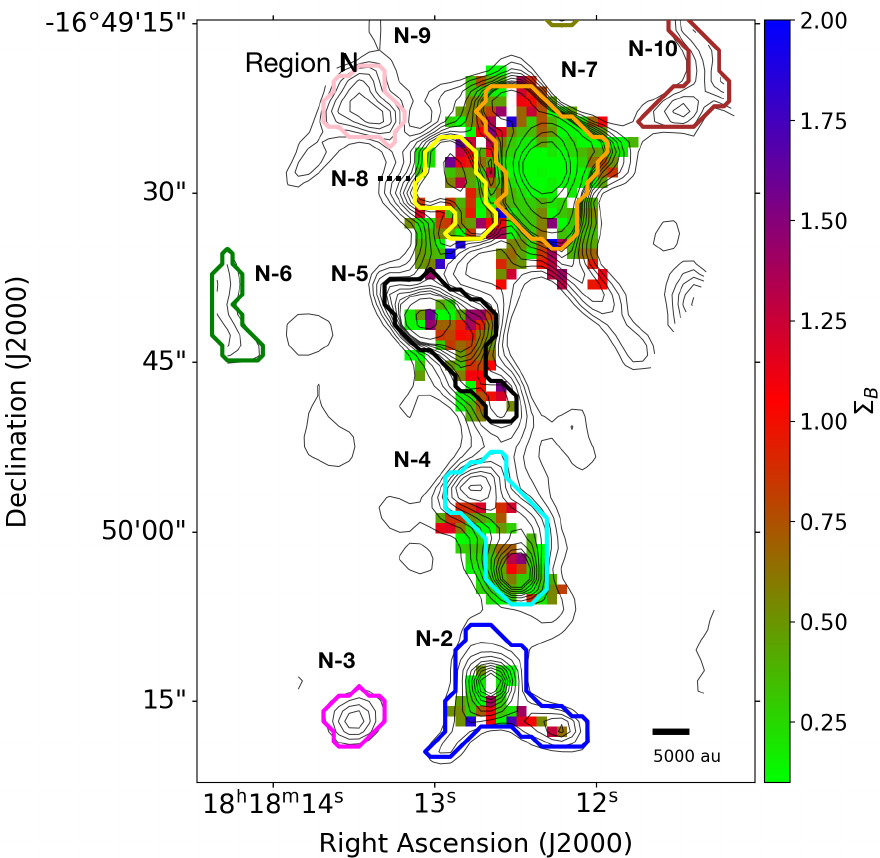}
     \includegraphics[trim = 0cm 0cm 0cm 0cm,clip,width=0.5\textwidth]{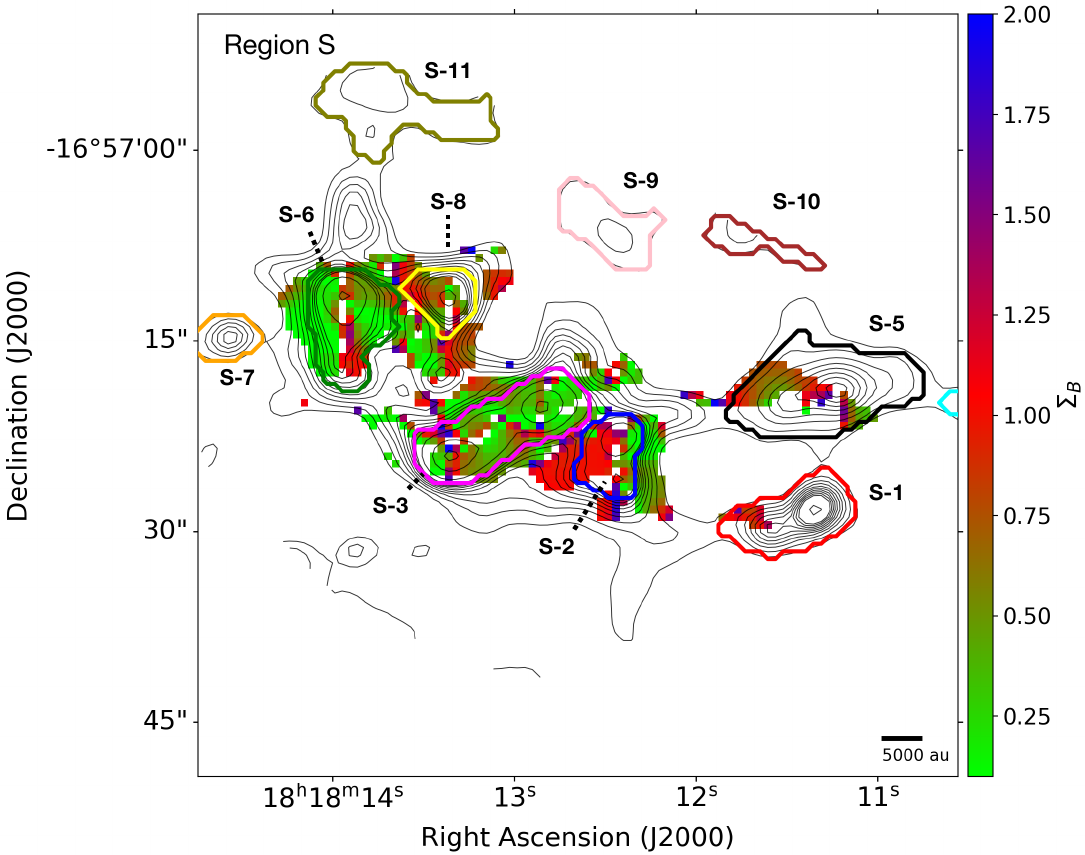}
    \caption{\sigb~map in color toward region N. Black and colored contours show stokes I  and the leaves found by dendrogram analysis.}
    \label{fig:sigma_N}\label{fig:sigmab_S}
\end{figure*}

We computed the relative significance of the magnetic field and gravitational potential in order to develop an analysis analogous to that originally proposed by \cite{Koch2012} and later refined and tested by \cite{Koch2013, Koch2014}.
This method, from now on IG-method, assumes that emission intensity is due to the transport of matter, which is driven by a combination of magnetohydrodynamic forces (MHD). Therefore, the intensity gradient would measure the motion direction. Then, solving the MHD equations for the magnetic field, the authors provide an expression for the magnetic field in the POS as a function of measurable variables on a 2D map: 

\begin{equation}
   \Bigg[\dfrac{B_{\mathrm{pos}}}{G} \Bigg] = \sqrt{\frac{sin \psi}{sin \alpha} \left(\nabla P + \rho \bigtriangledown \phi \right) 4\pi R},
   \label{eq:bfieldKoch}
\end{equation}

\noindent where density ($\rho$) is in g cm$^{-3}$, gravitational potential gradient ($\nabla \phi$) is in cm s$^{-2}$, and B-field curvature radius (R) is in cm. The resulting B-field is in Gauss. For simplicity, we neglected the pressure gradient ($\nabla P$), which is typically small compared to gravity potential.
Here,~\sigb~is the magnetic field significance, according to the following equation:

\begin{equation}
    \begin{array}{cc}
    \Sigma_{B} = & \sin \Psi \sin \alpha^{-1} \\
    \end{array}
\end{equation}

\noindent where $\Psi$ and $\alpha$ are the angles in the POS between the orientation of local gravity and intensity gradient, and those between polarization and intensity gradient, respectively. 
Values of \sigb~greater than or less than unity indicate whether the magnetic field force is able to prevent gravitational collapse or not, respectively.
Local gravity is computed through the surrounding mass distribution in any position on the map, where the main assumption is to use the dust emission as a proxy for the gas mass. 

This technique, which is part of a set of methods for inferring the role of the magnetic field in the collapse based on the geometry of the magnetic field as well as on density structures, has been tested extensively and applied in various studies \citep[\eg
][]{Tang2009A, Koch2014, Koch2018, Koch2022ApJ}. 
Following this recipe, we have calculated \sigb-maps towards Regions N, S, and F1, as well as the average magnetic field strength for each condensation. 

Figure~\ref{fig:sigma_N}-left presents the \sigb~map of region N, revealing that most values are below 1. This suggests that gravitational collapse is the dominant force over magnetic tension in the region.

Figure~\ref{fig:sigmab_S}-right shows the \sigb-map toward the region S, which appears to be dominated by values lower than 1. 
Lower \sigb~averaged values, typically under 0.5, are found toward S3, S6, and S8 condensations. Overall, every condensation shows gravitational collapse dominance. 
Table~\ref{tab:dendrogramN} summarizes \sigb~averaged values, standard deviation, and percentage of pixels that exhibit values under 1 for every condensation. 

\begin{figure}[!h]
    \centering
    \includegraphics[trim = 0cm 0cm 0cm 0cm, clip, width=0.4\textwidth]{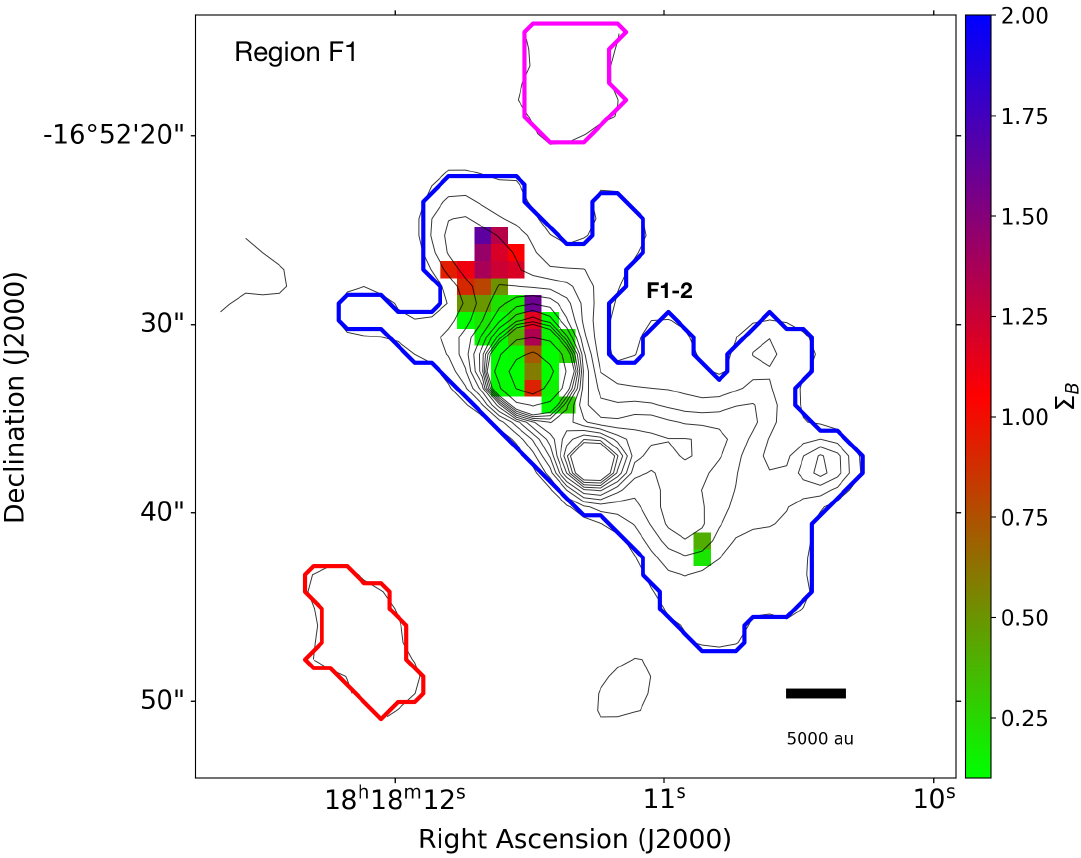}
    \caption{Similar to Fig.\ref{fig:sigma_N}, but for region F1.}
    \label{fig:sigmab_F1}
\end{figure}

Figure~\ref{fig:sigmab_F1} presents the \sigb-map toward region F1 (F1-2), revealing a predominance of values below unity, which indicates gravitational dominance over magnetic support.

Table~\ref{tab:Bpos} includes the calculation of the magnetic field strength using the method presented above, as well as the one calculated using the DCF-C method \citep{DavisPhysRev,ChandrasekharFermi1953} by means of Equation~2 of \citet{Crutcher2004}:

\begin{equation}
    B_{\text{pos}}^{C} = Q \sqrt{4\pi \rho} \frac{\sigma_{\mathrm{v}}}{\sigma_{\mathrm{PA}}}\approx 5.2 \frac{\sqrt{n(\text{H}_2}) \Delta\,v}{\langle \sigma_{\mathrm{PA}}\rangle}~\mu G, 
    \label{eq:DCF}
\end{equation}

\noindent where $\rho$ is the gas density (g/cm$^3$), $\sigma_{\mathrm{v}}$ is the velocity dispersion (km/s), $\sigma_{PA}$ is the polarization angle dispersion, n(H$_2$) is the gas number density (cm$^{-3}$), $\Delta v = \sqrt{8\ln{2}}~\sigma_v $ is the linewidth (km/s), and Q is a factor of order unity. We also computed the version presented by \cite{Falceta-Goncalves2008} (DCF-FC), which allows avoiding the limitation regarding the angle dispersion,

\begin{equation}
    B_{\text{pos}}^{FG} = Q \sqrt{4\pi \rho} \frac{\sigma_{\mathrm{v}}}{tan(\sigma_{\mathrm{P.A.}})}~G
    \label{eq:FG2008}
\end{equation}

In addition, we applied the correction presented in \cite{Heitsch2001} (DCF-H), that reflects equipartition between kinetic and magnetic energy.

\begin{equation}
    B_{\text{pos}}^{H} = Q \sqrt{4\pi \rho} \frac{\sigma_{\mathrm{v}}}{\langle \sigma ({tan( \mathrm{PA}}))\rangle} \sqrt{  (1 + 3\sigma tan (PA)^2}~G.
    \label{eq:H2001}
\end{equation}

In all three cases we have considered a Q factor of 0.28, which is well suited for the present small scales and high densities \citep{Liu2021, Liu2022}.
We estimated column density and mass for every condensation from the current flux observation and the kinematic temperature of ammonia estimated in \cite{Busquet2013} by applying the optically thin approximation \citep{Hildegrand1983, Andre1993}:
\begin{equation}
    M_{\mathrm{dust}} = \frac{S_{\nu}  d^2}{\kappa_{\nu}  B_{\nu}(T_d)},
\end{equation}

\noindent where we assumed a dust opacity, $\kappa_{\nu}$, of 0.9 cm$^2$ g$^{-1}$ \cite{Ossenkopf&Henning1994} and also a gas-to-dust ratio of 100 to compute the total mass. 

Finally, following \cite{Hildebrand2009, Houde2009, Houde2011}, in Appendix \ref{App:sf} we show the fitted square root of the Structure Function (SF$^2$) of the polarization orientation angles toward each region as a function of the separation between segments.

\begin{equation}
    SF^2(l) = b^2 + m^2l^2 + \sigma^2(l),
\end{equation}

where b, the intercept with the y-axis, shows the turbulent contribution to the angular dispersion, m$^2$ is a constant, and $\sigma^2(l)$ is the uncertainty on the polarization angles. We consider a separation no smaller than one third of the beam, \ie one pixel. Then we can estimate the magnetic field strength as follows (DCF-SF):

\begin{equation}
    B_{\text{pos}}^{SF} = Q \sqrt{4\pi \rho} \frac{\sigma_{\mathrm{v}}}{b}~G. 
    \label{eq:SF}
\end{equation}

We omit the DCF-C method (Equation \ref{eq:DCF}) results in cases where the standard deviation of polarization angles exceeds 25$^{\circ}$, as this violates the small-angle approximation underlying the tangent-angle equivalence.
While the four DCF methods give similar results ranging from 0.2 to 2.3~mG, the method developed by \cite{Koch2012} reflects a wider range from 0.8 to 15.9~mG (see Table \ref{tab:Bpos}). Later we will discuss the validity of these methods at current scales.

\begin{equation}
    \lambda = \frac{(M/\theta)_{observed}}{(M/\theta)_{crit}} = 2.5 \times 10^{-21} \frac{n(H_2)}{B}
    \label{eq:lambda}
\end{equation}

The mass-to-flux ratios ($\lambda$) computed using equation \ref{eq:lambda}, presented in  \cite{Crutcher2004}, are shown in Table~\ref{tab:Bpos}. The mass-to-flux ratio can also be calculated independently of the magnetic field strength, as a function of the radius and the significance of the magnetic field (\sigb) \citep[see][for details]{Koch2012b} (see also Table~\ref{tab:Bpos}).

Towards the most massive condensations, \eg N-7, the mass-to-flux ratio exceeds unity, indicating a supercritical state, regardless of the method we use to estimate the magnetic field. However, in other cases (\eg N-2, N-4, S-6), we observed discrepancies depending on the method used, which we will discuss below (see Table \ref{tab:Bpos}).

\section{Discussion} \label{sec:discussion}

\subsection{Mass-to-flux ratios}

Quantifying the amount of mass per unit magnetic flux, known as the mass-to-flux ratio, is essential for understanding the influence of magnetic fields on star formation.

In this study, we compute the mass-to-flux ratio using the IG method \ie independently of the strength of the magnetic field, and also from the magnetic field strength obtained using several DCF method variants (see Eq. \ref{eq:lambda}).
For the more massive condensation studied during this work (N-7), both IG, DCF-FG, DCF-SF, and DCF-H methods agree on a supercritical state. However, for some other condensations (N-2, N-4, S-6), we find discrepancies among methods.

To understand these discrepancies, first we look at the magnetic field strength. 
In all cases, the IG method provides larger magnetic fields, with the difference ranging from a factor of two to one order of magnitude.
\cite{Liu2021} suggest that the DCF method tends to overestimate the field. If the DCF method does indeed overestimate the magnetic field strength, leading to a lower mass-to-flux ratio, this could explain the mass-to-flux discrepancy across methods. 
However, this would also imply an even larger divergence in the field strength estimates between IG and DCF family methods. 

Second, we must also consider uncertainties in the mass estimates. Our calculations exclude mass from the innermost tens of astronomical units, due to optical thickness, as well as mass already accreted onto the protostar. Both omissions could contribute to an underestimation of the total mass, and thus the mass-to-flux ratio, which would reduce the discrepancies mentioned above.

Third, we must consider additional sources of uncertainty. For example, the estimation of the velocity dispersion has been identified as one of the greatest sources of uncertainty in the DCF analysis \citep{Chen2022}.
In addition, geometrical effects, due to an inclined system, could also impact the DCF results \citep{Goncalves2005}.
On the other hand, the IG method is affected by the local turbulence field that could act as a contaminant. In addition, significant changes in temperature over a map would lead to a wrong estimation of the mass that would further lead to wrong gravity direction and therefore to wrong magnetic field strength, causing the method to fail \citep{Koch2012}.

Finally, despite numerous sources of uncertainty that impact all existing methods for estimating the magnetic field, and therefore the mass-to-flux ratio, almost all methods agree in pointing towards gravitational collapse in the most massive condensations (N-7 and S-3), where we also observe a greater distortion of the magnetic field compared to the field at larger scales.
Furthermore, \sigb~analysis indicates that the magnetic tension force is not able to hold the gravitational pull toward the condensation that dominates gravitationally in the region. 
All together supports the hypothesis that collapse has already begun, and the consequent flow of matter towards sinks begins to have an impact on the morphology of the field that is drawn in \citep{Gomez2018, Suin2025}.

\subsection{Multiscale collapse}

$\bullet$ Hub-N. 
\cite{Santos2016} present the magnetic field morphology in the IRDC G14, around the cloud, in optical, and also at filamentary cloud scales in NIR.
\cite{AnezLopez2020} analyze the magnetic field at filament scales (CSO). On these scales, where N-2, N-4, N-5, N-7, and N-8 are together embedded in the $\sim$0.3~pc Hub-N region, they observe that the orientation of the magnetic field becomes mostly perpendicular to the filament system, although it shows slight alterations in the peak emission environment.
During the current work, we zoom in on Hub-N, where the magnetic field on scales of $\sim$0.05~pc exhibits a largely uniform orientation in the southern half. In contrast, the northern half, where the bulk of the mass is concentrated, shows clear signs of magnetic field distortion. 
Overall, we can clearly see the transition of the magnetic field from the scales around the cloud ($\sim$ 141$^{\circ}$), to filamentary cloud scales ($\sim$ 139$^{\circ}$), to filament scales ($\sim$ 105$^{\circ}$), reaching Hub scales in the current observations ($\sim$ 95 $^{\circ}$). We even see sub-regions with dramatic transitions of almost 90$^{\circ}$ (N7).

A similar behavior is reported by \cite{Koch2022ApJ}, who present envelope-to-core scale polarization observations of the high-mass star-forming region W51. They find a predominantly uniform field at large scales, with the first distortions appearing near the intensity peak, and higher-resolution data revealing additional substructure.

This overall magnetic morphology is consistent with theoretical models in which accreting material travels along magnetic field lines that lie perpendicular to the filament \citep{Gomez2018}. These models also predict an eventual flip in the field orientation as material flowing along the filament drags the field.
For example, the N-7 region displays a bimodal distribution that was not present at larger scales: segments in the northern portion align with the large-scale magnetic field, whereas those in the southern portion are oriented perpendicular to it. This configuration would be compatible with accretion from the south, moving parallel to the filament’s major axis, that would shape the field.

$\bullet$ Hub-S.
The scarce large-scale magnetic field detection prevents a complete comparison. The field substructure appears more complex, and in principle its orientation does not correspond to the direction either perpendicular or parallel to the filaments. 
On the other hand, if we analyze the magnetic field in the Hub-S as a whole, we see an evident bimodal distribution that resembles that previously observed at filament scales with the CSO telescope (see Fig. \ref{fig:paHistoALMA}). 
However, individually, the condensations show magnetic field morphologies that, although uniform, deviate from the orientation at larger scales.
One possible explanation is that we are resolving the common envelope of the condensations, and therefore, the region is dominated by the collapse of multiple cores that drag the field within their area of influence. The influence of protostar feedback in the region may also be impacting the field morphology \citep{Hull2020}.

$\bullet$ F1.
Region F1 shows a magnetic field aligned with the elongated continuum emission, which roughly follows the orientation of the large-scale filament as expected in a scenario where filaments transport material to the hubs \citep{Gomez2018}, as for example \cite{Peretto2014} shows toward the infrared dark filament SDC13.

In conclusion, we note that both hubs show a bimodal distribution of polarization angles, and in the case of Hub-N, it was not present at larger scales, with peaks separated by approximately 90~$^{\circ}$. In region S, this pattern has been previously observed with CSO data at $\sim$10'' angular resolution. 
In region-N, the emergence of a second peak aligns with regions where the magnetic field becomes distorted, providing further evidence that gravity is starting to influence and bend the field away from its initial, large-scale orientation.
In short, while at filament scales we see magnetic field morphology compatible with accretion towards the filament itself, when we go down the analysis point to collapse towards the individual cores, which is in agreement with the GHC scheme \citep{Gomez2018, Vazquez-Semadeni2019}.

\begin{figure}
    \centering
    \includegraphics[width=0.4\textwidth]{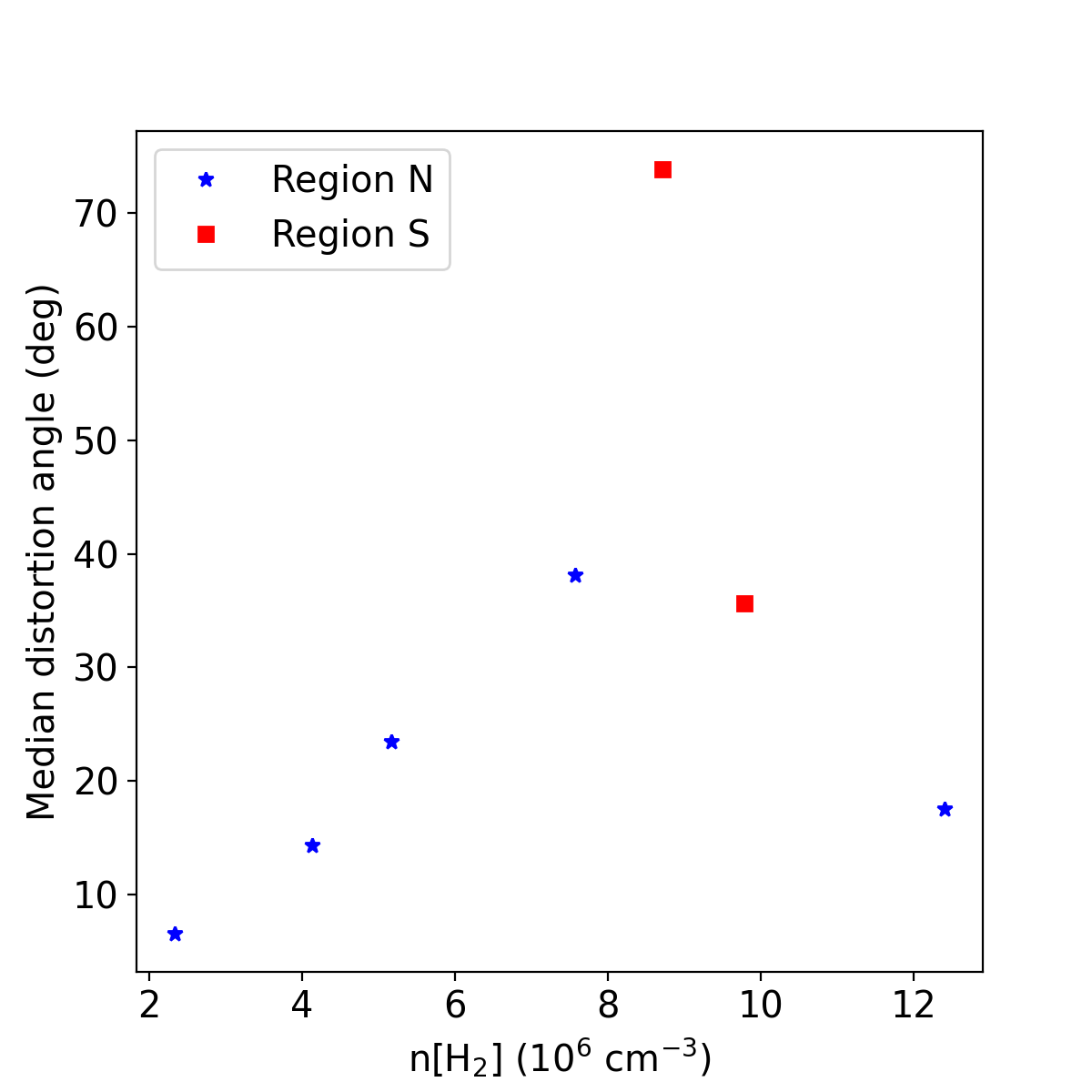}
    \caption{Median distortion of the magnetic field as a function of the averaged number density computed for each individual condensation in both Hub-N and Hub-S, showed by blue stars and red squares respectively.}
    \label{fig:disto-densi}
\end{figure}

Finally, we observed a positive correlation between the median magnetic field distortion and the averaged density of the individual condensation in the range 2-9 $\times$ 10$^{-6}$ cm$^{-3}$; from this density onward, the correlation reverses (see Fig. \ref{fig:disto-densi}). 
In the Hub-filaments system, observations show a similar alignment transition when the density increases (\eg \cite{Pillai2020, Arzoumanian2021} ) as a consequence of material flows \citep{Gomez2018, Suin2025}.
In addition, it is also plausible that the outflows present in the region \citep{Zhao2025arXiv250216413Z} have an impact on both the magnetic field morphology and also the filamentary structures \citep{Suin2025}. So we cannot rule out a combination of factors affecting the morphology of the field.

\section{Conclusion}\label{sec:conclusion}

In the present work, we analyze ALMA polarization observations at 1.39~mm with 2'' (3200 ~AU) angular resolution toward the infrared dark cloud G14.225-0.502, a region composed of two hub-like structures connected by a filament. Previous studies have suggested that these are twin hubs with differing magnetic field strength and fragmentation properties. By investigating the morphology and strength of the magnetic field at small scales, we reached several key conclusions.

First, polarized emission was detected in three out of the five observed fields. Specifically in Regions N, S, and F1 associated with Hub-N, Hub-S, and the Hub-C candidate, multiple emission enhancements were resolved across all five fields. 
On average, the magnetic field strength is a factor of two larger in the Hub-S. However, at the present resolution, Hubs N and S do not exhibit significant differences in their fragmentation levels according to the number of leaves reported by the dendrogram analysis.

Second, in region N the magnetic field morphology is consistent with current models (GHC), where lateral material accretion shapes the magnetic field perpendicularly to the filament. In addition, the magnetic field distortion found toward the more massive condensations could be explained by the flow of material toward the main gravitational sinks within the Hub-N. However, star formation feedback cannot be ruled out as an ingredient, especially in the Hub-S, where we see a generally less homogeneous magnetic field.

Finally, the derived \sigb~values, together with the mass-to-flux ratios, strongly support the interpretation that gravity is the prevailing force in these distorted regions. 
The areas associated with the second polarization angle peak not only show \sigb $< 1$, indicating a supercritical state, but also coincide with zones of maximum magnetic field distortion. 
Crucially, we are able to trace a clear transition in the magnetic field morphology from a uniform configuration to a more complex, gravity-influenced structure. 
This direct link between field morphology and the \sigb~parameter provides compelling observational evidence that gravity increasingly dominates magnetic support at smaller spatial scales, shaping the magnetic field during the early stages of core collapse.

\begin{acknowledgements}
We would like to thank the anonymous referee for their contribution to the improvement of the manuscript through their constructive comments.
We acknowledge Dr. F. Alves for the constructive discussion and for sharing his perspective on the article.
N.A.L. acknowledge financial support from the European Research Council (ERC) via the ERC Synergy Grant
ECOGAL (grant 855130).
G.B. and J.M.G. acknowledges support from the PID2020-117710GB-I00 grant funded by MCIN AEI/10.13039/501100011033 and from the PID2023-146675NB-I00 (MCI-AEI-FEDER, UE) program.
J.M.G. and A.M. has been supported by the program Unidad de Excelencia María de Maeztu, awarded to the Institut de Ciències de l'Espai (CEX2020-001058-M).
J.L. is partially supported by Grant-in-Aid for Scientific Research (KAKENHI Number 23H01221 and 25K17445) of the Japan Society for the Promotion of Science (JSPS).
H.B.L. is supported by the National Science and Technology Council (NSTC) of Taiwan (Grant Nos. 111-2112-M-110-022-MY3, 113-2112-M-110-022-MY3). 
Z.Y.L. is supported in part by NASA 80NSSC20K0533, NSF AST-2307199, and the Virginia Institute of Theoretical Astronomy (VITA).
H.V.C. acknowledges financial support from National Science and Technology Council (NSTC) of Taiwan grants NSTC 112-2112-M-007-041.
S.P.L. acknowledges support from the National Science and Technology Council (NSTC) of Taiwan under grants 109-2112-M-007-010-MY3, 112-2112-M-007-011, and 113-2112-M-007-004.

This research made use of astrodendro, a Python package to compute dendrograms of Astronomical data (http://www.dendrograms.org/) and Astropy (http://www.astropy.org), also CASA version 6.5.5.
\end{acknowledgements}

\bibliography{bibliografia}{}

@ARTICLE{Andre1993,
       author = {{Andre}, Philippe and {Ward-Thompson}, Derek and {Barsony}, Mary},
        title = "{Submillimeter Continuum Observations of rho Ophiuchi A: The Candidate Protostar VLA 1623 and Prestellar Clumps}",
      journal = {\apj},
         year = 1993,
        month = mar,
       volume = {406},
        pages = {122},
          doi = {10.1086/172425},
       adsurl = {https://ui.adsabs.harvard.edu/abs/1993ApJ...406..122A}
}

@ARTICLE{AnezLopez2020,
       author = {{A{\~n}ez-L{\'o}pez}, N. and {Busquet}, G. and {Koch}, P.~M. and {Girart}, J.~M. and {Liu}, H.~B. and {Santos}, F. and {Chapman}, N.~L. and {Novak}, G. and {Palau}, A. and {Ho}, P.~T.~P. and {Zhang}, Q.},
        title = "{Role of the magnetic field in the fragmentation process: the case of G14.225-0.506}",
      journal = {\aap},
         year = 2020,
        month = dec,
       volume = {644},
          eid = {A52},
        pages = {A52},
          doi = {10.1051/0004-6361/202039152},
archivePrefix = {arXiv},
       eprint = {2010.13503},
 primaryClass = {astro-ph.GA},
       adsurl = {https://ui.adsabs.harvard.edu/abs/2020A&A...644A..52A}
}

@ARTICLE{Arzoumanian2021,
       author = {{Arzoumanian}, D. and {Furuya}, R.~S. and {Hasegawa}, T. and {Tahani}, M. and {Sadavoy}, S. and {Hull}, C.~L.~H. and {Johnstone}, D. and {Koch}, P.~M. and {Inutsuka}, S. and {Doi}, Y. and {Hoang}, T. and {Onaka}, T. and {Iwasaki}, K. and {Shimajiri}, Y. and {Inoue}, T. and {Peretto}, N. and {Andr{\'e}}, P. and {Bastien}, P. and {Berry}, D. and {Chen}, H. -R.~V. and {Di Francesco}, J. and {Eswaraiah}, C. and {Fanciullo}, L. and {Fissel}, L.~M. and {Hwang}, J. and {Kang}, J. -h. and {Kim}, G. and {Kim}, K. -T. and {Kirchschlager}, F. and {Kwon}, W. and {Lee}, C.~W. and {Liu}, H. -L. and {Lyo}, A. -R. and {Pattle}, K. and {Soam}, A. and {Tang}, X. and {Whitworth}, A. and {Ching}, T. -C. and {Coud{\'e}}, S. and {Wang}, J. -W. and {Ward-Thompson}, D. and {Lai}, S. -P. and {Qiu}, K. and {Bourke}, T.~L. and {Byun}, D. -Y. and {Chen}, M. and {Chen}, Z. and {Chen}, W.~P. and {Cho}, J. and {Choi}, Y. and {Choi}, M. and {Chrysostomou}, A. and {Chung}, E.~J. and {Dai}, S. and {Diep}, P.~N. and {Duan}, H. -Y. and {Duan}, Y. and {Eden}, D. and {Fiege}, J. and {Franzmann}, E. and {Friberg}, P. and {Fuller}, G. and {Gledhill}, T. and {Graves}, S. and {Greaves}, J. and {Griffin}, M. and {Gu}, Q. and {Han}, I. and {Hatchell}, J. and {Hayashi}, S. and {Houde}, M. and {Jeong}, I. -G. and {Kang}, M. and {Kang}, S. -j. and {Kataoka}, A. and {Kawabata}, K. and {Kemper}, F. and {Kim}, M. -R. and {Kim}, K.~H. and {Kim}, J. and {Kim}, S. and {Kirk}, J. and {Kobayashi}, M.~I.~N. and {K{\"o}nyves}, V. and {Kusune}, T. and {Kwon}, J. and {Lacaille}, K. and {Law}, C. -Y. and {Lee}, C. -F. and {Lee}, Y. -H. and {Lee}, S. -S. and {Lee}, H. and {Lee}, J. -E. and {Li}, H. -b. and {Li}, D. and {Li}, D.~L. and {Liu}, J. and {Liu}, T. and {Liu}, S. -Y. and {Lu}, X. and {Mairs}, S. and {Matsumura}, M. and {Matthews}, B. and {Moriarty-Schieven}, G. and {Nagata}, T. and {Nakamura}, F. and {Nakanishi}, H. and {Ngoc}, N.~B. and {Ohashi}, N. and {Park}, G. and {Parsons}, H. and {Pyo}, T. -S. and {Qian}, L. and {Rao}, R. and {Rawlings}, J. and {Rawlings}, M. and {Retter}, B. and {Richer}, J. and {Rigby}, A. and {Saito}, H. and {Savini}, G. and {Scaife}, A. and {Seta}, M. and {Shinnaga}, H. and {Tamura}, M. and {Tang}, Y. -W. and {Tomisaka}, K. and {Tram}, L.~N. and {Tsukamoto}, Y. and {Viti}, S. and {Wang}, H. and {Xie}, J. and {Yen}, H. -W. and {Yoo}, H. and {Yuan}, J. and {Yun}, H. -S. and {Zenko}, T. and {Zhang}, G. and {Zhang}, C. -P. and {Zhang}, Y. and {Zhou}, J. and {Zhu}, L. and {de Looze}, I. and {Dowell}, C.~D. and {Eyres}, S. and {Falle}, S. and {Friesen}, R. and {Robitaille}, J. -F. and {van Loo}, S.},
        title = "{Dust polarized emission observations of NGC 6334. BISTRO reveals the details of the complex but organized magnetic field structure of the high-mass star-forming hub-filament network}",
      journal = {\aap},
         year = 2021,
        month = mar,
       volume = {647},
          eid = {A78},
        pages = {A78},
          doi = {10.1051/0004-6361/202038624},
archivePrefix = {arXiv},
       eprint = {2012.13060},
 primaryClass = {astro-ph.GA},
       adsurl = {https://ui.adsabs.harvard.edu/abs/2021A&A...647A..78A}
}

@ARTICLE{Bonnel1998,
       author = {{Bonnell}, Ian A. and {Bate}, Matthew R. and {Zinnecker}, Hans},
        title = "{On the formation of massive stars}",
      journal = {\mnras},
         year = 1998,
        month = jul,
       volume = {298},
       number = {1},
        pages = {93-102},
          doi = {10.1046/j.1365-8711.1998.01590.x},
archivePrefix = {arXiv},
       eprint = {astro-ph/9802332},
 primaryClass = {astro-ph},
       adsurl = {https://ui.adsabs.harvard.edu/abs/1998MNRAS.298...93B}
}

@ARTICLE{Busquet2013,
   author = {{Busquet}, G. and {Zhang}, Q. and {Palau}, A. and {Liu}, H.~B. and 
	{S{\'a}nchez-Monge}, {\'A}. and {Estalella}, R. and {Ho}, P.~T.~P. and 
	{de Gregorio-Monsalvo}, I. and {Pillai}, T. and {Wyrowski}, F. and 
	{Girart}, J.~M. and {Santos}, F.~P. and {Franco}, G.~A.~P.},
    title = "{Unveiling a Network of Parallel Filaments in the Infrared Dark Cloud G14.225-0.506}",
  journal = {\apjl},
archivePrefix = "arXiv",
   eprint = {1212.5917},
     year = 2013,
    month = feb,
   volume = 764,
      eid = {L26},
    pages = {L26},
      doi = {10.1088/2041-8205/764/2/L26},
   adsurl = {http://adsabs.harvard.edu/abs/2013ApJ...764L..26B}
}

@ARTICLE{Busquet2016,
   author = {{Busquet}, G. and {Estalella}, R. and {Palau}, A. and {Liu}, H.~B. and 
	{Zhang}, Q. and {Girart}, J.~M. and {de Gregorio-Monsalvo}, I. and 
	{Pillai}, T. and {Anglada}, G. and {Ho}, P.~T.~P.},
    title = "{What Is Controlling the Fragmentation in the Infrared Dark Cloud G14.225-0.506?: Different Levels of Fragmentation in Twin Hubs.}",
  journal = {\apj},
archivePrefix = "arXiv",
   eprint = {1602.02500},
     year = 2016,
    month = mar,
   volume = 819,
      eid = {139},
    pages = {139},
      doi = {10.3847/0004-637X/819/2/139},
   adsurl = {http://adsabs.harvard.edu/abs/2016ApJ...819..139B}
}

@ARTICLE{ChandrasekharFermi1953,
   author = {{Chandrasekhar}, S. and {Fermi}, E.},
    title = "{Magnetic Fields in Spiral Arms.}",
  journal = {\apj},
     year = 1953,
    month = jul,
   volume = 118,
    pages = {113},
      doi = {10.1086/145731},
   adsurl = {http://adsabs.harvard.edu/abs/1953ApJ...118..113C}
}

@ARTICLE{Chen2019,
       author = {{Chen}, Huei-Ru Vivien and {Zhang}, Qizhou and {Wright}, M.~C.~H. and {Busquet}, Gemma and {Lin}, Yuxin and {Liu}, Hauyu Baobab and {Olguin}, F.~A. and {Sanhueza}, Patricio and {Nakamura}, Fumitaka and {Palau}, Aina and {Ohashi}, Satoshi and {Tatematsu}, Ken'ichi and {Liao}, Li-Wen},
        title = "{Filamentary Accretion Flows in the Infrared Dark Cloud G14.225-0.506 Revealed by ALMA}",
      journal = {\apj},
         year = 2019,
        month = apr,
       volume = {875},
       number = {1},
          eid = {24},
        pages = {24},
          doi = {10.3847/1538-4357/ab0f3e},
archivePrefix = {arXiv},
       eprint = {1903.04376},
 primaryClass = {astro-ph.GA},
       adsurl = {https://ui.adsabs.harvard.edu/abs/2019ApJ...875...24C}
}

@ARTICLE{Chen2022,
       author = {{Chen}, Che-Yu and {Li}, Zhi-Yun and {Mazzei}, Renato R. and {Park}, Jinsoo and {Fissel}, Laura M. and {Chen}, Michael C. -Y. and {Klein}, Richard I. and {Li}, Pak Shing},
        title = "{The Davis-Chandrasekhar-Fermi method revisited}",
      journal = {\mnras},
         year = 2022,
        month = aug,
       volume = {514},
       number = {2},
        pages = {1575-1594},
          doi = {10.1093/mnras/stac1417},
archivePrefix = {arXiv},
       eprint = {2205.09134},
 primaryClass = {astro-ph.GA},
       adsurl = {https://ui.adsabs.harvard.edu/abs/2022MNRAS.514.1575C}
}

@ARTICLE{Cortes2021,
       author = {{Cort{\'e}s}, Paulo C. and {Sanhueza}, Patricio and {Houde}, Martin and {Mart{\'\i}n}, Sergio and {Hull}, Charles L.~H. and {Girart}, Josep M. and {Zhang}, Qizhou and {Fernandez-Lopez}, Manuel and {Zapata}, Luis A. and {Stephens}, Ian W. and {Li}, Hua-bai and {Wu}, Benjamin and {Olguin}, Fernando and {Lu}, Xing and {Guzm{\'a}n}, Andres E. and {Nakamura}, Fumitaka},
        title = "{Magnetic Fields in Massive Star-forming Regions (MagMaR). II. Tomography through Dust and Molecular Line Polarization in NGC 6334I(N)}",
      journal = {\apj},
         year = 2021,
        month = dec,
       volume = {923},
       number = {2},
          eid = {204},
        pages = {204},
          doi = {10.3847/1538-4357/ac28a1},
archivePrefix = {arXiv},
       eprint = {2109.09270},
 primaryClass = {astro-ph.GA},
       adsurl = {https://ui.adsabs.harvard.edu/abs/2021ApJ...923..204C}
}

@ARTICLE{Crutcher2004,
       author = {{Crutcher}, Richard M. and {Nutter}, D.~J. and {Ward-Thompson}, D. and
         {Kirk}, J.~M.},
        title = "{SCUBA Polarization Measurements of the Magnetic Field Strengths in the L183, L1544, and L43 Prestellar Cores}",
      journal = {\apj},
         year = 2004,
        month = jan,
       volume = {600},
       number = {1},
        pages = {279-285},
          doi = {10.1086/379705},
archivePrefix = {arXiv},
       eprint = {astro-ph/0305604},
 primaryClass = {astro-ph},
       adsurl = {https://ui.adsabs.harvard.edu/abs/2004ApJ...600..279C}
}

@article{DavisPhysRev,
  title = {The Strength of Interstellar Magnetic Fields},
  author = {Davis, Leverett},
  journal = {Phys. Rev.},
  volume = {81},
  issue = {5},
  pages = {890--891},
  numpages = {0},
  year = {1951},
  month = {Mar},
  publisher = {American Physical Society},
  doi = {10.1103/PhysRev.81.890.2},
  url = {https://link.aps.org/doi/10.1103/PhysRev.81.890.2}
}

@ARTICLE{Davis1951,
       author = {{Davis}, Leverett, Jr. and {Greenstein}, Jesse L.},
        title = "{The Polarization of Starlight by Aligned Dust Grains.}",
      journal = {\apj},
         year = 1951,
        month = sep,
       volume = {114},
        pages = {206},
          doi = {10.1086/145464},
       adsurl = {https://ui.adsabs.harvard.edu/abs/1951ApJ...114..206D}
}

@ARTICLE{DiazMarquez2024,
       author = {{D{\'\i}az-M{\'a}rquez}, Elena and {Grau}, Roger and {Busquet}, Gemma and {Girart}, Josep Miquel and {S{\'a}nchez-Monge}, {\'A}lvaro and {Palau}, Aina and {Povich}, Matthew S. and {A{\~n}ez-L{\'o}pez}, Nacho and {Liu}, Hauyu Baobab and {Zhang}, Qizhou and {Estalella}, Robert},
        title = "{Radio survey of the stellar population in the infrared dark cloud G14.225-0.506}",
      journal = {\aap},
         year = 2024,
        month = feb,
       volume = {682},
          eid = {A180},
        pages = {A180},
          doi = {10.1051/0004-6361/202348085},
archivePrefix = {arXiv},
       eprint = {2311.12542},
 primaryClass = {astro-ph.GA},
       adsurl = {https://ui.adsabs.harvard.edu/abs/2024A&A...682A.180D}
}

@ARTICLE{Falceta-Goncalves2008,
       author = {{Falceta-Gon{\c{c}}alves}, Diego and {Lazarian}, Alex and {Kowal}, Grzegorz},
        title = "{Studies of Regular and Random Magnetic Fields in the ISM: Statistics of Polarization Vectors and the Chandrasekhar-Fermi Technique}",
      journal = {\apj},
         year = 2008,
        month = may,
       volume = {679},
       number = {1},
        pages = {537-551},
          doi = {10.1086/587479},
archivePrefix = {arXiv},
       eprint = {0801.0279},
 primaryClass = {astro-ph},
       adsurl = {https://ui.adsabs.harvard.edu/abs/2008ApJ...679..537F}
}

@ARTICLE{Gomez2018,
       author = {{G{\'o}mez}, Gilberto C. and {V{\'a}zquez-Semadeni}, Enrique and
         {Zamora-Avil{\'e}s}, Manuel},
        title = "{The magnetic field structure in molecular cloud filaments}",
      journal = {\mnras},
         year = 2018,
        month = nov,
       volume = {480},
       number = {3},
        pages = {2939-2944},
          doi = {10.1093/mnras/sty2018},
archivePrefix = {arXiv},
       eprint = {1801.03169},
 primaryClass = {astro-ph.GA},
       adsurl = {https://ui.adsabs.harvard.edu/abs/2018MNRAS.480.2939G}
}

@ARTICLE{Goncalves2005,
       author = {{Gon{\c{c}}alves}, J. and {Galli}, D. and {Walmsley}, M.},
        title = "{Polarized dust emission of magnetized molecular cloud cores}",
      journal = {\aap},
         year = 2005,
        month = feb,
       volume = {430},
        pages = {979-985},
          doi = {10.1051/0004-6361:20041652},
archivePrefix = {arXiv},
       eprint = {astro-ph/0410390},
 primaryClass = {astro-ph},
       adsurl = {https://ui.adsabs.harvard.edu/abs/2005A&A...430..979G}
}

@ARTICLE{Hildegrand1983,
       author = {{Hildebrand}, R.~H.},
        title = "{The determination of cloud masses and dust characteristics from submillimetre thermal emission.}",
      journal = {\qjras},
         year = 1983,
        month = sep,
       volume = {24},
        pages = {267-282},
       adsurl = {https://ui.adsabs.harvard.edu/abs/1983QJRAS..24..267H}
}

@ARTICLE{Hildebrand2000,
   author = {{Hildebrand}, R.~H. and {Davidson}, J.~A. and {Dotson}, J.~L. and 
	{Dowell}, C.~D. and {Novak}, G. and {Vaillancourt}, J.~E.},
    title = "{A Primer on Far-Infrared Polarimetry}",
  journal = {\pasp},
     year = 2000,
    month = sep,
   volume = 112,
    pages = {1215-1235},
      doi = {10.1086/316613},
   adsurl = {http://adsabs.harvard.edu/abs/2000PASP..112.1215H}
}

@ARTICLE{Hildebrand2009,
   author = {{Hildebrand}, R.~H. and {Kirby}, L. and {Dotson}, J.~L. and 
	{Houde}, M. and {Vaillancourt}, J.~E.},
    title = "{Dispersion of Magnetic Fields in Molecular Clouds. I}",
  journal = {\apj},
archivePrefix = "arXiv",
   eprint = {0811.0813},
     year = 2009,
    month = may,
   volume = 696,
    pages = {567-573},
      doi = {10.1088/0004-637X/696/1/567},
   adsurl = {http://adsabs.harvard.edu/abs/2009ApJ...696..567H}
}

@ARTICLE{Heitsch2001,
       author = {{Heitsch}, Fabian and {Zweibel}, Ellen G. and {Mac Low}, Mordecai-Mark and
         {Li}, Pakshing and {Norman}, Michael L.},
        title = "{Magnetic Field Diagnostics Based on Far-Infrared Polarimetry: Tests Using Numerical Simulations}",
      journal = {\apj},
         year = 2001,
        month = nov,
       volume = {561},
       number = {2},
        pages = {800-814},
          doi = {10.1086/323489},
archivePrefix = {arXiv},
       eprint = {astro-ph/0103286},
 primaryClass = {astro-ph},
       adsurl = {https://ui.adsabs.harvard.edu/abs/2001ApJ...561..800H}
}

@ARTICLE{Houde2009,
       author = {{Houde}, Martin and {Vaillancourt}, John E. and {Hildebrand}, Roger H. and
         {Chitsazzadeh}, Shadi and {Kirby}, Larry},
        title = "{Dispersion of Magnetic Fields in Molecular Clouds. II.}",
      journal = {\apj},
         year = 2009,
        month = dec,
       volume = {706},
       number = {2},
        pages = {1504-1516},
          doi = {10.1088/0004-637X/706/2/1504},
archivePrefix = {arXiv},
       eprint = {0909.5227},
 primaryClass = {astro-ph.GA},
       adsurl = {https://ui.adsabs.harvard.edu/abs/2009ApJ...706.1504H}
}

@ARTICLE{Houde2011,
       author = {{Houde}, Martin and {Rao}, Ramprasad and {Vaillancourt}, John E. and {Hildebrand}, Roger H.},
        title = "{Dispersion of Magnetic Fields in Molecular Clouds. III.}",
      journal = {\apj},
         year = 2011,
        month = jun,
       volume = {733},
       number = {2},
          eid = {109},
        pages = {109},
          doi = {10.1088/0004-637X/733/2/109},
archivePrefix = {arXiv},
       eprint = {1103.4772},
 primaryClass = {astro-ph.GA},
       adsurl = {https://ui.adsabs.harvard.edu/abs/2011ApJ...733..109H}
}

@ARTICLE{Huang2025,
       author = {{Huang}, Bo and {Girart}, Josep M. and {Stephens}, Ian W. and {Myers}, Philip C. and {Zhang}, Qizhou and {Cortes}, Paulo and {S{\'a}nchez-Monge}, {\'A}lvaro and {Fern{\'a}ndez L{\'o}pez}, Manuel and {Le Gouellec}, Valentin J.~M. and {Megeath}, Tom and {Murillo}, Nadia M. and {Carpenter}, John M. and {Li}, Zhi-Yun and {Liu}, Junhao and {Looney}, Leslie W. and {Sadavoy}, Sarah and {Karnath}, Nicole and {Kwon}, Woojin},
        title = "{Characterizing Magnetic Properties of Young Protostars in Orion}",
      journal = {\apj},
         year = 2025,
        month = may,
       volume = {984},
       number = {1},
          eid = {29},
        pages = {29},
          doi = {10.3847/1538-4357/adc30b},
archivePrefix = {arXiv},
       eprint = {2503.14726},
 primaryClass = {astro-ph.GA},
       adsurl = {https://ui.adsabs.harvard.edu/abs/2025ApJ...984...29H}
}

@ARTICLE{Hull2020,
       author = {{Hull}, Charles L.~H. and {Le Gouellec}, Valentin J.~M. and {Girart}, Josep M. and {Tobin}, John J. and {Bourke}, Tyler L.},
        title = "{Understanding the Origin of the Magnetic Field Morphology in the Wide-binary Protostellar System BHR 71}",
      journal = {\apj},
         year = 2020,
        month = apr,
       volume = {892},
       number = {2},
          eid = {152},
        pages = {152},
          doi = {10.3847/1538-4357/ab5809},
archivePrefix = {arXiv},
       eprint = {1910.07290},
 primaryClass = {astro-ph.SR},
       adsurl = {https://ui.adsabs.harvard.edu/abs/2020ApJ...892..152H}
}

@ARTICLE{Junhao2020,
       author = {{Liu}, Junhao and {Zhang}, Qizhou and {Qiu}, Keping and {Liu}, Hauyu Baobab and {Pillai}, Thushara and {Girart}, Josep Miquel and {Li}, Zhi-Yun and {Wang}, Ke},
        title = "{Magnetic Fields in the Early Stages of Massive Star Formation as Revealed by ALMA}",
      journal = {\apj},
         year = 2020,
        month = jun,
       volume = {895},
       number = {2},
          eid = {142},
        pages = {142},
          doi = {10.3847/1538-4357/ab9087},
archivePrefix = {arXiv},
       eprint = {2005.01705},
 primaryClass = {astro-ph.GA},
       adsurl = {https://ui.adsabs.harvard.edu/abs/2020ApJ...895..142L}
}

@ARTICLE{Junhao2023,
       author = {{Liu}, Junhao and {Zhang}, Qizhou and {Koch}, Patrick M. and {Liu}, Hauyu Baobab and {Li}, Zhi-Yun and {Li}, Shanghuo and {Girart}, Josep Miquel and {Chen}, Huei-Ru Vivien and {Ching}, Tao-Chung and {Ho}, Paul T.~P. and {Lai}, Shih-Ping and {Qiu}, Keping and {Rao}, Ramprasad and {Tang}, Ya-wen},
        title = "{Multi-scale Physical Properties of NGC 6334 as Revealed by Local Relative Orientations between Magnetic Fields, Density Gradients, Velocity Gradients, and Gravity}",
      journal = {\apj},
         year = 2023,
        month = mar,
       volume = {945},
       number = {2},
          eid = {160},
        pages = {160},
          doi = {10.3847/1538-4357/acb540},
archivePrefix = {arXiv},
       eprint = {2211.00152},
 primaryClass = {astro-ph.GA},
       adsurl = {https://ui.adsabs.harvard.edu/abs/2023ApJ...945..160L}
}

@ARTICLE{Junhao2023b,
       author = {{Liu}, Junhao and {Zhang}, Qizhou and {Liu}, Hauyu Baobab and {Qiu}, Keping and {Li}, Shanghuo and {Li}, Zhi-Yun and {Ho}, Paul T.~P. and {Girart}, Josep Miquel and {Ching}, Tao-Chung and {Chen}, Huei-Ru Vivien and {Lai}, Shih-Ping and {Rao}, Ramprasad and {Tang}, Ya-wen},
        title = "{Deviation from a Continuous and Universal Turbulence Cascade in NGC 6334 due to Massive Star Formation Activity}",
      journal = {\apj},
         year = 2023,
        month = may,
       volume = {949},
       number = {1},
          eid = {30},
        pages = {30},
          doi = {10.3847/1538-4357/acc4c0},
archivePrefix = {arXiv},
       eprint = {2303.08170},
 primaryClass = {astro-ph.GA},
       adsurl = {https://ui.adsabs.harvard.edu/abs/2023ApJ...949...30L}
}

@ARTICLE{Junhao2024,
       author = {{Liu}, Junhao and {Zhang}, Qizhou and {Lin}, Yuxin and {Qiu}, Keping and {Koch}, Patrick M. and {Liu}, Hauyu Baobab and {Li}, Zhi-Yun and {Girart}, Josep Miquel and {Pillai}, Thushara G.~S. and {Li}, Shanghuo and {Chen}, Huei-Ru Vivien and {Ching}, Tao-Chung and {Ho}, Paul T.~P. and {Lai}, Shih-Ping and {Rao}, Ramprasad and {Tang}, Ya-Wen and {Wang}, Ke},
        title = "{Dark Dragon Breaks Magnetic Chain: Dynamical Substructures of IRDC G28.34 Form in Supported Environments}",
      journal = {\apj},
         year = 2024,
        month = may,
       volume = {966},
       number = {1},
          eid = {120},
        pages = {120},
          doi = {10.3847/1538-4357/ad3105},
archivePrefix = {arXiv},
       eprint = {2403.03437},
 primaryClass = {astro-ph.GA},
       adsurl = {https://ui.adsabs.harvard.edu/abs/2024ApJ...966..120L}
}

@ARTICLE{Koch2012,
   author = {{Koch}, P.~M. and {Tang}, Y.-W. and {Ho}, P.~T.~P.},
    title = "{Magnetic Field Strength Maps for Molecular Clouds: A New Method Based on a Polarization-Intensity Gradient Relation}",
  journal = {\apj},
archivePrefix = "arXiv",
   eprint = {1201.4263},
     year = 2012,
    month = mar,
   volume = 747,
      eid = {79},
    pages = {79},
      doi = {10.1088/0004-637X/747/1/79},
   adsurl = {http://adsabs.harvard.edu/abs/2012ApJ...747...79K}
}

@ARTICLE{Koch2012b,
       author = {{Koch}, Patrick M. and {Tang}, Ya-Wen and {Ho}, Paul T.~P.},
        title = "{Quantifying the Significance of the Magnetic Field from Large-scale Cloud to Collapsing Core: Self-similarity, Mass-to-flux Ratio, and Star Formation Efficiency}",
      journal = {\apj},
         year = "2012",
        month = "Mar",
       volume = {747},
       number = {1},
          eid = {80},
        pages = {80},
          doi = {10.1088/0004-637X/747/1/80},
archivePrefix = {arXiv},
       eprint = {1201.4313},
 primaryClass = {astro-ph.GA},
       adsurl = {https://ui.adsabs.harvard.edu/abs/2012ApJ...747...80K}
}

@ARTICLE{Koch2013,
       author = {{Koch}, Patrick M. and {Tang}, Ya-Wen and {Ho}, Paul T.~P.},
        title = "{Interpreting the Role of the Magnetic Field from Dust Polarization Maps}",
      journal = {\apj},
         year = "2013",
        month = "Sep",
       volume = {775},
       number = {1},
          eid = {77},
        pages = {77},
          doi = {10.1088/0004-637X/775/1/77},
archivePrefix = {arXiv},
       eprint = {1308.6185},
 primaryClass = {astro-ph.GA},
       adsurl = {https://ui.adsabs.harvard.edu/abs/2013ApJ...775...77K}
}

@ARTICLE{Koch2014,
       author = {{Koch}, Patrick M. and {Tang}, Ya-Wen and {Ho}, Paul T.~P. and
         {Zhang}, Qizhou and {Girart}, Josep M. and {Chen}, Huei-Ru Vivien and
         {Frau}, Pau and {Li}, Hua-Bai and {Li}, Zhi-Yun and
         {Liu}, Hau-Yu Baobab and {Padovani}, Marco and {Qiu}, Keping and
         {Yen}, Hsi-Wei and {Chen}, How-Huan and {Ching}, Tao-Chung and
         {Lai}, Shih-Ping and {Rao}, Ramprasad},
        title = "{The Importance of the Magnetic Field from an SMA-CSO-combined Sample of Star-forming Regions}",
      journal = {\apj},
         year = 2014,
        month = dec,
       volume = {797},
       number = {2},
          eid = {99},
        pages = {99},
          doi = {10.1088/0004-637X/797/2/99},
archivePrefix = {arXiv},
       eprint = {1411.3830},
 primaryClass = {astro-ph.GA},
       adsurl = {https://ui.adsabs.harvard.edu/abs/2014ApJ...797...99K}
}

@ARTICLE{Koch2018,
       author = {{Koch}, Patrick M. and {Tang}, Ya-Wen and {Ho}, Paul T.~P. and
         {Yen}, Hsi-Wei and {Su}, Yu-Nung and {Takakuwa}, Shigehisa},
        title = "{Polarization Properties and Magnetic Field Structures in the High-mass Star-forming Region W51 Observed with ALMA}",
      journal = {\apj},
         year = 2018,
        month = mar,
       volume = {855},
       number = {1},
          eid = {39},
        pages = {39},
          doi = {10.3847/1538-4357/aaa4c1},
archivePrefix = {arXiv},
       eprint = {1801.08264},
 primaryClass = {astro-ph.GA},
       adsurl = {https://ui.adsabs.harvard.edu/abs/2018ApJ...855...39K}
}

@ARTICLE{Koch2022ApJ,
       author = {{Koch}, Patrick M. and {Tang}, Ya-Wen and {Ho}, Paul T.~P. and {Hsieh}, Pei-Ying and {Wang}, Jia-Wei and {Yen}, Hsi-Wei and {Duarte-Cabral}, Ana and {Peretto}, Nicolas and {Su}, Yu-Nung},
        title = "{A Multiscale Picture of the Magnetic Field and Gravity from a Large-scale Filamentary Envelope to Core-accreting Dust Lanes in the High-mass Star-forming Region W51}",
      journal = {\apj},
         year = 2022,
        month = nov,
       volume = {940},
       number = {1},
          eid = {89},
        pages = {89},
          doi = {10.3847/1538-4357/ac96e3},
archivePrefix = {arXiv},
       eprint = {2210.07593},
 primaryClass = {astro-ph.GA},
       adsurl = {https://ui.adsabs.harvard.edu/abs/2022ApJ...940...89K}
}

@ARTICLE{Kwon2022,
       author = {{Kwon}, Woojin and {Pattle}, Kate and {Sadavoy}, Sarah and {Hull}, Charles L.~H. and {Johnstone}, Doug and {Ward-Thompson}, Derek and {Di Francesco}, James and {Koch}, Patrick M. and {Furuya}, Ray and {Doi}, Yasuo and {Le Gouellec}, Valentin J.~M. and {Hwang}, Jihye and {Lyo}, A. -Ran and {Soam}, Archana and {Tang}, Xindi and {Hoang}, Thiem and {Kirchschlager}, Florian and {Eswaraiah}, Chakali and {Fanciullo}, Lapo and {Kim}, Kyoung Hee and {Onaka}, Takashi and {K{\"o}nyves}, Vera and {Kang}, Ji-hyun and {Lee}, Chang Won and {Tamura}, Motohide and {Bastien}, Pierre and {Hasegawa}, Tetsuo and {Lai}, Shih-Ping and {Qiu}, Keping and {Berry}, David and {Arzoumanian}, Doris and {Bourke}, Tyler L. and {Byun}, Do-Young and {Chen}, Wen Ping and {Chen}, Huei-Ru Vivien and {Chen}, Mike and {Chen}, Zhiwei and {Ching}, Tao-Chung and {Cho}, Jungyeon and {Choi}, Yunhee and {Choi}, Minho and {Chrysostomou}, Antonio and {Chung}, Eun Jung and {Coud{\'e}}, Simon and {Dai}, Sophia and {Diep}, Pham Ngoc and {Duan}, Yan and {Duan}, Hao-Yuan and {Eden}, David and {Fiege}, Jason and {Fissel}, Laura M. and {Franzmann}, Erica and {Friberg}, Per and {Friesen}, Rachel and {Fuller}, Gary and {Gledhill}, Tim and {Graves}, Sarah and {Greaves}, Jane and {Griffin}, Matt and {Gu}, Qilao and {Han}, Ilseung and {Hatchell}, Jennifer and {Hayashi}, Saeko and {Houde}, Martin and {Inoue}, Tsuyoshi and {Inutsuka}, Shu-ichiro and {Iwasaki}, Kazunari and {Jeong}, Il-Gyo and {Kang}, Miju and {Karoly}, Janik and {Kataoka}, Akimasa and {Kawabata}, Koji and {Kemper}, Francisca and {Kim}, Kee-Tae and {Kim}, Gwanjeong and {Kim}, Mi-Ryang and {Kim}, Shinyoung and {Kim}, Jongsoo and {Kirk}, Jason and {Kobayashi}, Masato I.~N. and {Kusune}, Takayoshi and {Kwon}, Jungmi and {Lacaille}, Kevin and {Law}, Chi-Yan and {Lee}, Chin-Fei and {Lee}, Yong-Hee and {Lee}, Hyeseung and {Lee}, Jeong-Eun and {Lee}, Sang-Sung and {Li}, Dalei and {Li}, Di and {Li}, Hua-bai and {Lin}, Sheng-Jun and {Liu}, Sheng-Yuan and {Liu}, Hong-Li and {Liu}, Junhao and {Liu}, Tie and {Lu}, Xing and {Mairs}, Steve and {Matsumura}, Masafumi and {Matthews}, Brenda and {Moriarty-Schieven}, Gerald and {Nagata}, Tetsuya and {Nakamura}, Fumitaka and {Nakanishi}, Hiroyuki and {Ngoc}, Nguyen Bich and {Ohashi}, Nagayoshi and {Park}, Geumsook and {Parsons}, Harriet and {Peretto}, Nicolas and {Priestley}, Felix and {Pyo}, Tae-Soo and {Qian}, Lei and {Rao}, Ramprasad and {Rawlings}, Jonathan and {Rawlings}, Mark G. and {Retter}, Brendan and {Richer}, John and {Rigby}, Andrew and {Saito}, Hiro and {Savini}, Giorgio and {Seta}, Masumichi and {Shimajiri}, Yoshito and {Shinnaga}, Hiroko and {Tahani}, Mehrnoosh and {Tang}, Ya-Wen and {Tomisaka}, Kohji and {Tram}, Le Ngoc and {Tsukamoto}, Yusuke and {Viti}, Serena and {Wang}, Hongchi and {Wang}, Jia-Wei and {Whitworth}, Anthony and {Wu}, Jintai and {Xie}, Jinjin and {Yen}, Hsi-Wei and {Yoo}, Hyunju and {Yuan}, Jinghua and {Yun}, Hyeong-Sik and {Zenko}, Tetsuya and {Zhang}, Yapeng and {Zhang}, Chuan-Peng and {Zhang}, Guoyin and {Zhou}, Jianjun and {Zhu}, Lei and {de Looze}, Ilse and {Andr{\'e}}, Philippe and {Dowell}, C. Darren and {Eyres}, Stewart and {Falle}, Sam and {Robitaille}, Jean-Fran{\c{c}}ois and {Loo}, Sven van},
        title = "{B-fields in Star-forming Region Observations (BISTRO): Magnetic Fields in the Filamentary Structures of Serpens Main}",
      journal = {\apj},
         year = 2022,
        month = feb,
       volume = {926},
       number = {2},
          eid = {163},
        pages = {163},
          doi = {10.3847/1538-4357/ac4bbe},
archivePrefix = {arXiv},
       eprint = {2201.05059},
 primaryClass = {astro-ph.GA},
       adsurl = {https://ui.adsabs.harvard.edu/abs/2022ApJ...926..163K}
}

@INCOLLECTION{Lazarian2015,
       author = {{Lazarian}, Alexander and {Andersson}, B.-G. and {Hoang}, Thiem},
        title = "{Grain alignment: Role of radiative torques and paramagnetic relaxation}",
    booktitle = {Polarimetry of Stars and Planetary Systems},
         year = 2015,
       editor = {{Kolokolova}, Ludmilla and {Hough}, James and {Levasseur-Regourd}, Anny-Chantal},
        pages = {81},
          doi = {10.48550/arXiv.1511.03696},
       adsurl = {https://ui.adsabs.harvard.edu/abs/2015psps.book...81L}
}

@ARTICLE{Lee2025,
       author = {{Lee}, Han-Tsung and {Tang}, Ya-Wen and {Koch}, Patrick M. and {Wang}, Jia-Wei and {Clarke}, Seamus and {Fuller}, Gary A. and {Peretto}, Nicolas and {Kim}, Won-Ju and {Yen}, Hsi-Wei},
        title = "{From filament to clumps and cores: A multiscale study of fragmentation and the role of the magnetic field and gas velocity in the infrared dark cloud SDC18.624-0.070}",
      journal = {\aap},
         year = 2025,
        month = apr,
       volume = {696},
          eid = {A163},
        pages = {A163},
          doi = {10.1051/0004-6361/202452974},
       adsurl = {https://ui.adsabs.harvard.edu/abs/2025A&A...696A.163L}
}

@ARTICLE{Li2015,
       author = {{Li}, Hua-Bai and {Yuen}, Ka Ho and {Otto}, Frank and {Leung}, Po Kin and {Sridharan}, T.K. and {Zhang}, Qizhou and {Liu}, Hauyu and {Tang}, Ya-Wen and {Qiu}, Keping},
        title = "{Self-similar Fragmentation Regulated by Magnetic Fields
        in a Massive Star Forming Filament}",
      journal = {Nature},
         year = "2015",
        month = "March",
       volume = {20},
       number = {7548},
        pages = {422-430},
          doi = {10.1038/nature14291},
archivePrefix = {arXiv},
       eprint = {},
 primaryClass = {astro-ph},
       adsurl = {https://doi.org/10.1038/nature14291}
}

@ARTICLE{Liu2021,
       author = {{Liu}, Junhao and {Zhang}, Qizhou and {Commer{\c{c}}on}, Beno{\^\i}t and {Valdivia}, Valeska and {Maury}, Ana{\"e}lle and {Qiu}, Keping},
        title = "{Calibrating the Davis-Chandrasekhar-Fermi Method with Numerical Simulations: Uncertainties in Estimating the Magnetic Field Strength from Statistics of Field Orientations}",
      journal = {\apj},
         year = 2021,
        month = oct,
       volume = {919},
       number = {2},
          eid = {79},
        pages = {79},
          doi = {10.3847/1538-4357/ac0cec},
archivePrefix = {arXiv},
       eprint = {2106.09934},
 primaryClass = {astro-ph.GA},
       adsurl = {https://ui.adsabs.harvard.edu/abs/2021ApJ...919...79L}
}

@ARTICLE{Liu2022,
       author = {{Liu}, Junhao and {Zhang}, Qizhou and {Qiu}, Keping},
        title = "{Magnetic field properties in star formation: A review of their analysis methods and interpretation}",
      journal = {Frontiers in Astronomy and Space Sciences},
         year = 2022,
        month = sep,
       volume = {9},
          eid = {943556},
        pages = {943556},
          doi = {10.3389/fspas.2022.943556},
archivePrefix = {arXiv},
       eprint = {2208.06492},
 primaryClass = {astro-ph.GA},
       adsurl = {https://ui.adsabs.harvard.edu/abs/2022FrASS...9.3556L}
}

@ARTICLE{McKee&Tan2003,
       author = {{McKee}, Christopher F. and {Tan}, Jonathan C.},
        title = "{The Formation of Massive Stars from Turbulent Cores}",
      journal = {\apj},
         year = 2003,
        month = mar,
       volume = {585},
       number = {2},
        pages = {850-871},
          doi = {10.1086/346149},
archivePrefix = {arXiv},
       eprint = {astro-ph/0206037},
 primaryClass = {astro-ph},
       adsurl = {https://ui.adsabs.harvard.edu/abs/2003ApJ...585..850M}
}

@ARTICLE{Maury2022,
       author = {{Maury}, Ana{\"e}lle and {Hennebelle}, Patrick and {Girart}, Josep Miquel},
        title = "{Recent progress with observations and models to characterize the magnetic fields from star-forming cores to protostellar disks}",
      journal = {Frontiers in Astronomy and Space Sciences},
         year = 2022,
        month = oct,
       volume = {9},
          eid = {949223},
        pages = {949223},
          doi = {10.3389/fspas.2022.949223},
archivePrefix = {arXiv},
       eprint = {2209.01251},
 primaryClass = {astro-ph.GA},
       adsurl = {https://ui.adsabs.harvard.edu/abs/2022FrASS...9.9223M}
}

@ARTICLE{Ohashi2016,
       author = {{Ohashi}, Satoshi and {Sanhueza}, Patricio and {Chen}, Huei-Ru Vivien and
         {Zhang}, Qizhou and {Busquet}, Gemma and {Nakamura}, Fumitaka and
         {Palau}, Aina and {Tatematsu}, Ken'ichi},
        title = "{Dense Core Properties in the Infrared Dark Cloud G14.225-0.506 Revealed by ALMA}",
      journal = {\apj},
         year = 2016,
        month = dec,
       volume = {833},
       number = {2},
          eid = {209},
        pages = {209},
          doi = {10.3847/1538-4357/833/2/209},
archivePrefix = {arXiv},
       eprint = {1610.08581},
 primaryClass = {astro-ph.GA},
       adsurl = {https://ui.adsabs.harvard.edu/abs/2016ApJ...833..209O}
}

@ARTICLE{Ossenkopf&Henning1994,
       author = {{Ossenkopf}, V. and {Henning}, Th.},
        title = "{Dust opacities for protostellar cores.}",
      journal = {\aap},
         year = 1994,
        month = nov,
       volume = {291},
        pages = {943-959},
       adsurl = {https://ui.adsabs.harvard.edu/abs/1994A&A...291..943O}
}

@ARTICLE{Padoan2020,
       author = {{Padoan}, Paolo and {Pan}, Liubin and {Juvela}, Mika and {Haugb{\o}lle}, Troels and {Nordlund}, {\r{A}}ke},
        title = "{The Origin of Massive Stars: The Inertial-inflow Model}",
      journal = {\apj},
         year = 2020,
        month = sep,
       volume = {900},
       number = {1},
          eid = {82},
        pages = {82},
          doi = {10.3847/1538-4357/abaa47},
archivePrefix = {arXiv},
       eprint = {1911.04465},
 primaryClass = {astro-ph.GA},
       adsurl = {https://ui.adsabs.harvard.edu/abs/2020ApJ...900...82P}
}

@ARTICLE{Palau2021,
       author = {{Palau}, Aina and {Zhang}, Qizhou and {Girart}, Josep M. and {Liu}, Junhao and {Rao}, Ramprasad and {Koch}, Patrick M. and {Estalella}, Robert and {Chen}, Huei-Ru Vivien and {Liu}, Hauyu Baobab and {Qiu}, Keping and {Li}, Zhi-Yun and {Zapata}, Luis A. and {Bontemps}, Sylvain and {Ho}, Paul T.~P. and {Beuther}, Henrik and {Ching}, Tao-Chung and {Shinnaga}, Hiroko and {Ahmadi}, Aida},
        title = "{Does the Magnetic Field Suppress Fragmentation in Massive Dense Cores?}",
      journal = {\apj},
         year = 2021,
        month = may,
       volume = {912},
       number = {2},
          eid = {159},
        pages = {159},
          doi = {10.3847/1538-4357/abee1e},
archivePrefix = {arXiv},
       eprint = {2010.12099},
 primaryClass = {astro-ph.GA},
       adsurl = {https://ui.adsabs.harvard.edu/abs/2021ApJ...912..159P}
}

@INPROCEEDINGS{Pattle2023,
       author = {{Pattle}, K. and {Fissel}, L. and {Tahani}, M. and {Liu}, T. and {Ntormousi}, E.},
        title = "{Magnetic Fields in Star Formation: from Clouds to Cores}",
    booktitle = {Protostars and Planets VII},
         year = 2023,
       editor = {{Inutsuka}, S. and {Aikawa}, Y. and {Muto}, T. and {Tomida}, K. and {Tamura}, M.},
       series = {Astronomical Society of the Pacific Conference Series},
       volume = {534},
        month = jul,
        pages = {193},
          doi = {10.48550/arXiv.2203.11179},
archivePrefix = {arXiv},
       eprint = {2203.11179},
 primaryClass = {astro-ph.GA},
       adsurl = {https://ui.adsabs.harvard.edu/abs/2023ASPC..534..193P}
}

@ARTICLE{Peretto2014,
       author = {{Peretto}, N. and {Fuller}, G.~A. and {Andr{\'e}}, Ph. and
         {Arzoumanian}, D. and {Rivilla}, V.~M. and {Bardeau}, S. and
         {Duarte Puertas}, S. and {Guzman Fernandez}, J.~P. and {Lenfestey}, C. and
         {Li}, G. -X. and {Olguin}, F.~A. and {R{\"o}ck}, B.~R. and
         {de Villiers}, H. and {Williams}, J.},
        title = "{SDC13 infrared dark clouds: Longitudinally collapsing filaments?}",
      journal = {\aap},
         year = 2014,
        month = jan,
       volume = {561},
          eid = {A83},
        pages = {A83},
          doi = {10.1051/0004-6361/201322172},
archivePrefix = {arXiv},
       eprint = {1311.0203},
 primaryClass = {astro-ph.GA},
       adsurl = {https://ui.adsabs.harvard.edu/abs/2014A&A...561A..83P}
}

@ARTICLE{Pillai2020,
       author = {{Pillai}, Thushara G.~S. and {Clemens}, Dan P. and {Reissl}, Stefan and
         {Myers}, Philip C. and {Kauffmann}, Jens and
         {Lopez-Rodriguez}, Enrique and {Alves}, F.~O. and {Franco}, G.~A.~P. and
         {Henshaw}, Jonathan and {Menten}, Karl M. and {Nakamura}, Fumitaka and
         {Seifried}, Daniel and {Sugitani}, Koji and {Wiesemeyer}, Helmut},
        title = "{Magnetized filamentary gas flows feeding the young embedded cluster in Serpens South}",
      journal = {Nature Astronomy},
         year = 2020,
        month = aug,
          doi = {10.1038/s41550-020-1172-6},
       adsurl = {https://ui.adsabs.harvard.edu/abs/2020NatAs.tmp..159P}
}

@ARTICLE{PovichWhitney2010,
   author = {{Povich}, M.~S. and {Whitney}, B.~A.},
    title = "{Evidence for Delayed Massive Star Formation in the M17 Proto-OB Association}",
  journal = {\apjl},
archivePrefix = "arXiv",
   eprint = {1004.1712},
     year = 2010,
    month = may,
   volume = 714,
    pages = {L285-L289},
      doi = {10.1088/2041-8205/714/2/L285},
   adsurl = {http://adsabs.harvard.edu/abs/2010ApJ...714L.285P}
}

@ARTICLE{Povich2016,
   author = {{Povich}, M.~S. and {Townsley}, L.~K. and {Robitaille}, T.~P. and 
	{Broos}, P.~S. and {Orbin}, W.~T. and {King}, R.~R. and {Naylor}, T. and 
	{Whitney}, B.~A.},
    title = "{Rapid Circumstellar Disk Evolution and an Accelerating Star Formation Rate in the Infrared Dark Cloud M17 SWex}",
  journal = {\apj},
archivePrefix = "arXiv",
   eprint = {1604.06497},
     year = 2016,
    month = jul,
   volume = 825,
      eid = {125},
    pages = {125},
      doi = {10.3847/0004-637X/825/2/125},
   adsurl = {http://adsabs.harvard.edu/abs/2016ApJ...825..125P}
}

@ARTICLE{Rosolowsky2008,
       author = {{Rosolowsky}, E.~W. and {Pineda}, J.~E. and {Kauffmann}, J. and {Goodman}, A.~A.},
        title = "{Structural Analysis of Molecular Clouds: Dendrograms}",
      journal = {\apj},
         year = 2008,
        month = jun,
       volume = {679},
       number = {2},
        pages = {1338-1351},
          doi = {10.1086/587685},
archivePrefix = {arXiv},
       eprint = {0802.2944},
 primaryClass = {astro-ph},
       adsurl = {https://ui.adsabs.harvard.edu/abs/2008ApJ...679.1338R}
}

@ARTICLE{Santos2016,
       author = {{Santos}, F{\'a}bio P. and {Busquet}, Gemma and {Franco}, Gabriel A.~P. and
         {Girart}, Josep Miquel and {Zhang}, Qizhou},
        title = "{Magnetically Dominated Parallel Interstellar Filaments in the Infrared Dark Cloud G14.225-0.506}",
      journal = {\apj},
         year = "2016",
        month = "Dec",
       volume = {832},
       number = {2},
          eid = {186},
        pages = {186},
          doi = {10.3847/0004-637X/832/2/186},
archivePrefix = {arXiv},
       eprint = {1609.08052},
 primaryClass = {astro-ph.GA},
       adsurl = {https://ui.adsabs.harvard.edu/abs/2016ApJ...832..186S}
}

@ARTICLE{Suin2025,
       author = {{Suin}, P. and {Arzoumanian}, D. and {Zavagno}, A. and {Hennebelle}, P.},
        title = "{The role of magnetic field and stellar feedback in the evolution of filamentary structures in collapsing star-forming clouds}",
      journal = {\aap},
         year = 2025,
        month = jun,
       volume = {698},
          eid = {A119},
        pages = {A119},
          doi = {10.1051/0004-6361/202553795},
archivePrefix = {arXiv},
       eprint = {2505.02903},
 primaryClass = {astro-ph.GA},
       adsurl = {https://ui.adsabs.harvard.edu/abs/2025A&A...698A.119S}
}

@ARTICLE{Tang2009A,
       author = {{Tang}, Ya-Wen and {Ho}, Paul T.~P. and {Koch}, Patrick M. and
         {Girart}, Josep M. and {Lai}, Shih-Ping and {Rao}, Ramprasad},
        title = "{Evolution of Magnetic Fields in High-Mass Star Formation: Linking Field Geometry and Collapse for the W51 e2/e8 Cores}",
      journal = {\apj},
         year = 2009,
        month = jul,
       volume = {700},
       number = {1},
        pages = {251-261},
          doi = {10.1088/0004-637X/700/1/251},
archivePrefix = {arXiv},
       eprint = {0905.1996},
 primaryClass = {astro-ph.SR},
       adsurl = {https://ui.adsabs.harvard.edu/abs/2009ApJ...700..251T}
}

@ARTICLE{Tang2019,
       author = {{Tang}, Ya-Wen and {Koch}, Patrick M. and {Peretto}, Nicolas and
         {Novak}, Giles and {Duarte-Cabral}, Ana and {Chapman}, Nicholas L. and
         {Hsieh}, Pei-Ying and {Yen}, Hsi-Wei},
        title = "{Gravity, Magnetic Field, and Turbulence: Relative Importance and Impact on Fragmentation in the Infrared Dark Cloud G34.43+00.24}",
      journal = {\apj},
         year = 2019,
        month = jun,
       volume = {878},
       number = {1},
          eid = {10},
        pages = {10},
          doi = {10.3847/1538-4357/ab1484},
archivePrefix = {arXiv},
       eprint = {1903.12397},
 primaryClass = {astro-ph.GA},
       adsurl = {https://ui.adsabs.harvard.edu/abs/2019ApJ...878...10T}
}

@ARTICLE{Vazquez-Semadeni2019,
       author = {{V{\'a}zquez-Semadeni}, Enrique and {Palau}, Aina and
         {Ballesteros-Paredes}, Javier and {G{\'o}mez}, Gilberto C. and
         {Zamora-Avil{\'e}s}, Manuel},
        title = "{Global hierarchical collapse in molecular clouds. Towards a comprehensive scenario}",
      journal = {\mnras},
         year = 2019,
        month = dec,
       volume = {490},
       number = {3},
        pages = {3061-3097},
          doi = {10.1093/mnras/stz2736},
archivePrefix = {arXiv},
       eprint = {1903.11247},
 primaryClass = {astro-ph.GA},
       adsurl = {https://ui.adsabs.harvard.edu/abs/2019MNRAS.490.3061V}
}

@ARTICLE{Wang2006,
       author = {{Wang}, Yang and {Zhang}, Qizhou and {Rathborne}, Jill M. and
         {Jackson}, James and {Wu}, Yuefang},
        title = "{Water Masers Associated with Infrared Dark Cloud Cores}",
      journal = {\apjl},
         year = "2006",
        month = "Nov",
       volume = {651},
       number = {2},
        pages = {L125-L128},
          doi = {10.1086/508939},
       adsurl = {https://ui.adsabs.harvard.edu/abs/2006ApJ...651L.125W}
}

@ARTICLE{Wu2024A,
       author = {{Wu}, Jintai and {Qiu}, Keping and {Poidevin}, Fr{\'e}d{\'e}rick and {Bastien}, Pierre and {Liu}, Junhao and {Ching}, Tao-Chung and {Bourke}, Tyler L. and {Ward-Thompson}, Derek and {Pattle}, Kate and {Johnstone}, Doug and {Koch}, Patrick M. and {Arzoumanian}, Doris and {Lee}, Chang Won and {Fanciullo}, Lapo and {Onaka}, Takashi and {Hwang}, Jihye and {Le Gouellec}, Valentin J.~M. and {Soam}, Archana and {Tamura}, Motohide and {Tahani}, Mehrnoosh and {Eswaraiah}, Chakali and {Li}, Hua-Bai and {Berry}, David and {Furuya}, Ray S. and {Coud{\'e}}, Simon and {Kwon}, Woojin and {Lin}, Sheng-Jun and {Wang}, Jia-Wei and {Hasegawa}, Tetsuo and {Lai}, Shih-Ping and {Byun}, Do-Young and {Chen}, Zhiwei and {Chen}, Huei-Ru Vivien and {Chen}, Wen Ping and {Chen}, Mike and {Cho}, Jungyeon and {Choi}, Youngwoo and {Choi}, Yunhee and {Choi}, Minho and {Chrysostomou}, Antonio and {Chung}, Eun Jung and {Dai}, Sophia and {Di Francesco}, James and {Diep}, Pham Ngoc and {Doi}, Yasuo and {Duan}, Hao-Yuan and {Duan}, Yan and {Eden}, David and {Fiege}, Jason and {Fissel}, Laura M. and {Franzmann}, Erica and {Friberg}, Per and {Friesen}, Rachel and {Fuller}, Gary and {Gledhill}, Tim and {Graves}, Sarah and {Greaves}, Jane and {Griffin}, Matt and {Gu}, Qilao and {Han}, Ilseung and {Hayashi}, Saeko and {Hoang}, Thiem and {Houde}, Martin and {Inoue}, Tsuyoshi and {Inutsuka}, Shu-ichiro and {Iwasaki}, Kazunari and {Jeong}, Il-Gyo and {K{\"o}nyves}, Vera and {Kang}, Ji-hyun and {Kang}, Miju and {Karoly}, Janik and {Kataoka}, Akimasa and {Kawabata}, Koji and {Kim}, Shinyoung and {Kim}, Mi-Ryang and {Kim}, Kyoung Hee and {Kim}, Kee-Tae and {Kim}, Jongsoo and {Kim}, Hyosung and {Kim}, Gwanjeong and {Kirchschlager}, Florian and {Kirk}, Jason and {Kobayashi}, Masato I.~N. and {Kusune}, Takayoshi and {Kwon}, Jungmi and {Lacaille}, Kevin and {Law}, Chi-Yan and {Lee}, Hyeseung and {Lee}, Chin-Fei and {Lee}, Sang-Sung and {Lee}, Jeong-Eun and {Li}, Dalei and {Li}, Di and {Li}, Guangxing and {Liu}, Sheng-Yuan and {Liu}, Tie and {Liu}, Hong-Li and {Lu}, Xing and {Lyo}, A. -Ran and {Mairs}, Steve and {Matsumura}, Masafumi and {Matthews}, Brenda and {Moriarty-Schieven}, Gerald and {Nagata}, Tetsuya and {Nakamura}, Fumitaka and {Nakanishi}, Hiroyuki and {Ngoc}, Nguyen Bich and {Ohashi}, Nagayoshi and {Park}, Geumsook and {Parsons}, Harriet and {Peretto}, Nicolas and {Priestley}, Felix and {Pyo}, Tae-Soo and {Qian}, Lei and {Rao}, Ramprasad and {Rawlings}, Jonathan and {Rawlings}, Mark and {Retter}, Brendan and {Richer}, John and {Rigby}, Andrew and {Sadavoy}, Sarah and {Saito}, Hiro and {Savini}, Giorgio and {Seta}, Masumichi and {Sharma}, Ekta and {Shimajiri}, Yoshito and {Shinnaga}, Hiroko and {Tang}, Ya-Wen and {Tang}, Xindi and {Thuong}, Hoang Duc and {Tomisaka}, Kohji and {Tram}, Le Ngoc and {Tsukamoto}, Yusuke and {Viti}, Serena and {Wang}, Hongchi and {Whitworth}, Anthony and {Xie}, Jinjin and {Yang}, Meng-Zhe and {Yen}, Hsi-Wei and {Yoo}, Hyunju and {Yuan}, Jinghua and {Yun}, Hyeong-Sik and {Zenko}, Tetsuya and {Zhang}, Guoyin and {Zhang}, Chuan-Peng and {Zhang}, Yapeng and {Zhou}, Jianjun and {Zhu}, Lei and {Looze}, Ilse de and {Andr{\'e}}, Philippe and {Dowell}, C. Darren and {Eyres}, Stewart and {Falle}, Sam and {Robitaille}, Jean-Fran{\c{c}}ois and {van Loo}, Sven},
        title = "{A Tale of Three: Magnetic Fields along the Orion Integral-shaped Filament as Revealed by the JCMT BISTRO Survey}",
      journal = {\apjl},
         year = 2024,
        month = dec,
       volume = {977},
       number = {2},
          eid = {L31},
        pages = {L31},
          doi = {10.3847/2041-8213/ad93d2},
archivePrefix = {arXiv},
       eprint = {2412.17716},
 primaryClass = {astro-ph.GA},
       adsurl = {https://ui.adsabs.harvard.edu/abs/2024ApJ...977L..31W}
}

@ARTICLE{Xu2011,
       author = {{Xu}, Y. and {Moscadelli}, L. and {Reid}, M.~J. and {Menten}, K.~M. and
         {Zhang}, B. and {Zheng}, X.~W. and {Brunthaler}, A.},
        title = "{Trigonometric Parallaxes of Massive Star-forming Regions. VIII. G12.89+0.49, G15.03-0.68 (M17), and G27.36-0.16}",
      journal = {\apj},
         year = "2011",
        month = "May",
       volume = {733},
       number = {1},
          eid = {25},
        pages = {25},
          doi = {10.1088/0004-637X/733/1/25},
archivePrefix = {arXiv},
       eprint = {1103.3139},
 primaryClass = {astro-ph.GA},
       adsurl = {https://ui.adsabs.harvard.edu/abs/2011ApJ...733...25X}
}

@ARTICLE{Zhang2014,
       author = {{Zhang}, Qizhou and {Qiu}, Keping and {Girart}, Josep M. and
         {Liu}, Hauyu Baobab and {Tang}, Ya-Wen and {Koch}, Patrick M. and
         {Li}, Zhi-Yun and {Keto}, Eric and {Ho}, Paul T.~P. and
         {Rao}, Ramprasad and {Lai}, Shih-Ping and {Ching}, Tao-Chung and
         {Frau}, Pau and {Chen}, How-Huan and {Li}, Hua-Bai and
         {Padovani}, Marco and {Bontemps}, Sylvain and {Csengeri}, Timea and
         {Ju{\'a}rez}, Carmen},
        title = "{Magnetic Fields and Massive Star Formation}",
      journal = {\apj},
         year = 2014,
        month = sep,
       volume = {792},
       number = {2},
          eid = {116},
        pages = {116},
          doi = {10.1088/0004-637X/792/2/116},
archivePrefix = {arXiv},
       eprint = {1407.3984},
 primaryClass = {astro-ph.GA},
       adsurl = {https://ui.adsabs.harvard.edu/abs/2014ApJ...792..116Z}
}

@ARTICLE{Zhang2025,
       author = {{Zhang}, Qizhou and {Liu}, Junhao and {Zeng}, Lingzhen and {Soler}, J.~D. and {Chen}, Huei-Ru Vivien and {Ching}, Tao-Chung and {Ho}, Paul T.~P. and {Girart}, Josep Miquel and {Koch}, Patrick M. and {Lai}, Shih-Ping and {Li}, Shanghuo and {Li}, Zhi-Yun and {Liu}, Hauyu Baobab and {Qiu}, Keping and {Rao}, Ramprasad},
        title = "{Impact of Gravity on Changing Magnetic Field Orientations in a Sample of Massive Protostellar Clusters Observed with ALMA}",
      journal = {\apj},
         year = 2025,
        month = oct,
       volume = {992},
       number = {1},
          eid = {103},
        pages = {103},
          doi = {10.3847/1538-4357/adfdcb},
archivePrefix = {arXiv},
       eprint = {2508.18538},
 primaryClass = {astro-ph.GA},
       adsurl = {https://ui.adsabs.harvard.edu/abs/2025ApJ...992..103Z}
}

@ARTICLE{Zhao2025arXiv250216413Z,
       author = {{Zhao}, Yanhanle and {Zhang}, Qizhou and {Liu}, Junhao and {Pan}, Xing and {Zeng}, Lingzhen},
        title = "{Dense Cores in IRDC G14.225-0.506 revealed by ALMA Observations}",
      journal = {arXiv e-prints},
         year = 2025,
        month = feb,
          eid = {arXiv:2502.16413},
        pages = {arXiv:2502.16413},
          doi = {10.48550/arXiv.2502.16413},
archivePrefix = {arXiv},
       eprint = {2502.16413},
 primaryClass = {astro-ph.GA},
       adsurl = {https://ui.adsabs.harvard.edu/abs/2025arXiv250216413Z}
}

@ARTICLE{Zucker2020,
       author = {{Zucker}, Catherine and {Speagle}, Joshua S. and {Schlafly}, Edward F. and {Green}, Gregory M. and {Finkbeiner}, Douglas P. and {Goodman}, Alyssa and {Alves}, Jo{\~a}o},
        title = "{A compendium of distances to molecular clouds in the Star Formation Handbook}",
      journal = {\aap},
         year = 2020,
        month = jan,
       volume = {633},
          eid = {A51},
        pages = {A51},
          doi = {10.1051/0004-6361/201936145},
archivePrefix = {arXiv},
       eprint = {2001.00591},
 primaryClass = {astro-ph.GA},
       adsurl = {https://ui.adsabs.harvard.edu/abs/2020A&A...633A..51Z}
}
\bibliographystyle{aasjournal}

\begin{appendix}
\section{Condensations identified in continuum emission}
\begin{table}[!ht]
\tiny
\centering
\caption{Coordinates of condensations and equivalences with previous detections}
\begin{threeparttable}
\begin{tabular}{c|c|c|l}
\toprule

ID & RA\tnote{a}  & DEC &  Previous detections \\
            &   (hh:mm:ss.ss)       &  (dd:mm:ss.ss) &   and nomenclature. \\
\hline
\hline

N-2 & 18:18:12.60 & -16:50:14.52 &  a[N7], b[MM2e], \\
    &             &              &  c, d[4, 6, 7] \\
N-3 & 18:18:13.49 & -16:50:16.54 &  d[3] \\
N-4 & 18:18:12.58 & -16:49:59.97 &  a[N9, N11], d[14, 16, 18] \\ 
N-5 & 18:18:12.94 & -16:49:42.10 &  a[N3, N23], b[MM3], \\
    &             &              &  d[22, 24, 26, 36, 41] \\  
N-6 & 18:18:14.28 & -16:49:40.78 & d[34] \\
N-7 & 18:18:12.33 & -16:49:27.56 & a[N1, N4], b[MM1a, \\
    &             &              & MM1bd, MM1cd, \\
    &             &              & MM1dd, MM1f, MM1ed, c, \\ 
    &             &              & d[63, 67, 70]\\
N-8 & 18:18:12.87 & -16:49:29.48 &  a[N5], b, d[55, 58, 61, 62]\\
N-9 & 18:18:13.44 & -16:49:22.34 &  d[1, 4, 68]\\
N-10 & 18:18:11.4 & -16:49:19.75 &  d[69]\\
F1 & 18:18:11.27 & -16:52:33.88 & a[N2, N6], d[3, 4, 6, 9, \\ 
   &              &              & 10, 12, 15, 16, 17], \\
   &              &              & e[Hub-C]\\
F2-1 & 18:18:07.57 & -16:51:29.52 & c, d[2, 3]\\
F2-2 & 18:18:06.08 & -16:51:25.71 & d[5, 7] \\
F2-3 & 18:18:06.98 & -16:51:18.75 & a[N13] d[9, 10] \\
F3-1 & 18:18:11.02 & -16:51:49.14 & d[1] \\
F3-2 &18:18:11.61 & -16:51:45.00 &  d[2]\\
F3-3 &18:18:11.58 & -16:51:35.95 &  d[4, 6]\\
F3-4 & 18:18:11.58 & -16:51:20.72 & d[7]\\
S-1 &18:18:11.43 & -16:57:28.77 & a[S5], b[MM1a, MM1b], \\
    &            &              &  d[46, 48] \\
S-2 &18:18:12.46 & -16:57:23.94 & a[S4, S7], b[MM3, MM3a], \\
    &            &              & d[54, 56, 61]\\
S-3 &18:18:13.08 & -16:57:21.99 & a[S1, S2], b[MM5a, MM5b,\\
    &            &              &  MM4], c, d[59, 62, 65]\\
S-5 &18:18:11.30 & -16:57:18.93 & a[S17], d[60, 67, 68] \\
S-6 &18:18:13.92 & -16:57:13.17 & a[S3, S11], b[MM7a, \\ 
    &            &              & MM7bd, MM7cd], c, \\
    &            &              & d[74, 77] \\
S-7 &18:18:14.64 & -16:57:14.90 &   a[S19], d[76]\\
S-8 &18:18:13.39 & -16:57:11.56 & a[S6], b[MM6a, MM6bd],\\
    &            &              & d[78, 80] \\
S-9 &18:18:12.48 & -16:57:06.06 &  d [82, 83, 92, 94, \\
    &            &              &  95, 98, 99]\\
S-10 &18:18:11.70 & -16:57:07.13 &   \\
S-11 &18:18:13.67 & -16:56:56.57 &  d [102, 104] \\
\bottomrule
\end{tabular}
\begin{tablenotes}
\footnotesize
\item[a] \cite{Ohashi2016} \item[b] \cite{Busquet2016} \item[c] \cite{DiazMarquez2024} \item [d] \cite{Zhao2025arXiv250216413Z} \item [e] \cite{Chen2019}.
\end{tablenotes}
\end{threeparttable}
\label{tab:loc-cores}
\end{table}
\section{Regions F2 and F3}

\begin{figure}
    \centering
    \includegraphics[trim= 0cm 0cm 0cm 0cm, clip, width=0.4\linewidth]{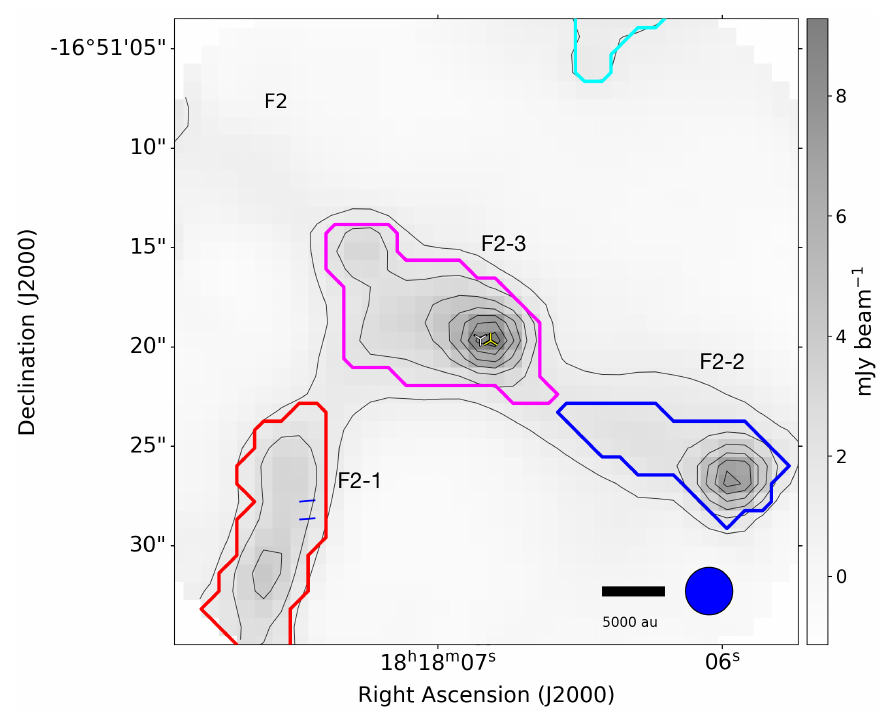}
    \includegraphics[trim= 0cm 0cm 0cm 0cm, clip, width=0.4\linewidth]{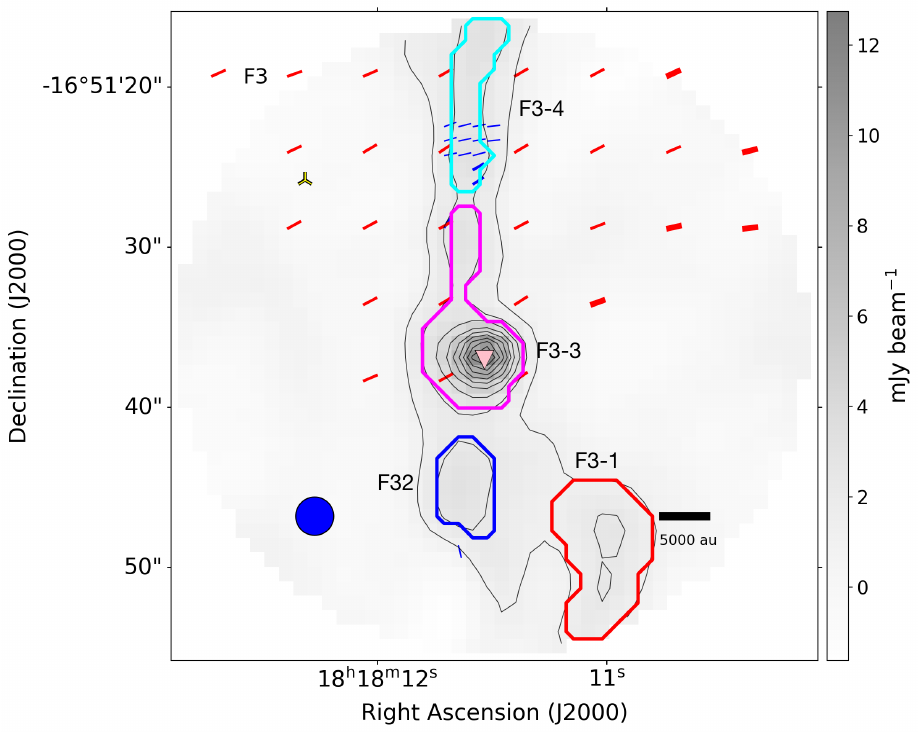}
    \caption{1.39~mm continuum emission, toward region F2 (left panel) and region F3 (right panel), in contours that depict [10, 20, 30, 40, 50, 60, 70, 80, 90, 100, 150, 300, 500, 700] times rms noise (0.12 \mjybeam), overlapped with magnetic field segments in blue, where thin and thick segments depict segments within the inner 50$\%$ and 40$\%$ of the primary beam model respectively. 
    At 350 \mum, the red segments show the CSO magnetic field. The thin segments have a polarization intensity between 2 and 3 sigmas, and the thick segments have an intensity greater than 3 times sigma \citep[CSO sigma = 0.02~\jybeam][]{AnezLopez2020}. Color contours show leaves from dendrogram analysis, also black label present the corresponding ID label. Blue solid circle depict the ALMA beam size. The Pink triangle shows the H$_2$O maser detected in \cite{Wang2006}. Yellow tripods depict radio sources detected at 6~cm (C-band) by \cite{DiazMarquez2024}. White tripods show $NH_3$ cores detected by \cite{Ohashi2016}.}
    \label{fig:leavesF2}
\end{figure}

We present maps of the areas where polarization was either undetectable or barely noticeable in this section (see Fig. \ref{fig:leavesF2}).

\section{Local gravity maps}

\begin{figure}
    \centering
    \includegraphics[trim=0cm 0cm 0cm 0cm, clip,width=0.49\linewidth]{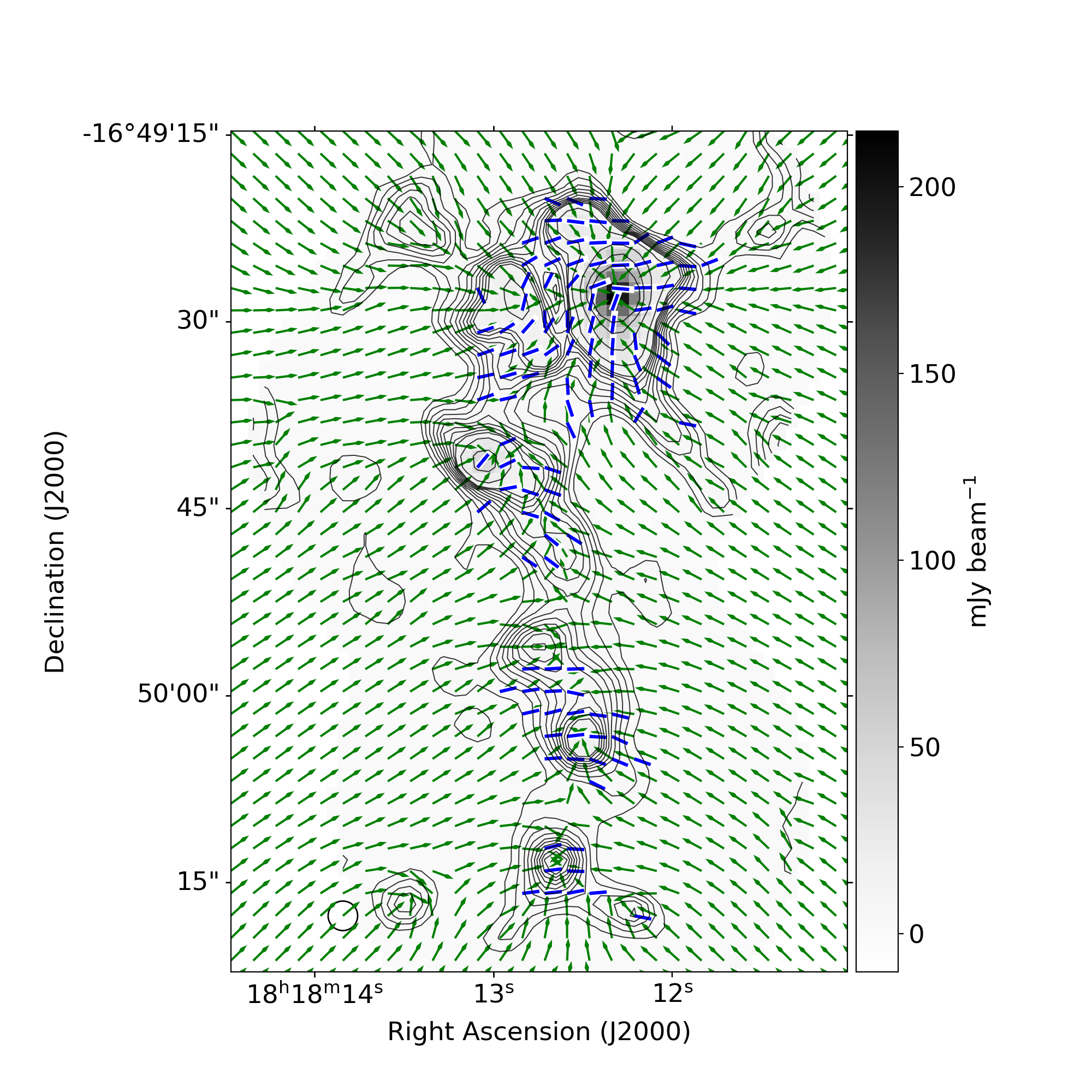}
    \includegraphics[trim=0cm 0cm 0cm 0cm, clip,width=0.49\linewidth]{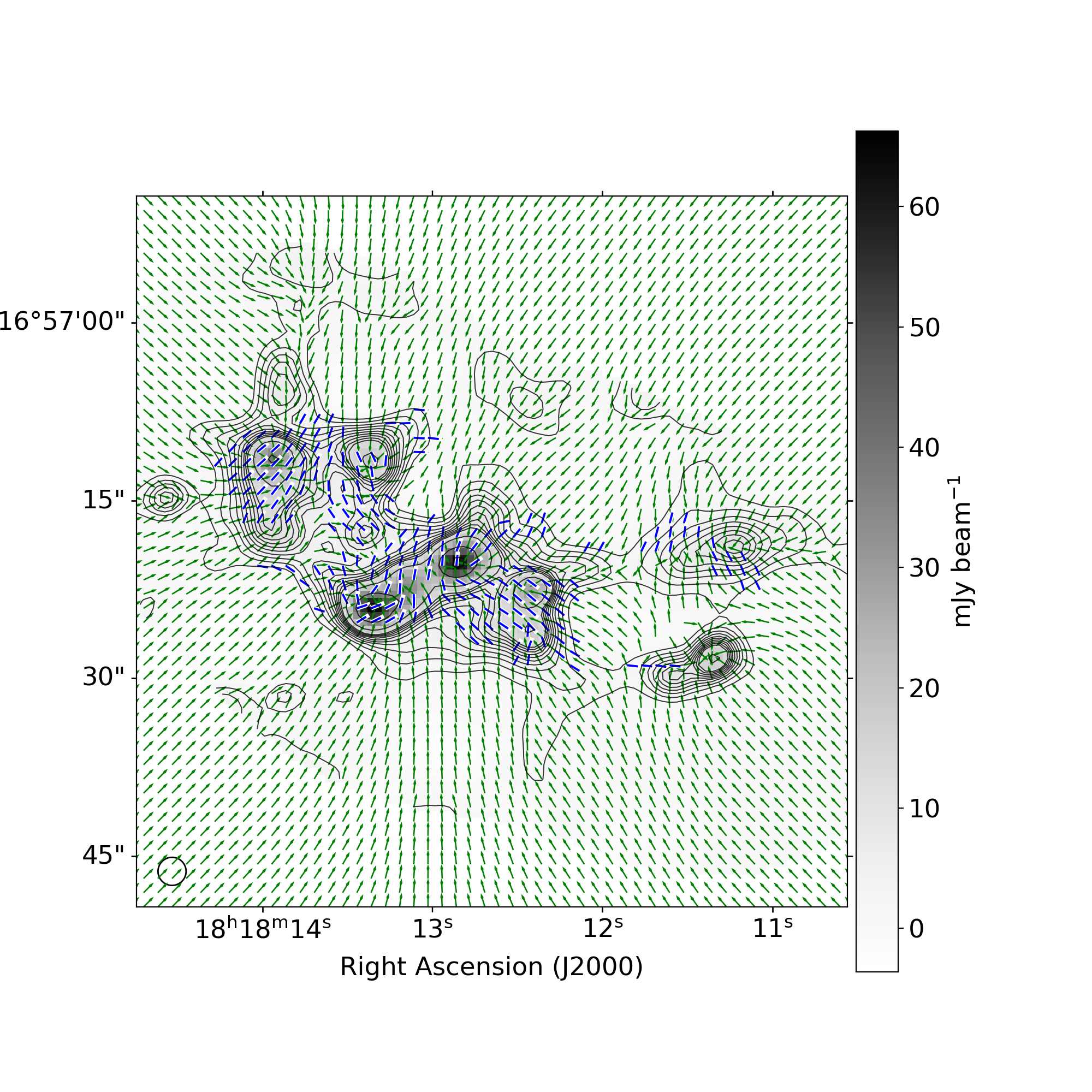}
    \caption{Local gravity normalized vectors overlapped with magnetic field.}
    \label{fig:LocGravALMA_N}
\end{figure}

\begin{figure}
    \centering
    \includegraphics[trim=0cm 0cm 0cm 0cm, clip,width=0.5\linewidth]{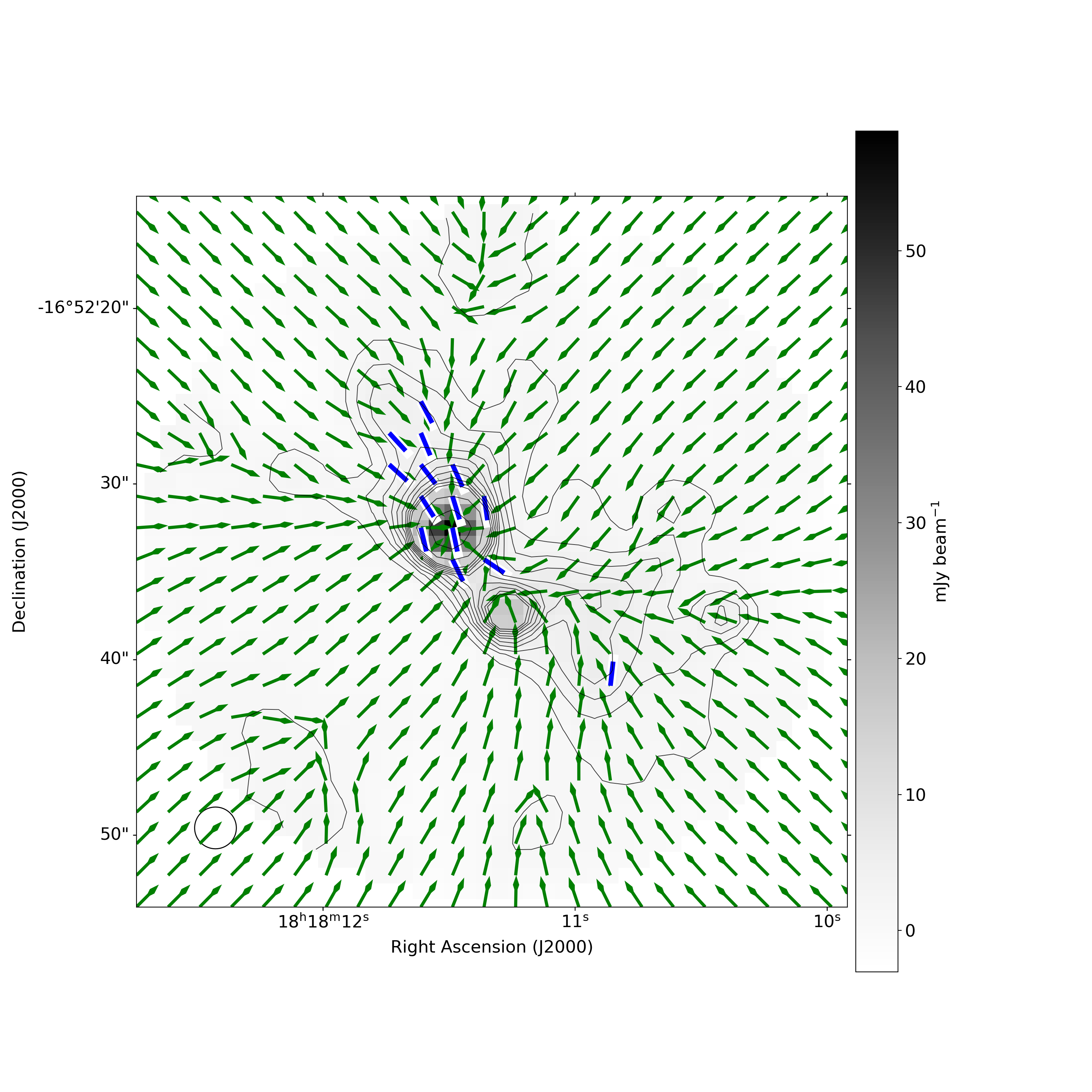}
    \caption{Local gravity normalized vectors}
    \label{fig:LocGravALMA_N1}
\end{figure}

As part of the IG method, we calculated an estimate of the gravitational potential. In this section, we present maps of the N, S, and F1 regions, showing the local gravitational field vectors as well as the magnetic field segments (see Fig. \ref{fig:LocGravALMA_N} and \ref{fig:LocGravALMA_N1}).

\section{Polarization fraction}

\begin{figure*}[!h]
    \centering
    \includegraphics[trim= 0cm 0cm 0cm 0cm ,clip,width=0.24\linewidth]{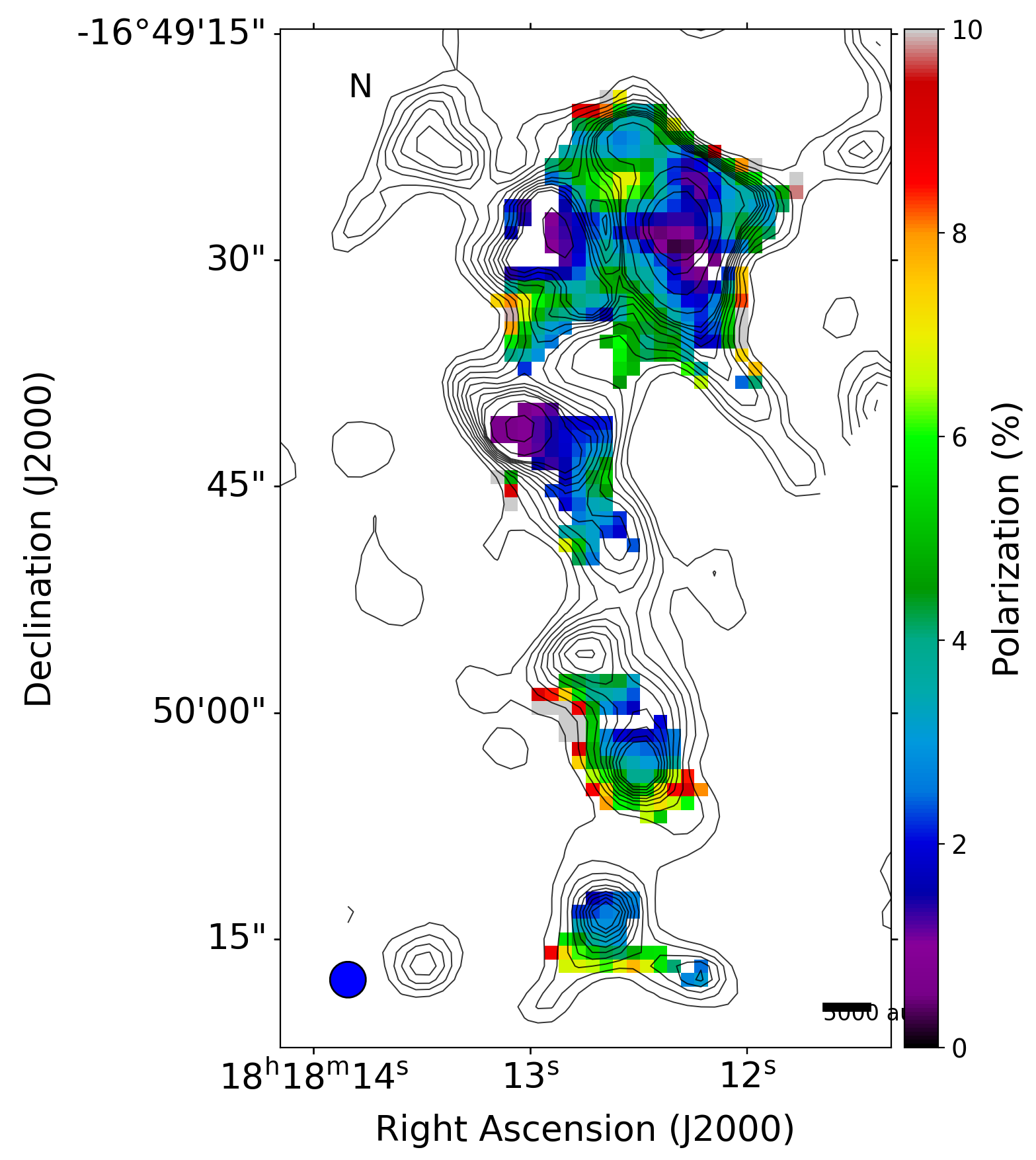}
     \includegraphics[trim= 0cm 0cm 0cm 0cm ,clip,width=0.24\linewidth]{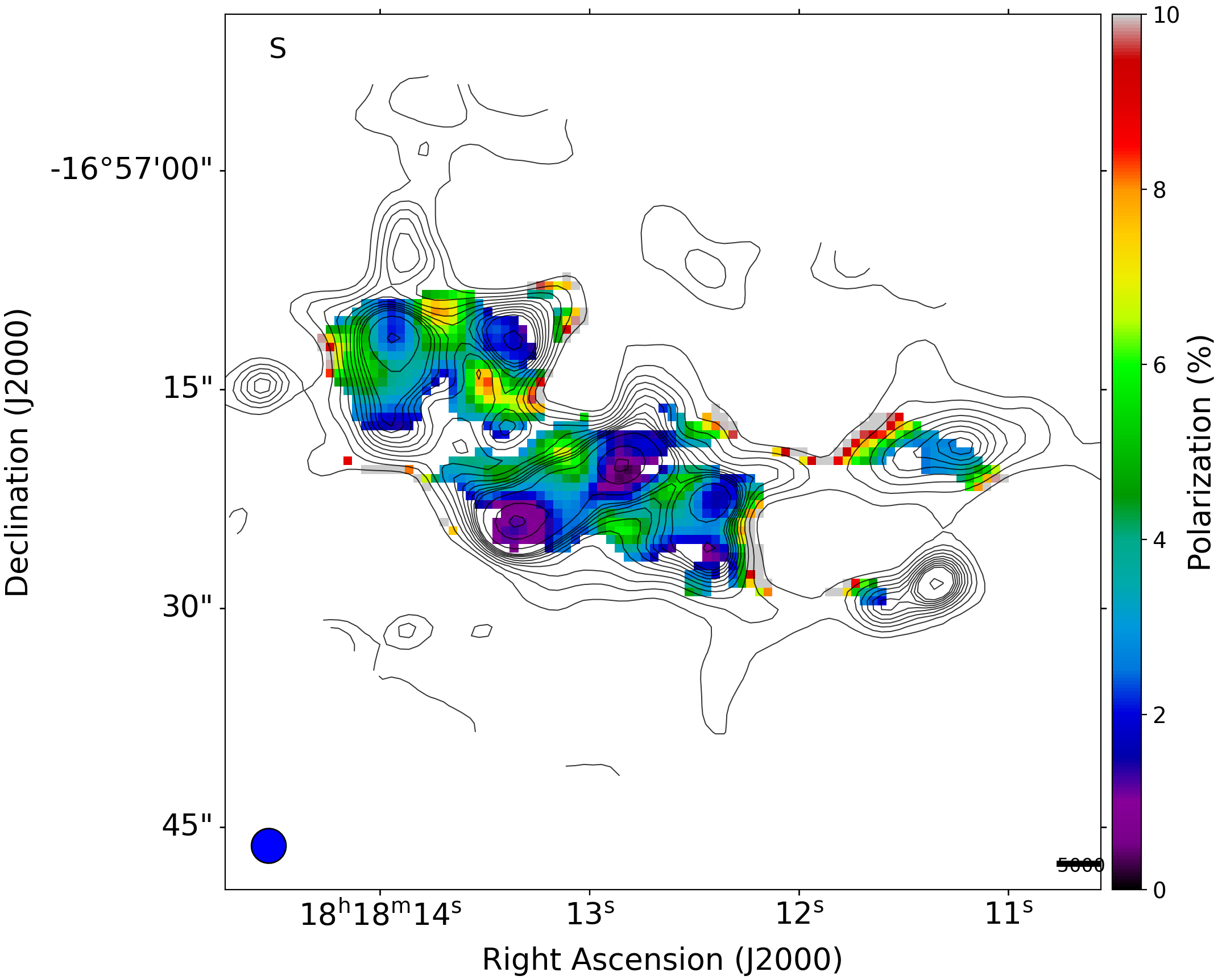}
     \includegraphics[trim= 0cm 0cm 0cm 0cm ,clip,width=0.24\linewidth]{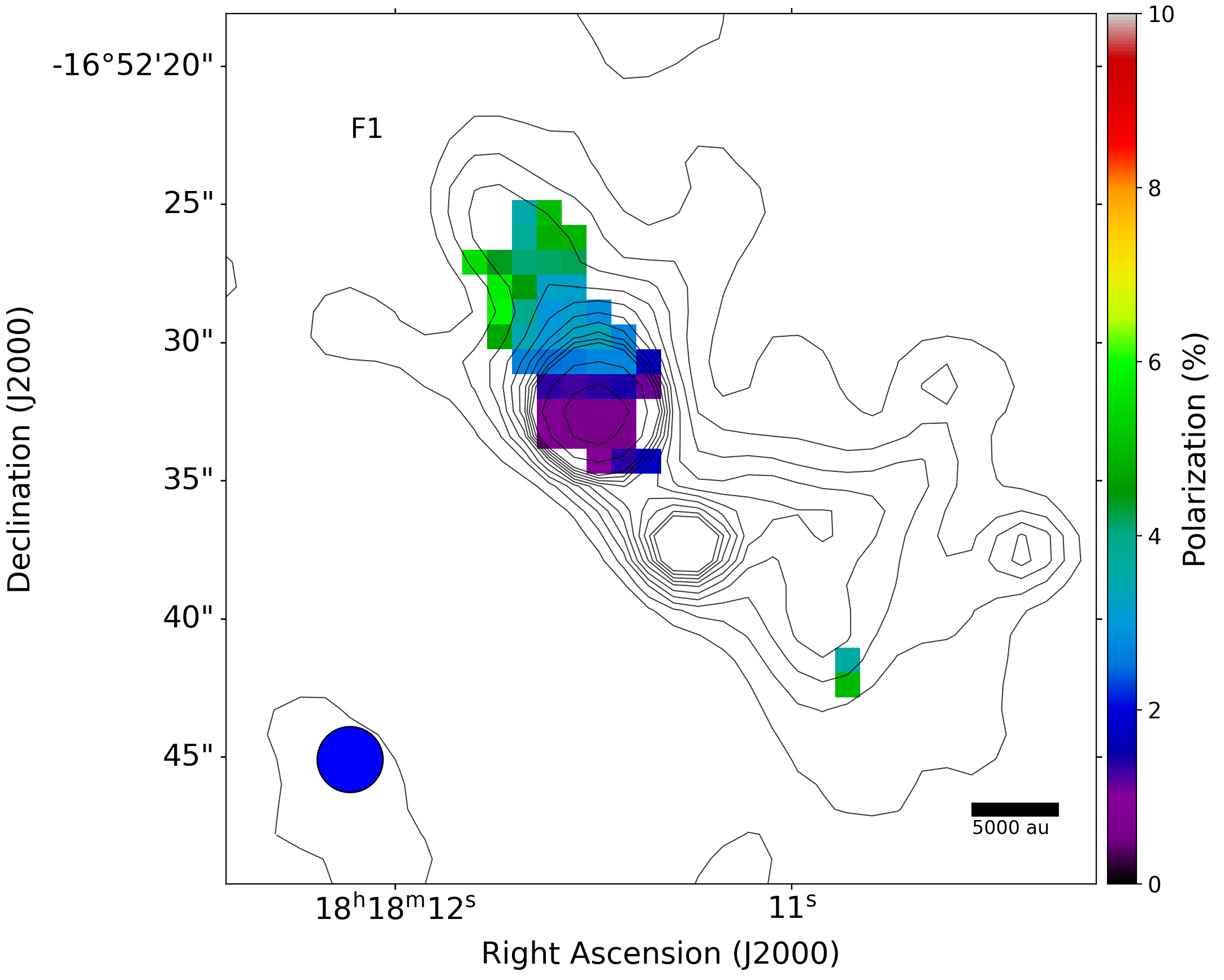}
    \caption{Polarization fraction in colors overlapped with continuum emission in contours.}
    \label{fig:polpercent}
\end{figure*}

In this section we present the maps that display the percentage of polarization detected towards the N, S and F1 regions (see Fig. \ref{fig:polpercent}). 

\section{Structure function}\label{App:sf}

\begin{figure*}
    \centering
    \includegraphics[width=0.25\linewidth]{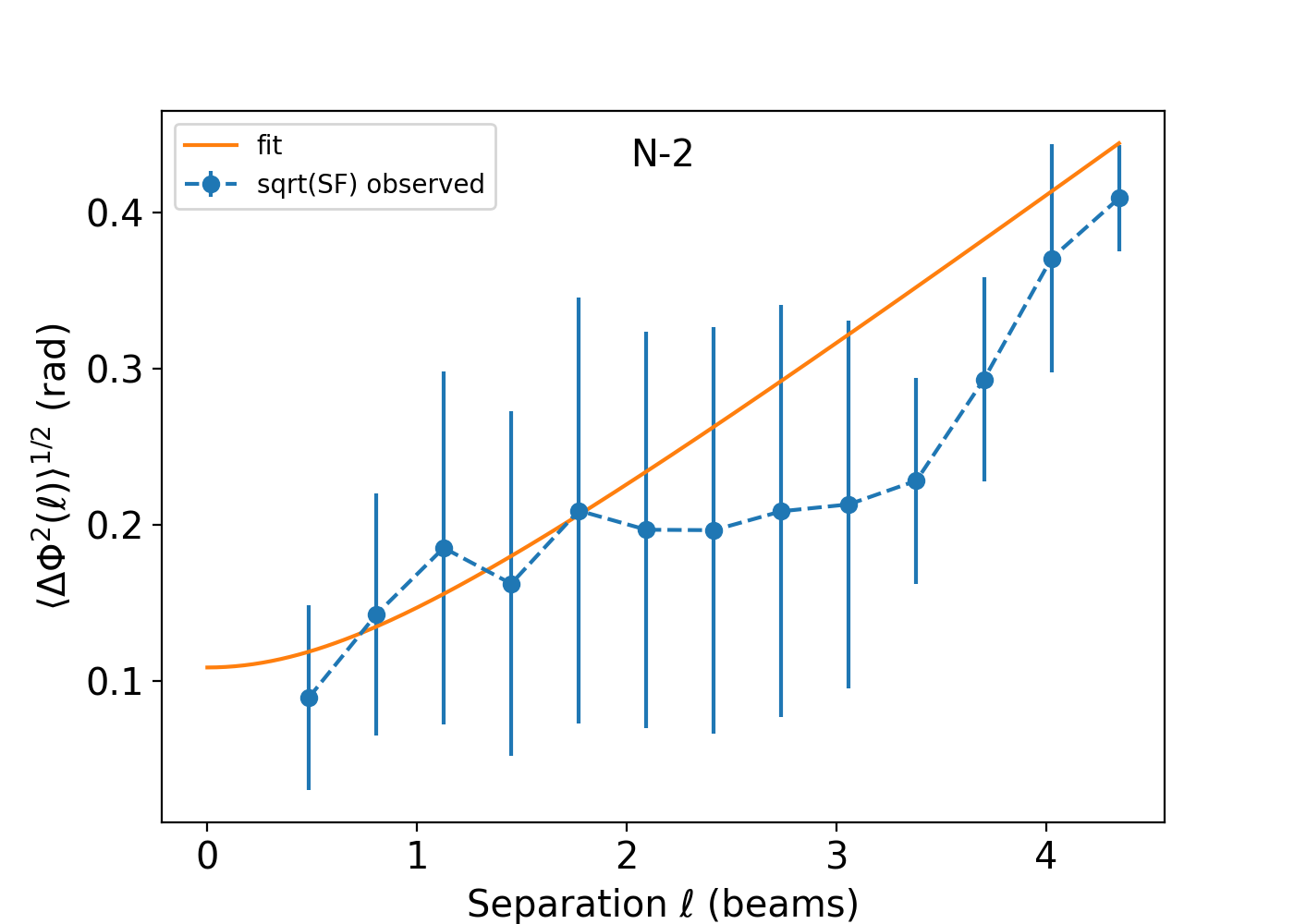}
    \includegraphics[width=0.25\linewidth]{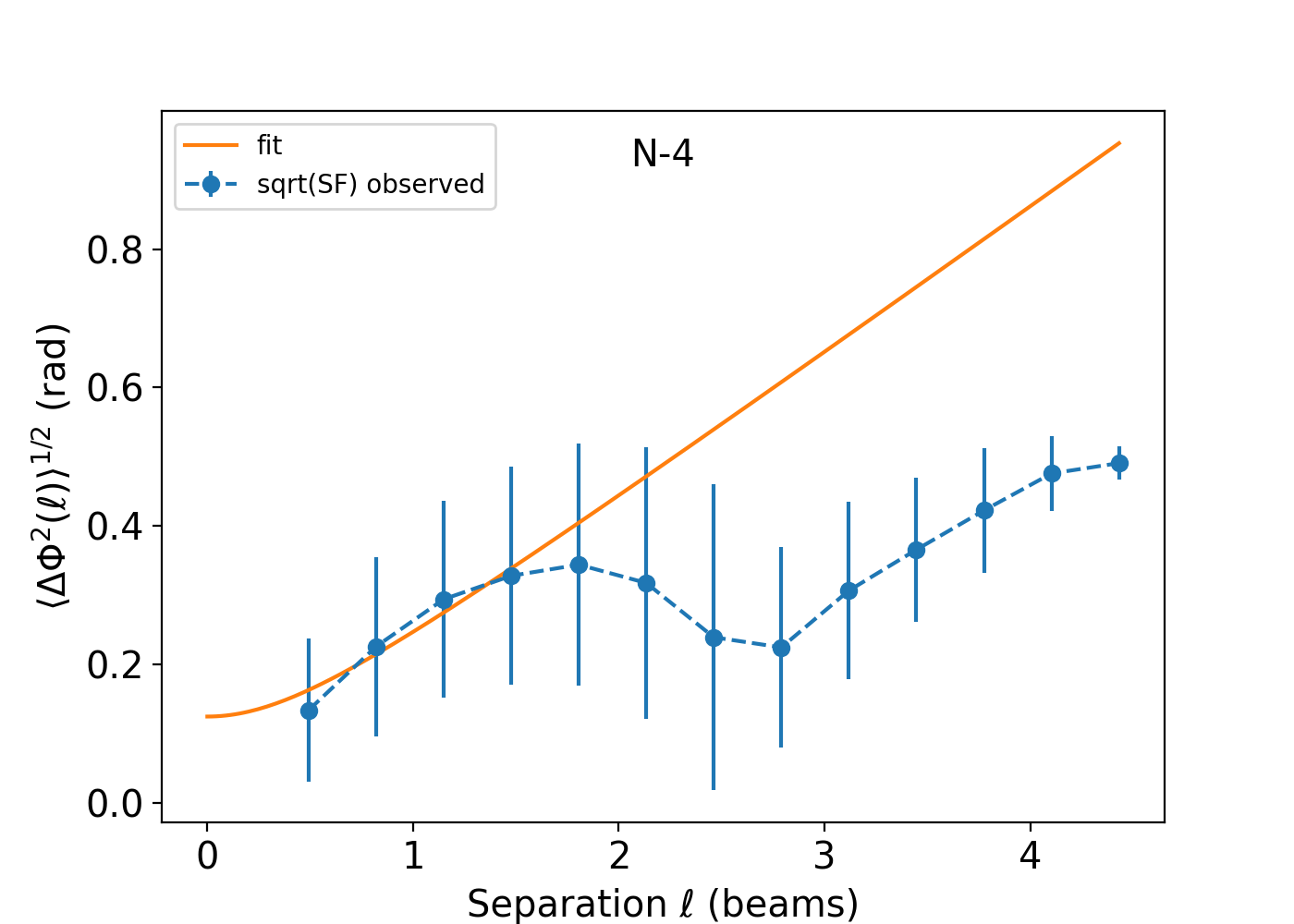}
    \includegraphics[width=0.25\linewidth]{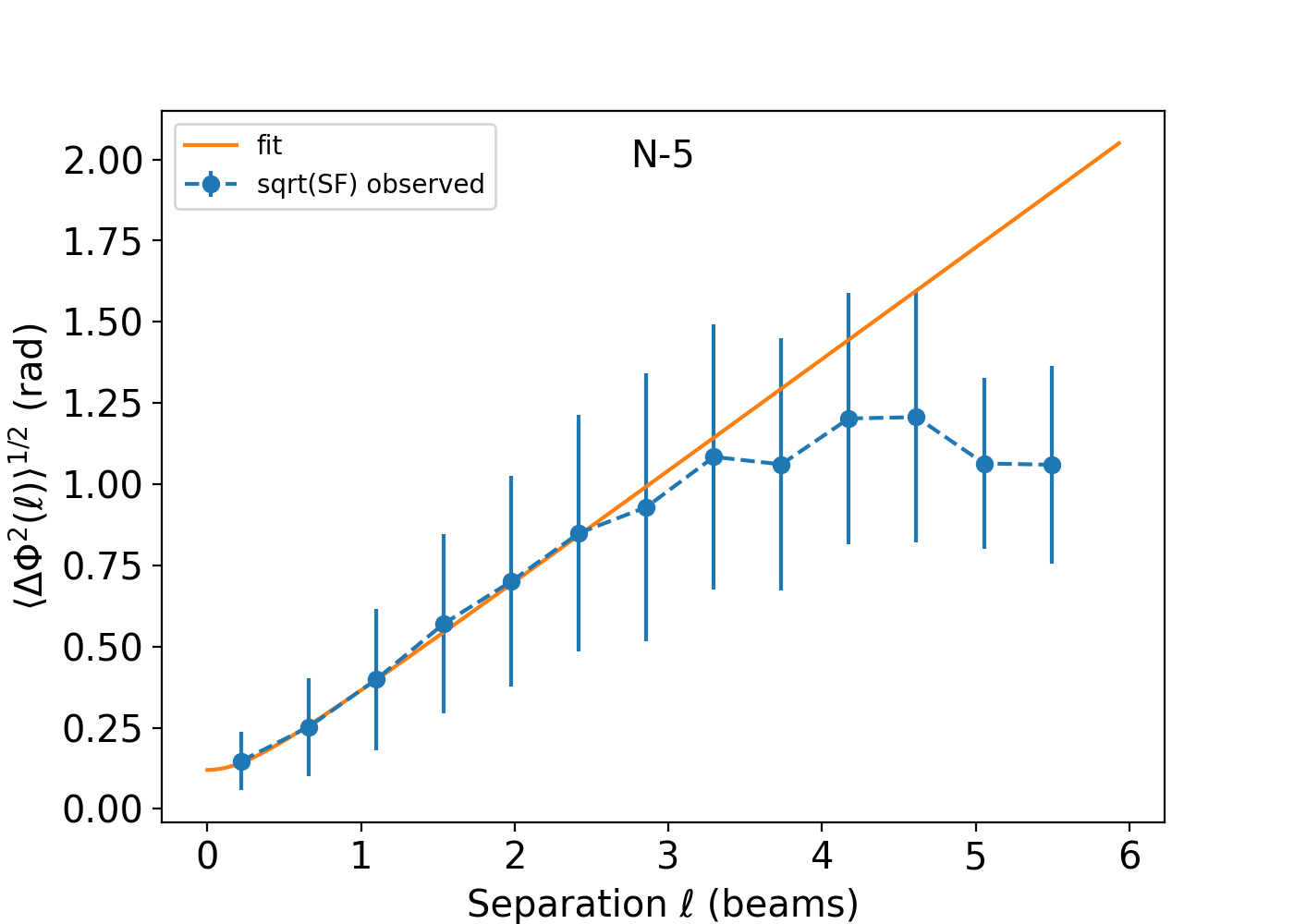}
    
    \includegraphics[width=0.25\linewidth]{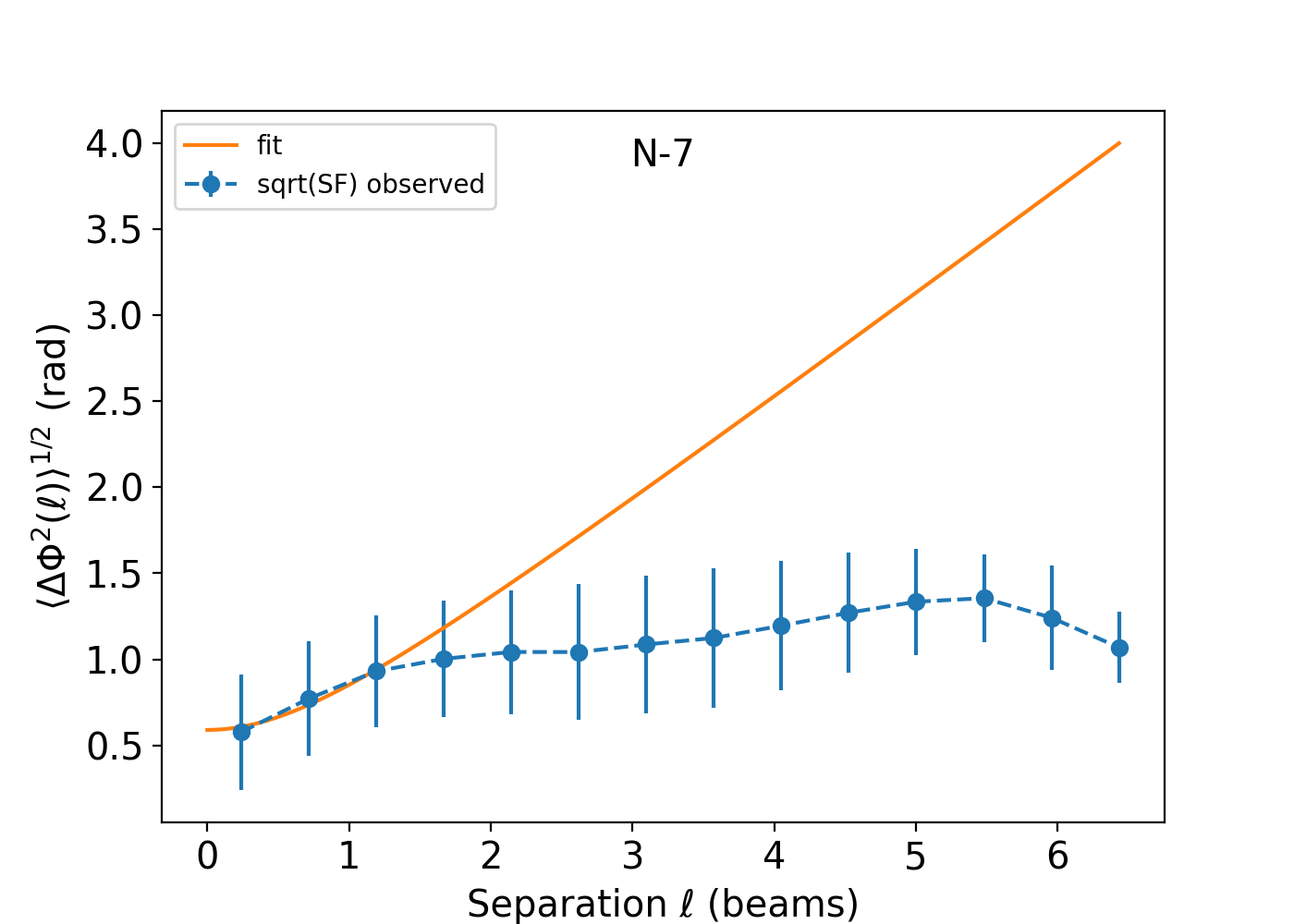}
    \includegraphics[width=0.25\linewidth]{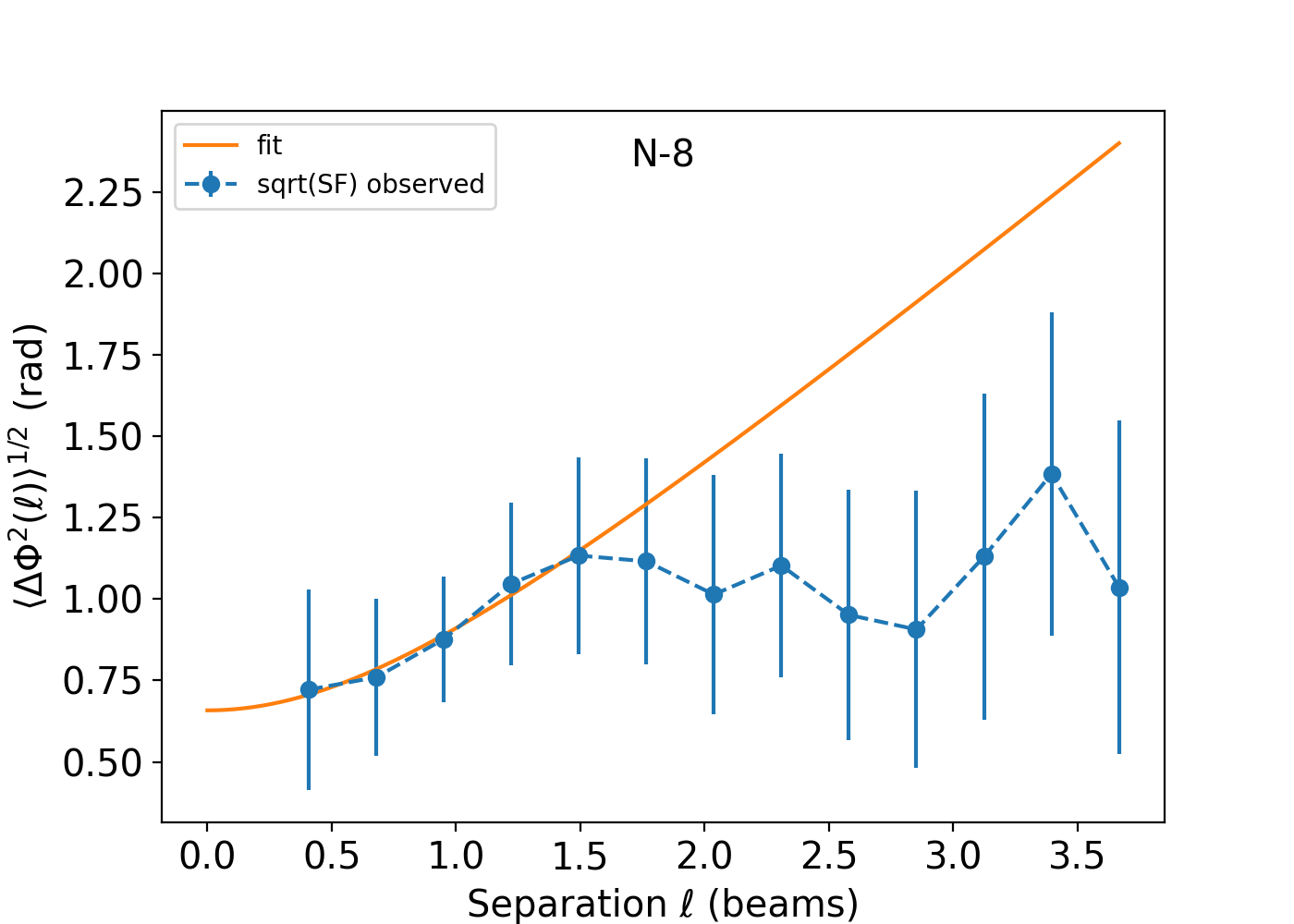}
    \caption{Structure functions and fits for regions N2, N4, N5, N7, and N8, where fittings are for l<1.5. The separation is expressed in terms of the beam size.  Error bars are the standard deviations of angle differences at a given distance. The fitted intercepts are 9.5$^{\circ}$, 17.2$^{\circ}$, 17.1$^{\circ}$, 53.4$^{\circ}$ and 48.1$^{\circ}$, respectively.}
    \label{fig:sfN_fitting}
\end{figure*}

\begin{figure*}
    \centering
    \includegraphics[width=0.24\linewidth]{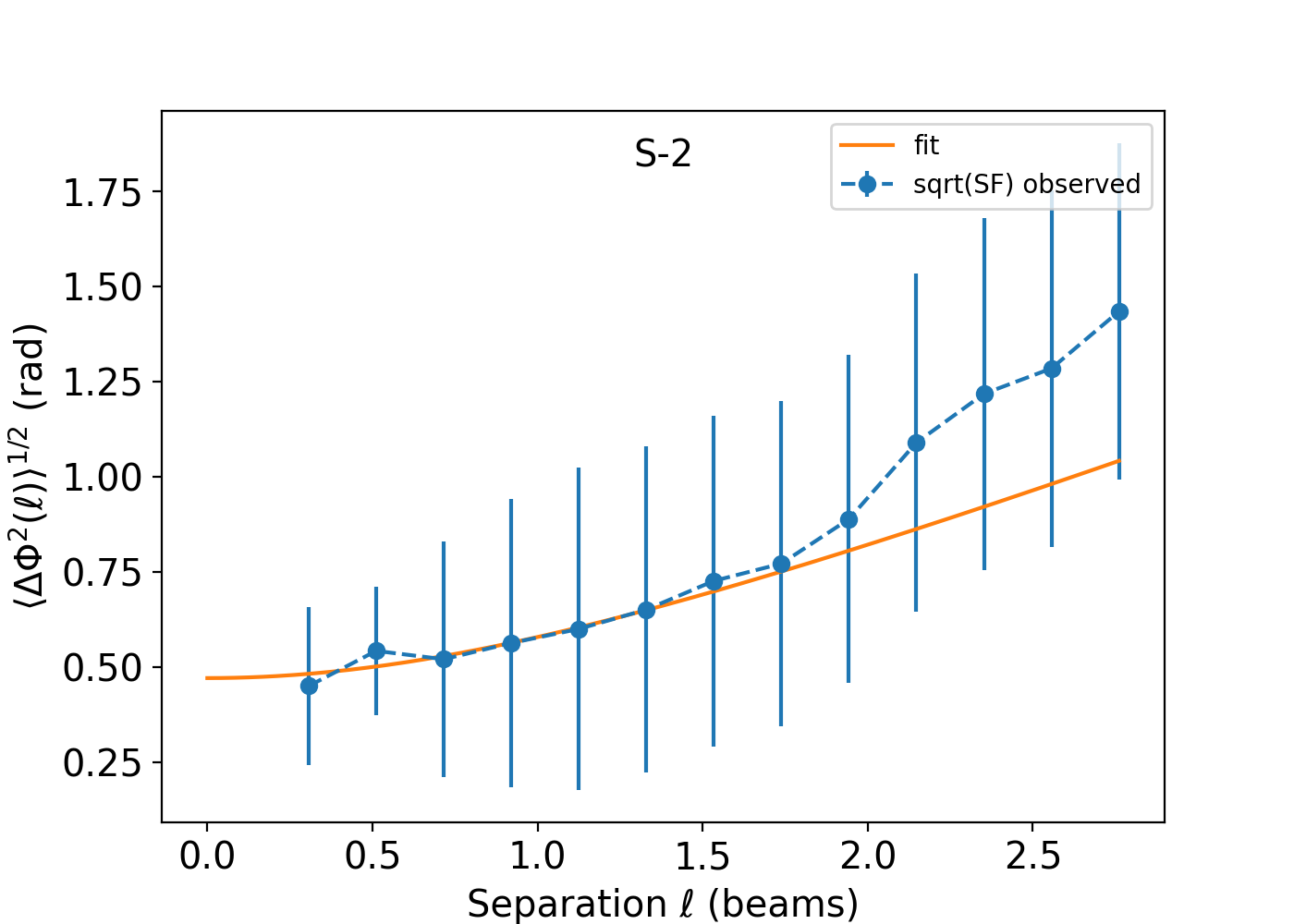}
    \includegraphics[width=0.24\linewidth]{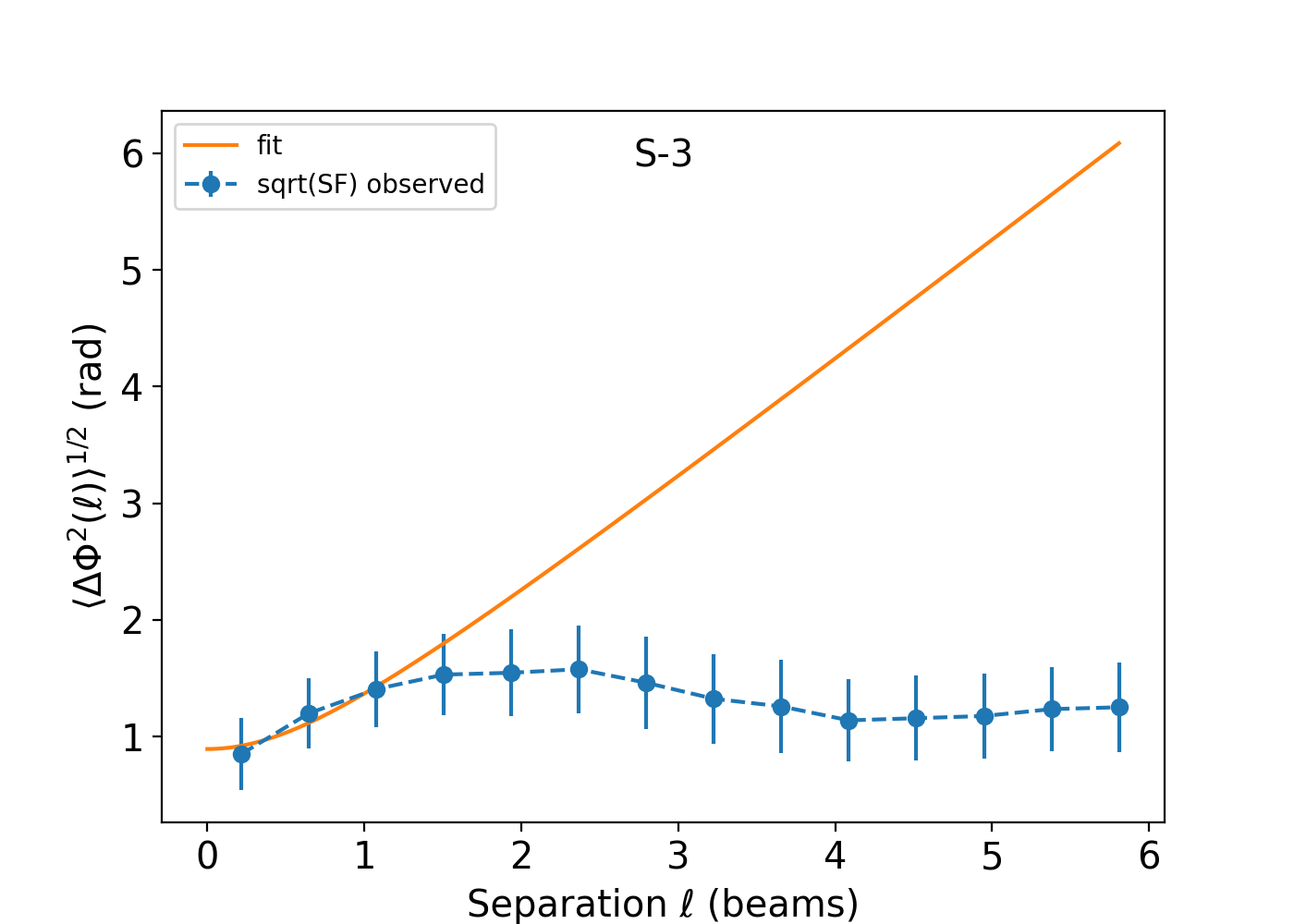}
    \includegraphics[width=0.24\linewidth]{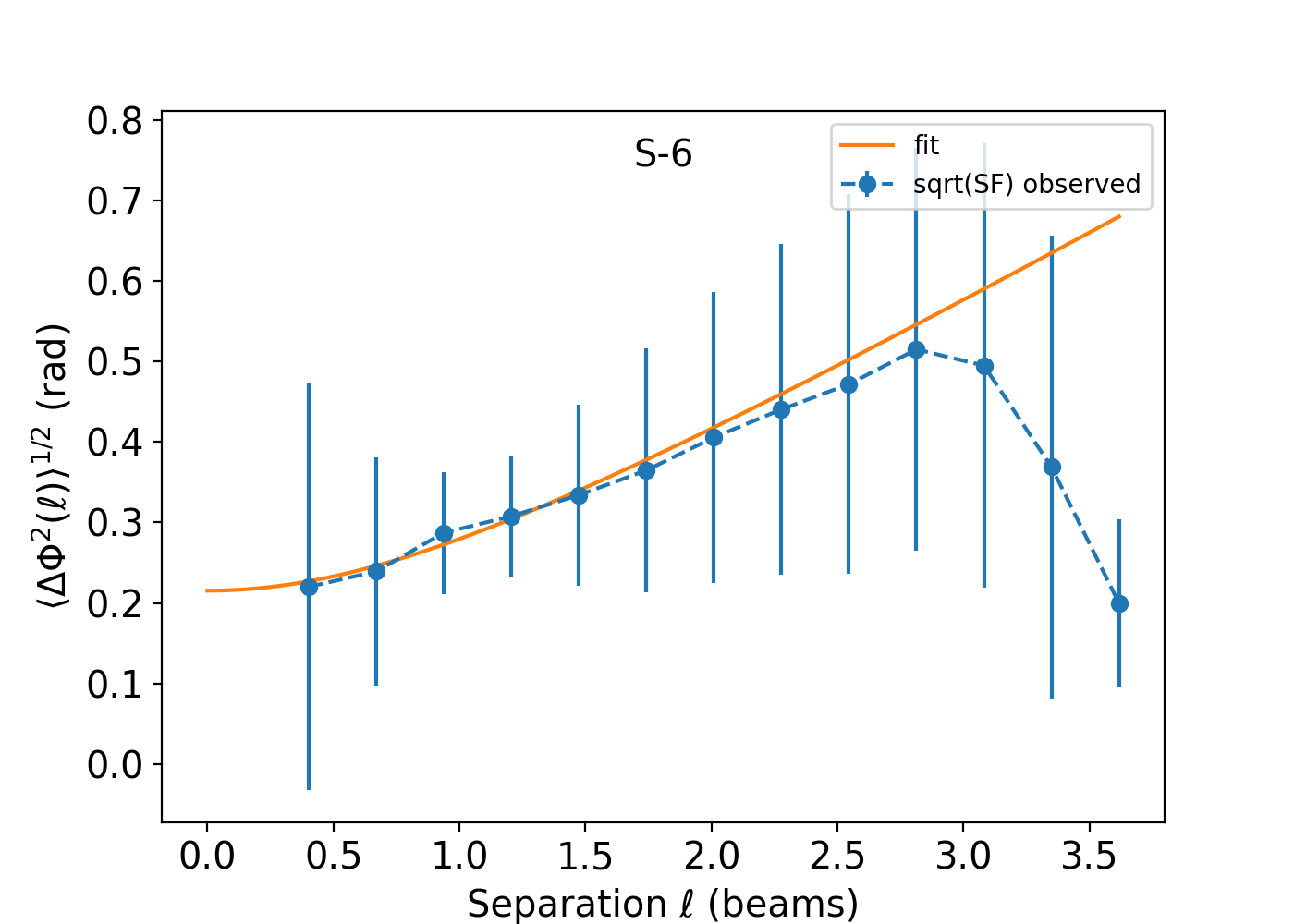}
    \includegraphics[width=0.24\linewidth]{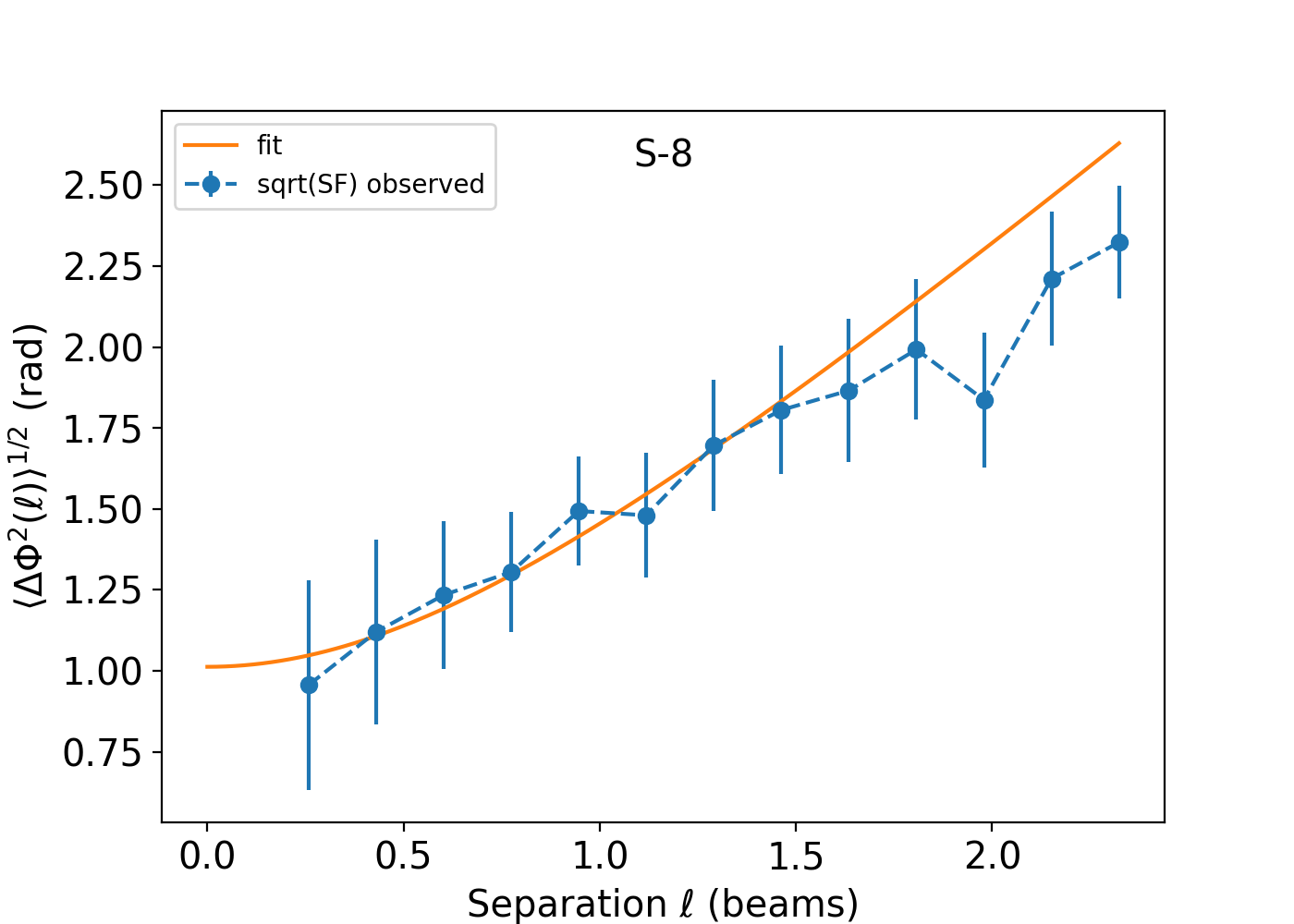}
    \caption{Structure functions and fits for regions S2, S3, S6, and S8, where fittings are for l<1.5. The separation is expressed in terms of the beam size.  Error bars are the standard deviations of angle differences at a given distance. The fitted intercepts are 26.9$^{\circ}$, 65.2$^{\circ}$, 31.7$^{\circ}$, and 57.7$^{\circ}$, respectively.}
    \label{fig:sfS_fitting}
\end{figure*}

\begin{figure}
    \centering
    \includegraphics[width=0.5\linewidth]{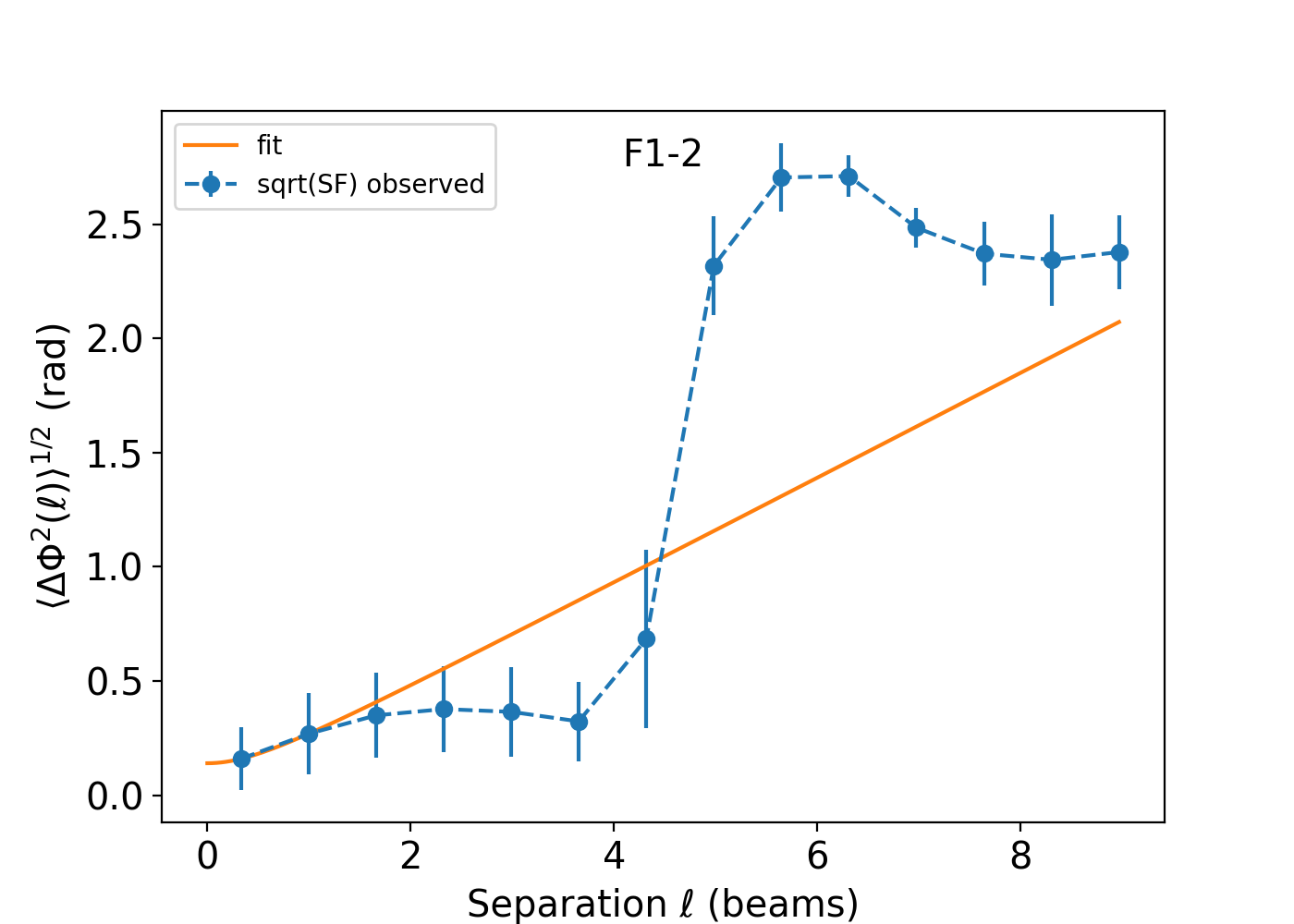}
    \caption{Structure functions and fits for regions F1-2, where fittings are for l<1.5. The separation is expressed in terms of the beam size.  Error bars are the standard deviations of angle differences at a given distance. The fitted intercept is 15$^{\circ}$.}
    \label{fig:sfF1_fitting}
\end{figure}

\end{appendix}

\end{document}